\documentclass[prd,twocolumn,showpacs,amsmath,amssymb]{revtex4}
\usepackage{bm}
\usepackage{graphicx}

\newcommand{\ud}{\mathrm{d}} 
\newcommand{\ui}{\mathrm{i}} 
\newcommand{\nb}{n_\mathrm{b}}

\newcommand{\GF}{G_\mathrm{F}}
\newcommand{\Hb}{\mathbf{H}}

\newcommand{\sB}{\bm \varsigma}
\newcommand{\eff}{\mathrm{eff}}
\newcommand{\HV}{\Hb_\mathrm{V}}
\newcommand{\muV}[1]{\mu_{\mathrm{V} #1}}
\newcommand{\HA}{\Hb_e}
\newcommand{\muA}{\mu_e}

\newcommand{\Esync}{E_\mathrm{sync}}
\newcommand{\rMSW}{r_\mathrm{MSW}}
\newcommand{\basef}[1]{{\bf \hat{e}}^\mathrm{f}_{#1}}

\newcommand{\basev}[1]{{\bf \hat{e}}^\mathrm{v}_{#1}}

\newcommand{\rX}{r_{\mathrm{X}}}
\newcommand{\rXo}{r_{\mathrm{X/MSW}}}
\newcommand{\rBS}{r_{\mathrm{BS}}}
\newcommand{\rBE}{r_{\mathrm{BE}}}
\newcommand{\rC}{r_{\mathrm{C}}}
\newcommand{\EC}{E_{\mathrm{C}}}
\newcommand{\EL}{E_{\mathrm{L}}}
\newcommand{\EM}{E_{\mathrm{M}}}
\newcommand{\EH}{E_{\mathrm{H}}}
\newcommand{\nuI}[1]{\nu_{\underline{#1}}}
\newcommand{\anuI}[1]{\bar\nu_{\underline{#1}}}
\newcommand{\avgE}[1]{\langle E_{\nuI{#1}} \rangle}
\newcommand{\avgaE}[1]{\langle E_{\anuI{#1}} \rangle}
\newcommand{\Pee}{P_{\nuI{e}\nu_e}}
\newcommand{\Paeae}{P_{\anuI{e}\bar\nu_e}}
\newcommand{\Ptt}{P_{\nuI{\tau}\nu_\tau}}
\newcommand{\tpb}{t_{\mathrm{PB}}}
\newcommand{\geoD}{D}

\newcommand{\myfigsep}{0.04 \textwidth}
\newcommand{\myfigwid}{0.48 \textwidth}

\begin{document}

\title{Simulation of Coherent Nonlinear Neutrino Flavor Transformation \\
in the Supernova Environment: Correlated Neutrino Trajectories}
\newcommand*{\UCSD}{Department of Physics, %
University of California, San Diego, %
La Jolla, CA 92093-0319}
\affiliation{\UCSD}
\newcommand*{\LANL}{Theoretical Division, Los Alamos National Laboratory, %
Los Alamos, NM 87545}
\affiliation{\LANL}
\newcommand*{\UMN}{School of Physics and Astronomy, %
University of Minnesota, Minneapolis, MN 55455}
\affiliation{\UMN}

\author{Huaiyu Duan}
\email{hduan@ucsd.edu}
\affiliation{\UCSD}
\author{George M.~Fuller}
\email{gfuller@ucsd.edu}
\affiliation{\UCSD}
\author{J.~Carlson}
\email{carlson@lanl.gov}
\affiliation{\LANL}
\author{Yong-Zhong Qian}
\email{qian@physics.umn.edu}
\affiliation{\UMN}

\date{\today}

\begin{abstract}
We present results of large-scale numerical simulations of the evolution
of neutrino and antineutrino flavors in the region above the late-time
post-supernova-explosion proto-neutron star. Our calculations are the first
to allow explicit flavor evolution histories on different neutrino
trajectories and to self-consistently couple flavor development on these
trajectories through forward scattering-induced quantum coupling.
Employing the atmospheric-scale neutrino mass-squared difference 
($|\delta m^2|\simeq 3\times 10^{-3}\,\mathrm{eV}^2$) and
values of $\theta_{13}$ allowed by current bounds, we find transformation
of neutrino and antineutrino flavors over broad ranges of energy and luminosity
in roughly the ``bi-polar'' collective mode. We find that 
this large-scale flavor
conversion, largely driven by the flavor off-diagonal neutrino-neutrino 
forward scattering potential, sets in much closer to the proto-neutron
star than simple estimates based on flavor-diagonal potentials and
Mikheyev-Smirnov-Wolfenstein evolution would indicate. In turn,
this suggests that models of \textit{r}-process nucleosynthesis sited
in the neutrino-driven wind could be affected substantially by
active-active neutrino flavor mixing, even with the small measured
neutrino mass-squared differences.
\end{abstract}
 
\pacs{14.60.Pq, 97.60.Bw}

\maketitle
                                                              
\section{Introduction%
\label{sec:introduction}}

In this paper we employ large-scale computational techniques to tackle
the vexing problem of neutrino flavor transformation
in the core collapse supernova environment. Neutrinos and the
weak interaction play pivotal roles in the core collapse/explosion
phenomenon. The Chandrasekhar mass core of iron-peak material
left at the end of the hydrostatic evolution of a massive
star goes dynamically unstable and collapses in $\sim 1$ s to
a proto-neutron star configuration at nuclear density.
The amount of
gravitational energy promptly converted into trapped seas of neutrinos
 is $\sim 1\%$ ($\sim 10^{52}$ erg) of the core mass.
Within a few
seconds after bounce 
 $\sim 10\%$ ($\sim 10^{53}$ erg) of the core mass (the gravitational binding
energy) will be emitted as neutrinos.

Nearly all of this gravitational energy is converted into
seas of $\nu_e$, $\bar\nu_e$, $\nu_\mu$, $\bar\nu_\mu$,
$\nu_\tau$ and $\bar\nu_\tau$ neutrinos in rough energy
equipartition. Though these neutrinos diffuse with short mean
free paths in the proto-neutron star, they decouple near the
stellar surface where the matter density falls off steeply,
the so-called neutrino sphere. Neutrinos propagate nearly
coherently above this point, though neutrino-matter interactions,
especially the charged current capture reactions
$\nu_e+n\rightarrow p+e^-$ and $\bar\nu_e+p\rightarrow n+e^+$,
can deposit energy and set the local neutron-to-proton 
ratio, $n/p$.

For this reason, and because the fluxes and energy spectra may
be different for $\nu_e$, $\bar\nu_e$ and 
$\nu_\mu \bar\nu_\mu \nu_\tau \bar\nu_\tau$,
the flavor content of the neutrino field
above the proto-neutron star and its evolution in time
and space can be important
\cite{Fuller:1992aa,Qian:1993dg,Qian:1994wh}. 
This can be true both for the supernova shock
reheating epoch (where the time post core-bounce
is $\tpb\lesssim 0.5$ s) and in the later hot bubble,
neutrino-driven wind epoch ($\tpb\gtrsim 3$ s). In this paper
we concentrate on the latter epoch.

Following the development of neutrino and antineutrino
flavors in the coherent regime above the proto-neutron star
surface is challenging. The potential governing the 
effective neutrino mass differences in this environment
will have contributions from charged current neutrino-electron
forward scattering and neutral-current neutrino-neutrino
forward scattering. The former contribution \cite{Wolfenstein:1977ue}
is diagonal in the flavor basis, while the latter
neutrino-neutrino potential has both flavor-diagonal
\cite{Fuller:1987aa} and flavor off-diagonal
\cite{Pantaleone:1992xh,Sigl:1992fn} components.
The neutrino-neutrino forward scattering potential
renders the neutrino flavor evolution problem nonlinear
in the sense that the potential which governs neutrino
flavor transformation is itself dependent on the flavor
evolution histories of the neutrinos.

Furthermore, neutrinos propagating on intersecting world
lines can have their flavor evolution
subsequently quantum mechanically coupled by forward
scattering (see Fig.~8 in Ref.~\cite{Qian:1994wh} and the
text beneath it).
We sometimes will refer to this coupling as ``entanglement''.
By this terminology we do not mean \textit{quantum entanglement}
of momentum states, a phenomenon which has been argued to
be unimportant in the supernova environment \cite{Friedland:2006ke}.
In any case, neutrino trajectories coming off the proto-neutron
star surface at different angles in general will have different
flavor evolution histories which must be self-consistently
calculated.

Another issue revolves around the efficacy of a mean field
Schr\"{o}dinger-like or Boltzmann kinetic equation approach
to the evolution of neutrino flavors
\cite{Qian:1994wh,Bell:2003mg,Friedland:2003eh,Friedland:2006ke}.
In this paper we will assume that higher order correlations
in neutrino-neutrino scattering are unimportant in
the coherent regime above the proto-neutron star.

Previous attempts to model neutrino flavor evolution in the
coherent regime \cite{Fuller:1987aa,Fuller:1992aa,Qian:1993dg,Qian:1994wh,%
Qian:1995ua,Pastor:2002we,Balantekin:2004ug} have made the approximation
that all neutrinos evolve in flavor space the way a radially-propagating
neutrino does. This we will term the ``single-angle'' approximation.
Since the neutrino-neutrino forward scattering
potential is intersection angle dependent, this is not always
a good approximation, especially for regions close to the proto-neutron star.

However, these previous studies done with the single-angle approximation
have found that it is possible to have large-scale collective
behavior in neutrino flavor evolution,
where all, or some significant subset of, neutrinos experience similar
time/space flavor evolution histories. They also have
shown that neutrino flavor transformation can differ significantly
from the Mikheyev-Smirnov-Wolfenstein (MSW) 
\cite{Wolfenstein:1977ue,Wolfenstein:1979ni,Mikheyev:1985aa}
paradigm. Recent work \cite{Fuller:2005ae} has shown that the
expected neutrino fluxes in both the shock reheating
and hot bubble epochs could provide the ``necessary'' conditions
for large-scale simultaneous collective neutrino and antineutrino
flavor transformation over broad ranges of neutrino energy. Whether
these expected neutrino fluxes are actually ``sufficient'' to
obtain these collective modes has remained an open question,
to be answered with an appropriately sophisticated numerical
simulation. Likewise, the range of possible collective
neutrino behavior \cite{Duan:2005cp}, be it the ``synchronized''
mode \cite{Pastor:2001iu} or the ``bi-polar'' mode
\cite{Duan:2005cp}, may depend sensitively on the neutrino flux
conditions and on the geometry.

It should be noted that many previous numerical studies
employing the single-angle approximation have also
used unphysically large values of neutrino mass-squared difference.
This is because with straight MSW, and without taking account
of the neutrino-neutrino scattering-induced flavor off-diagonal
potential, it requires $|\delta m^2|\gtrsim 1\,\mathrm{eV}^2$
to have significant neutrino flavor transformation deep
enough in the supernova envelope to affect shock reheating
or the \textit{r}-process 
\cite{Fuller:1992aa,Qian:1993dg,Qian:1994wh,Sigl:1994hc,Mezzacappa:1999co}.

However, recent observations/experiments 
(see, \textit{e.g.}, Ref.~\cite{Fogli:2005cq} for a review)
have revealed much about the fundamental flavor mixing parameters
of the three known ``active'' neutrinos. (In this paper,
we will ignore the effects of speculative additional ``sterile''
neutrino states.) We know the two mass-squared differences,
the atmospheric scale, 
$\delta m^2_{\mathrm{atm}}\simeq 3\times 10^{-3}\,\mathrm{eV}^2$,
and the solar scale, 
$\delta m^2_\odot\simeq 8\times 10^{-5}\,\mathrm{eV}^2$.
We as yet do not know the neutrino mass hierarchy related
to the atmospheric mixing and we do not know
the absolute neutrino mass eigenvalues.
Of the four mixing parameters in the unitary transformation
between the flavor (weak interaction) eigenstates and
the mass (energy) eigenstates, we know two of the three vacuum mixing
angles, $\theta_{12}$ and $\theta_{23}$, and we have a firm
upper limit on $\theta_{13}$, $\sin^22\theta_{13}\lesssim 0.1$.
We do not know the \textit{CP}-violating phase.

In this paper we study $2\times2$ neutrino and antineutrino flavor
transformation at the $\delta m^2_{\mathrm{atm}}$ scale, explicitly
following the coupled flavor evolution on neutrino trajectories
ranging from radially-directed to those tangential to the neutron star
surface. (In other words, we perform ``multiangle'' calculations
with many trajectory/angle zones.) Our goal is to study the nonlinear
behavior of the neutrino field in the coherent regime and to find
out if large-scale (collective mode) neutrino/antineutrino transformation
can occur in the late-time supernova environment.
We specialize to late time for two reasons: (1) this epoch is when
there may be significant differences in flux or energy spectrum
between $\nu_e$, $\bar\nu_e$, and the mu and tau flavor neutrinos;
and (2) this epoch may have a simpler, more compact, matter density
profile near the neutron star surface. We follow 
Refs.~\cite{Balantekin:1999dx,Caldwell:1999zk} and argue that
$2\times 2$ mixing is adequate because the $\nu_\mu$ and $\nu_\tau$
neutrinos are nearly maximally-mixed in vacuum 
($\theta_{23}\simeq \pi/4$) and these species experience nearly
identical interactions everywhere in the late-time supernova environment.

In Sec.~\ref{sec:background} we summarize the physical and geometric
assumptions in our numerical simulations in what we call
the ``neutrino bulb model''. In this section we also present the basic physics
of neutrino flavor transformation in the practical formalism 
used in our numerical simulations. We also review the spin analogy
for neutrino flavors, and estimate the (MSW) resonance locations
in the hot bubble using
both the standard MSW and synchronization mechanisms.
In Sec.~\ref{sec:multiangle} we explain some details of
our numerical codes and discuss the particular numerical difficulties 
and potential pitfalls in multiangle simulations. We also
present the main results of
our multiangle simulations. The simulations show large-scale
flavor transformation
different from what would be predicted if the conventional MSW or 
synchronization mechanisms apply.
In Sec.~\ref{sec:analysis} we identify the flavor transformation
in our results as being of  the bi-polar
type \cite{Duan:2005cp}, and we analyze this behavior 
with the help of single-angle simulations. 
In Sec.~\ref{sec:conclusions} we give our conclusions.

\section{Background Physics%
\label{sec:background}}
\subsection{Neutrino Bulb Model%
\label{sec:nu-bulb}}
At $\tpb\gtrsim 3$ s, the inner core of the progenitor star has settled
down into a proto-neutron star with a radius of about 10 km. In the following
$\sim 10$ s, the nascent neutron star radiates away its 
gravitational binding energy as outlined above. 
During this time, neutrinos could deposit energy into the
matter above the neutron star and create a high-entropy ``hot bubble''
between the proto-neutron star surface and the shock. 
Inside the hot bubble, a quasistatic and near adiabatic
mass outflow, the so-called ``neutrino-driven wind'', may be established
at this epoch as a result of neutrino/antineutrino heating
\cite{Duncan:1986aa,Qian:1996xt}.
To simplify the numerical
calculations of the flavor transformations of neutrinos and antineutrinos inside
the hot bubble, we approximate the physical and geometric conditions
of the post-shock supernova by a ``neutrino bulb model''.
This model is characterized by the following assumptions:
\begin{enumerate}
\item\label{it:nubulb-1} 
The neutron star emits neutrinos uniformly and isotropically
from the surface of a sphere (neutrino sphere) of radius $R_\nu$;
[Note that the neutrino flux emitted at angle $\vartheta_0$ with
respect to the normal direction at the neutrino sphere comes with a
geometric factor $\cos\vartheta_0$. See Eq.~\protect{\eqref{eq:dnnu}}.]
\item\label{it:nubulb-2} 
At any point outside the neutrino sphere, the physical conditions, 
such as baryon density $\nb$, temperature $T$, \textit{etc.}, 
depend only on the distance $r$ from this point to the center of the neutron
star;
\item Neutrinos are emitted from the neutron star surface
in pure flavor eigenstates and with Fermi-Dirac 
type energy spectra.
\end{enumerate}

\begin{figure}
\begin{center}
\includegraphics*[width=\myfigwid, keepaspectratio]{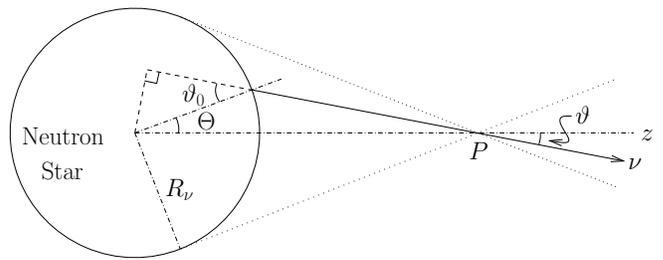}
\end{center}
\caption{\label{fig:nubulb}%
The geometric picture of the neutrino bulb model. 
An arbitrary neutrino beam (solid line) is shown emanating from 
a point on the neutrino sphere
with polar angle $\Theta$. This beam intersects the $z$--axis
at point $P$ with angle $\vartheta$. Because
neutrinos are emitted from the neutrino sphere of radius $R_\nu$,
point $P$ sees only neutrinos traveling
within the cone delimited by the dotted lines.
One of the most important geometric characteristics of a neutrino beam is
its emission angle $\vartheta_0$, defined with respect to
the normal direction at the point of emission on the neutrino sphere
($\vartheta_0=\Theta+\vartheta$). 
All other geometric properties of a neutrino
beam may be calculated using radius $r$ and $\vartheta_0$.}
\end{figure}

The neutrino bulb model, as illustrated in Fig.~\ref{fig:nubulb}, has
multifold symmetries. It is clearly spherically symmetric.
This means that one only
need study the physical conditions at a series of points along one
radial direction, which we choose to be the $z$--axis. It is also obvious
that the neutrino flux seen at any given point on the $z$--axis has a
cylindrical symmetry. As a result, different neutrino beams possessing the same
polar angle with respect to the $z$--axis
and with the same initial physical properties (flavor, energy, \textit{etc.})
should be completely equivalent. In other words, they will have
identical flavor evolution histories. One may
choose this polar angle to be $\vartheta$, the angle between the
direction of the beam and the $z$--axis. Alternatively,
a beam could be specified by the polar angle $\Theta$
giving the emission position of the beam on the neutrino sphere
(see Fig.~\ref{fig:nubulb}). 
A third option, which we have found to be most useful in our
numerical calculations, is to label the beam by emission
angle $\vartheta_0$. This is defined to be the angle
with respect to the normal direction at the point
of emission on  the neutrino sphere (see Fig.~\ref{fig:nubulb}).
This emission angle $\vartheta_0$ is an intrinsic
geometric property of the beam, and does not vary along the 
neutrino trajectory. Moreover, because of assumptions
\ref{it:nubulb-1} and \ref{it:nubulb-2} in the neutrino bulb model,
all the neutrino beams with the same emission
angle $\vartheta_0$ and the same initial physical properties
must be equivalent. In simulating the flavor transformations
of neutrinos in the neutrino bulb model, 
it is only necessary  to follow a
group of neutrinos which are uniquely indexed by their initial flavors, 
energies and emission angles.

At any given radius $r$, all the geometric properties of a neutrino beam
may be calculated using $r$ and $\vartheta_0$. For example, $\vartheta$ and
$\Theta$ are related $\vartheta_0$ through the following identity:
\begin{equation}
\frac{\sin\vartheta}{R_\nu} = \frac{\sin\Theta}{l - l_0} =
\frac{\sin\vartheta_0}{r},
\label{eq:equal-ratios}
\end{equation}
where
\begin{equation}
l \equiv r \cos\vartheta,
\label{eq:prop-distance}
\end{equation}
and
\begin{equation}
l_0 \equiv  R_\nu \cos\vartheta_0.
\end{equation}
Length $l - l_0$ in Eq.~\eqref{eq:equal-ratios} is also the total
propagation distance along the neutrino beam. At a point at radius $r$,
the neutrino beams are restricted to be within a cone of half-angle
\begin{equation}
\vartheta_{\max} = \arcsin \left(\frac{R_\nu}{r}\right)
\label{eq:max-theta}
\end{equation}
(see Fig.~\ref{fig:nubulb}). 

One can integrate flux over all neutrino beams (angles) 
and calculate the neutrino number density $n_\nu$ at radius $r$.
In this paper we use the symbol $\nu$ in the general sense, 
denoting either a neutrino or an antineutrino.
We use $\nu_\alpha$ ($\bar\nu_\alpha$) to denote a neutrino
(antineutrino) in flavor state $\alpha$, and 
$\nuI{\alpha}$ ($\anuI{\alpha}$) to denote a neutrino
(antineutrino) created at the neutrino sphere \textit{initially}
 in flavor state $\alpha$.
As an example, we shall calculate the differential number density
$\ud n_{\nuI{\alpha}}(\mathbf{q})$ at radius $r$: this
will have contributions from all
$\nuI{\alpha}$ with energy $q$
which propagate in directions within the range between 
${\bf \hat{q}}$ and ${\bf \hat{q}}+\ud {\bf \hat{q}}$.
Here a hatted vector ${\bf \hat{n}}$ denotes the direction
of vector $\mathbf{n}$, and is defined as 
${\bf \hat{n}}\equiv\mathbf{n}/|\mathbf{n}|$.
The differential number density
$\ud n_{\anuI{\alpha}}(\mathbf{q})$ of $\anuI{\alpha}$ can
be calculated in a similar way.  One finds that
\begin{subequations}
\label{eq:dnnu}
\begin{align}
\ud n_{\nuI{\alpha}}(\mathbf{q}) 
&= \frac{j_{\nuI{\alpha}}(q) 
\cos\vartheta_0 R_\nu^2 \,\ud(\cos\Theta)\ud\Phi}{(l-l_0)^2} 
\label{eq:dnnu-Theta}\\
&= j_{\nuI{\alpha}}(q)
\,\ud(\cos\vartheta)\ud\phi,
\label{eq:dnnu-theta}
\end{align}
\end{subequations}
where the velocity of a neutrino is taken to be the speed of light
($c=1$), and $j_{\nuI{\alpha}}(q)$ is the number flux of 
$\nu_{\alpha}$ with energy $q$
emitted in any direction at the neutrino sphere.
In Eq.~\eqref{eq:dnnu-Theta}, $R_\nu^2 \,\ud(\cos\Theta)\ud\Phi$ is
the differential area on the neutrino sphere which emits neutrinos
in the directions  within the range between 
${\bf \hat{q}}$ and ${\bf \hat{q}}+\ud {\bf \hat{q}}$,
and the factor $(l-l_0)^{-2}$ accounts for
the geometric dilution of the neutrino density.
In  Eq.~\eqref{eq:dnnu-theta}, $\vartheta$ and $\phi$ are the
polar and azimuthal angles of $\mathbf{q}$ and, in deriving
the equation, we have used 
Eq.~\eqref{eq:equal-ratios} and the identities
\begin{align}
\ud \Phi &= \ud \phi ,\\ 
\cos\vartheta_0 R_\nu \,\ud\Theta &= (l-l_0) \,\ud\vartheta.
\end{align}

As an added check on Eq.~\eqref{eq:dnnu}, note that
the total number of $\nuI{\alpha}$ with energy $q$
passing through the sphere of radius $r$ per unit time is
\begin{subequations}
\begin{align}
&4\pi r^2 \int \cos\vartheta \, \ud n_{\nuI{\alpha}}(\mathbf{q})
\nonumber\\
& = 8\pi^2 r^2 j_{\nuI{\alpha}}(q)
\int_{\cos\vartheta_{\max}}^1 \cos\vartheta \, \ud(\cos\vartheta)\\
& = 4\pi^2 R_\nu^2 j_{\nuI{\alpha}}(q).
\end{align}
\end{subequations}
This is indeed equal to the number of $\nu_\alpha$
with energy $q$ emitted per unit time from the neutrino sphere, 
\begin{equation}
4\pi R_\nu^2 \int_0^1 2\pi j_{\nuI{\alpha}}(q) \cos\vartheta_0 
\,\ud(\cos\vartheta_0)
= 4\pi^2 R_\nu^2 j_{\nuI{\alpha}}(q).
\end{equation}
We note that this flux can also be expressed as
\begin{equation}
4\pi R_\nu^2 \int_0^1 2\pi j_{\nuI{\alpha}}(q) \cos\vartheta_0 
\,\ud(\cos\vartheta_0)
=\frac{L_{\nuI{\alpha}}}{\avgE{\alpha}} f_{\nuI{\alpha}}(q),
\end{equation}
where $L_{\nuI{\alpha}}$, $\avgE{\alpha}$ and  $f_{\nuI{\alpha}}(q)$
are the energy luminosity, average energy and  normalized energy distribution
function of $\nuI{\alpha}$, respectively. Therefore one has
\begin{equation}
j_{\nuI{\alpha}}(q)=\frac{L_{\nuI{\alpha}}}%
{4\pi^2 R_\nu^2\avgE{\alpha}} f_{\nuI{\alpha}}(q).
\label{eq:Lj}
\end{equation}
We take $f_\nu(q)$
to be of the Fermi-Dirac form with two parameters $(T_\nu, \eta_\nu)$,
\begin{equation}
f_\nu (q) \equiv \frac{1}{F_2(\eta_\nu)} \frac{1}{T_\nu^3}
\frac{q^2}{\exp(q/T_\nu-\eta_\nu)+1},
\end{equation}
where $\eta_\nu$ is the degeneracy parameter, $T_\nu$ is the
 neutrino temperature, and
\begin{equation}
F_k (\eta) \equiv \int_0^\infty \frac{x^k \,\ud x}{\exp(x-\eta)+1}.
\end{equation}
For numerical calculations, we will take $\avgE{e} = 11$ MeV, 
$\avgaE{e} = 16$ MeV,
$\avgE{\mu} = \avgaE{\mu} = \avgE{\tau} = \avgaE{\tau} = 25$ MeV, 
and $\eta_{\nuI{e}}=\eta_{\anuI{e}}=\eta_{\nuI{\mu}}=\eta_{\anuI{\mu}}=
\eta_{\nuI{\tau}}=\eta_{\anuI{\tau}}=3$.
With these choices, we have
$T_{\nuI{e}}\simeq 2.76$ MeV, $T_{\anuI{e}} \simeq 4.01$ MeV,
and $T_{\nuI{\mu}} = T_{\anuI{\mu}} = T_{\nuI{\tau}} = T_{\anuI{\tau}}\simeq 6.26$ 
MeV.

In principle, one could use the profiles of baryon density $\nb$,
temperature $T$ and electron fraction $Y_e$ (net number of electrons
per baryon) obtained from numerical
simulations of core collapse supernovae. 
Here we will use a simple analytical density 
profile, and approximate the envelope
above the neutron star as a quasistatic
configuration with a constant entropy per baryon $S$
(see, \textit{e.g.}, Ref.~\cite{Fuller:2005ae}).
Taking the enthalpy per baryon, $TS$, as roughly the gravitational
binding energy of a baryon, one has the following temperature profile
\begin{equation}
T \simeq \frac{M_\mathrm{NS} m_N}{m_\mathrm{Pl}^2} S^{-1} r^{-1},
\label{eq:T-profile}
\end{equation}
where $M_\mathrm{NS}$ is the mass of the neutron star, $m_N$ is
the mass of a nucleon, and 
$m_\mathrm{Pl}\simeq 1.221\times 10^{22}\,\mathrm{MeV}$
is the Planck mass. 
We assume that the entropy per baryon $S$ in the hot bubble is
dominated by relativistic degrees of freedom, 
\begin{equation}
S \simeq \frac{2\pi^2}{45}g_\mathrm{s}\frac{T^3}{\nb},
\label{eq:entropy}
\end{equation}
where we have taken the Boltzmann constant $k_\mathrm{B}$ and
the reduced Planck constant $\hbar$ both to equal 1,
and $g_\mathrm{s}$ is the statistical weight in relativistic particles.
Combining Eqs.~\eqref{eq:T-profile} and \eqref{eq:entropy}, one
obtains the baryon density profile as
\begin{subequations}
\label{eq:nb-profile}
\begin{align}
\nb &\simeq \frac{2\pi^2}{45} g_\mathrm{s} 
\left(\frac{M_\mathrm{NS} m_N}{m_\mathrm{Pl}^2}\right)^3 S^{-4} r^{-3}\\
&\simeq (4.2\times 10^{30} \,\mathrm{cm}^{-3})
g_\mathrm{s} \left(\frac{M_{\mathrm{NS}}}{1.4 M_\odot}\right)^3
\left(\frac{100}{S}\right)^{4}
\left(\frac{10\,\mathrm{km}}{r}\right)^{3}
\end{align}
\end{subequations}

\begin{figure}
\begin{center}
\includegraphics*[width=\myfigwid, keepaspectratio]{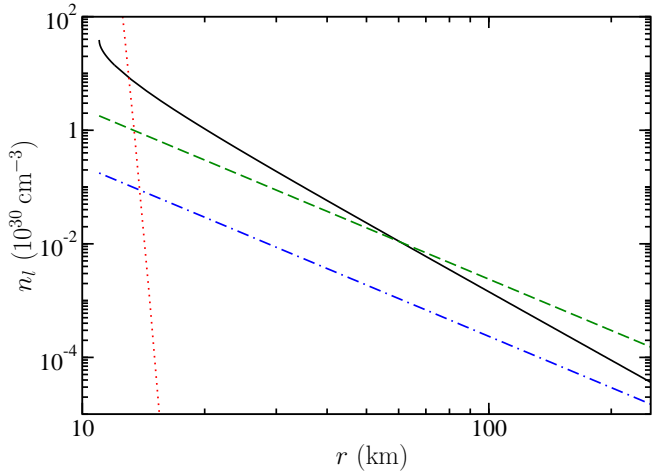}
\end{center}
\caption{\label{fig:nl-r} (Color online)
Plot of (effective) net lepton number density $n_l(r)$. 
The dashed and dot-dashed
lines are for the net electron density $n_e=Y_e n_b$ using the baryon density
profile in Eq.~\eqref{eq:nb-profile} with $S=140$ and 250, respectively.
The dotted line is for the net electron density assuming the baryon density
profile in Eq.~\eqref{eq:nb-profile-exp} only. The solid line is for the
effective net $\nuI{e}$ density along the radial trajectory
[Eq.~\eqref{eq:nue-eff}].}
\end{figure}

However, we note that, in reality, the baryon density $\nb$ near the
neutrino sphere is much higher than
that estimated from Eq.~\eqref{eq:nb-profile}.
In fact, near the neutrino sphere the density profile is better
represented by
\begin{equation}
\nb^\prime \simeq n_{\mathrm{b}0} \exp\left(-\frac{r-R_\nu}{h_{\mathrm{NS}}}\right),
\label{eq:nb-profile-exp}
\end{equation}
where $n_{\mathrm{b}0}$ is the baryon density at the neutrino sphere, and
\begin{subequations}
\begin{align}
h_{\mathrm{NS}} &\simeq R_\nu^2 
\left(\frac{ m_\mathrm{Pl}^2}{M_\mathrm{NS} m_N} \right)
T_{\mathrm{matt}}(R_\nu)\\
&\simeq (0.052\,\mathrm{km})\left(\frac{R_\nu}{10\,\mathrm{km}}\right)^2
\left(\frac{1.4 M_\odot}{M_{\mathrm{NS}}}\right)
\left[\frac{T_{\mathrm{matt}}(R_\nu)}{1\,\mathrm{MeV}}\right]
\end{align}
\end{subequations}
is the scale height with $T_{\mathrm{matt}}(R_\nu)$
being the matter temperature at the neutrino sphere.
This exponential fall-off in density is expected on
general physical grounds and is found in, \textit{e.g.},
the Mayle and Wilson supernova simulations
\cite{Qian:1993dg}. As discussed in Refs. \cite{Burrows:1982aa,Qian:1996xt},
a steady state between neutrino heating and cooling results in near
isothermal conditions in the vicinity of the neutron
star surface. This, coupled with the expected
very low electron fraction $Y_e$ near the neutron star surface,
implies that the baryon density must have this exponential
dependence on radius, at least for a radius interval $\sim h_\mathrm{NS}$.

It turns out that addition of this exponential density
profile near the neutrino sphere
facilitates the multiangle simulations of neutrino
flavor transformation. 
In Fig.~\ref{fig:nl-r} we plot the net electron number density
\begin{equation}
n_e = Y_e \nb
\end{equation}
obtained from the exponential profile in Eq.~\eqref{eq:nb-profile-exp}.
For comparison, we also plot $n_e(r)$ obtained
from the constant entropy profile [Eq.~\eqref{eq:nb-profile}] with
entropy per baryon $S=140$ and 250.
In both Fig.~\ref{fig:nl-r}
and in the rest of the paper, we take $M_{\mathrm{NS}}=1.4\,M_\odot$,
$R_\nu=11$ km, $Y_e=0.4$, $g_\mathrm{s}=11/2$,
$n_{\mathrm{b}0}=1.63\times 10^{36}\,\mathrm{cm}^{-3}$ and
$h_{\mathrm{NS}}=0.18\,\mathrm{km}$.
Note that once we have specified $n_{\mathrm{b}0}$  and $h_{\mathrm{NS}}$
our model for  the physical environment 
in the hot bubble is completely determined by the
choice of  entropy per baryon $S$. 
In units of Boltzmann constant
per baryon, we expect $S\sim 100$ in the hot bubble
\cite{Qian:1996xt}. 

\subsection{Neutrino Flavor Transformation in Supernovae%
\label{sec:flavor-transformation}}

Our objective is to study the flavor evolution of the neutrino
field when $\nu_e$ and
$\bar\nu_e$ mix with neutrinos and antineutrinos of another active
flavor (say $\nu_\tau$ and $\bar\nu_\tau$). We write the wave function
of the flavor doublet of a neutrino (or antineutrino) as
\begin{equation}
\psi_\nu = 
\left(\begin{array}{c} a \\ b \end{array}\right),
\label{eq:wv-func}
\end{equation}
where $a$ and $b$ are the amplitudes for a neutrino
to be in the $\nu_e$ ($\bar\nu_e$) and $\nu_\tau$ ($\bar\nu_\tau$)
flavor states, respectively.
The flavor evolution of $\psi_{\nuI{\alpha}}$ is determined by
the Schr\"{o}dinger equation (see, \textit{e.g.}, Ref.~\cite{Qian:1994wh})
\begin{widetext}
\begin{equation}
\label{eq:schroedinger-eq}
\ui \frac{\ud}{\ud t} \psi_{\nuI{\alpha}} = H \psi_{\nuI{\alpha}} 
= \frac{1}{2} \left(\begin{array}{cc}
-\Delta\cos 2\theta + A + B & \Delta\sin 2\theta + B_{e\tau} \\
\Delta\sin 2\theta + B_{e\tau}^* & \Delta\cos 2\theta - A - B
\end{array}\right) \psi_{\nuI{\alpha}},
\end{equation}
\end{widetext}
where $\theta$ is the vacuum mixing angle, $\Delta$, $A$ and 
$B_{(e\tau)}$ are the potentials induced by neutrino mass difference,
matter, and background neutrinos, respectively. 
One obtains the appropriate Hamiltonian for antineutrinos
by making the transformation
\begin{equation}
A \longrightarrow -A,\qquad
B \longrightarrow -B,\qquad
B_{e\tau} \longrightarrow -B_{e\tau}^*.
\end{equation} 

The vacuum potential is defined as
\begin{equation}
\Delta \equiv \frac{\delta m^2}{2E_\nu},
\end{equation}
where $\delta m^2$ is the neutrino mass-squared difference, and $E_\nu$
is the energy of the neutrino. (Note that we also use $q$ as the
energy or the magnitude of the momentum of a neutrino in this section,
which is the same as $E_\nu$.)
We define the mass-squared difference in terms of the appropriate
neutrino mass eigenvalues $m_1$ and $m_3$ to be $\delta m^2\equiv m_3^2-m_1^2$.
In what follows we employ the normal ($\delta m^2=\delta m_\mathrm{atm}^2$)
and inverted ($\delta m^2=-\delta m_\mathrm{atm}^2$) mass hierarchies.
The matter potential is
\begin{equation}
A = \sqrt{2} \GF n_e = \sqrt{2} \GF Y_e \nb,
\end{equation}
where $\GF$ is the Fermi coupling constant.
We define a reduced density matrix $\varrho_\nu$ 
(in the flavor basis) from $\psi_\nu$ as
\begin{equation}
\varrho_\nu \equiv \frac{1}{2}\left(\begin{array}{cc}
|a|^2-|b|^2 & 2ab^* \\
2a^* b & -|a|^2+|b|^2
\end{array}\right).
\label{eq:density-matrix}
\end{equation}
Note that this definition applies for neutrinos and antineutrinos.
This is, however, different from the convention adopted in 
Ref.~\cite{Sigl:1992fn}.
Using Eq.~\eqref{eq:density-matrix}, the neutrino-neutrino
forward scattering part of the Hamiltonian
in Eq.~\eqref{eq:schroedinger-eq} can be written as
\begin{widetext}
\begin{subequations}
\begin{align}
H_{\nu\nu} &= \frac{1}{2}\left(\begin{array}{cc}
B & B_{e\tau} \\ B_{e\tau}^* & -B
\end{array}\right) \\
&= \sqrt{2}\GF\sum_\alpha \Big[
\int (1-{\bf \hat{q}} \cdot {\bf \hat{q}^\prime})
\varrho_{\nuI{\alpha}}({\bf q^\prime}) 
\,\ud n_{\nuI{\alpha}}({\bf q^\prime}) \ud q^\prime
-\int (1-{\bf \hat{q}} \cdot {\bf \hat{q}^\prime})
\varrho_{\anuI{\alpha}}^*({\bf q^\prime}) 
\,\ud n_{\anuI{\alpha}}({\bf q^\prime}) \ud q^\prime \Big],
\label{eq:Hnunu}
\end{align}
\end{subequations}
where ${\bf q}$ and ${\bf q^\prime}$ are the momentum of the neutrino
of interest and that of the background neutrino, respectively, and the flavor
index is $\alpha=e$ or $\tau$. As mentioned above,
neutrinos of the same initial flavor, energy and emission angle have
 identical flavor evolution. Consequently one must have
\begin{equation}
\varrho_\nu({\bf q}) 
= \varrho_\nu(q, \vartheta).
\label{eq:density-matrix-dependence}
\end{equation}

We note that 
\begin{subequations}
\label{eq:int-phi}
\begin{align}
\int {\bf \hat{q}} \cdot {\bf \hat{q}^\prime} F(\vartheta^\prime) 
\,\ud {\bf \hat{q}^\prime}
&= \int [
\sin\vartheta\sin\vartheta^\prime
(\sin\phi\sin\phi^\prime+\cos\phi\cos\phi^\prime)
+\cos\vartheta\cos\vartheta^\prime]
F(\vartheta^\prime) \,\ud(\cos\vartheta^\prime)\ud\phi^\prime\\
&= 2\pi\int \cos\vartheta\cos\vartheta^\prime F(\vartheta^\prime)
\,\ud(\cos\vartheta^\prime),
\label{eq:int-phi-b}
\end{align}
\end{subequations}
where $F(\vartheta)$ is an arbitrary function of $\vartheta$,
and we have used the cylindrical symmetry around the $z$--axis
in deriving Eq.~\eqref{eq:int-phi-b}.
Using Eqs.~\eqref{eq:dnnu}, \eqref{eq:Lj} and \eqref{eq:int-phi},
one can rewrite Eq.~\eqref{eq:Hnunu} as
\begin{equation}
H_{\nu\nu} = \frac{\sqrt{2}\GF}{2\pi R_\nu^2} \sum_\alpha
\int (1-\cos\vartheta\cos\vartheta^\prime)
\left[\varrho_{\nuI{\alpha}}(q^\prime,\vartheta^\prime)
f_{\nuI{\alpha}}(q^\prime)
\frac{L_{\nuI{\alpha}}}{\avgE{\alpha}}
-\varrho_{\anuI{\alpha}}^*(q^\prime,\vartheta^\prime)
f_{\anuI{\alpha}}(q^\prime)
\frac{L_{\anuI{\alpha}}}{\avgaE{\alpha}}
\right]
\,\ud(\cos\vartheta^\prime)\ud q^\prime.
\tag{\ref{eq:Hnunu}$^\prime$}
\label{eq:Hnunu-multiangle}
\end{equation}
As noted in the introduction, previous simulations have
used the single-angle approximation, wherein one assumes
that the flavor evolution history of a neutrino is trajectory
independent,
\begin{equation}
\varrho_\nu({\bf q}) = \varrho_\nu(q),
\tag{\ref{eq:density-matrix-dependence}$^\prime$}
\end{equation}
and neutrinos on any trajectory transform in the same way as
neutrinos propagating in the radial direction.
Using the single-angle approximation, Eq.~\eqref{eq:Hnunu-multiangle}
can be further simplified to
\begin{equation}
H_{\nu\nu} = \frac{\sqrt{2}\GF}{2\pi R_\nu^2} \geoD(r/R_\nu)
\sum_\alpha \int
\left[\varrho_{\nuI{\alpha}}(q^\prime)
f_{\nuI{\alpha}}(q^\prime)
\frac{L_{\nuI{\alpha}}}{\avgE{\alpha}}
-\varrho_{\anuI{\alpha}}^*(q^\prime)
f_{\anuI{\alpha}}(q^\prime)
\frac{L_{\anuI{\alpha}}}{\avgaE{\alpha}}
\right]\,\ud q^\prime,
\tag{\ref{eq:Hnunu}$^{\prime\prime}$}
\label{eq:Hnunu-single-angle}
\end{equation}
\end{widetext}
where the geometric factor $\geoD(r/R_\nu)$ is defined as
\begin{equation}
\geoD(r/R_\nu) \equiv \frac{1}{2}\left[
1-\sqrt{1- \left( \frac{R_\nu}{r} \right) ^2 }  
\right]^2.
\label{eq:Psi}
\end{equation}

Although our simulations are carried out by solving 
Eq.~\eqref{eq:schroedinger-eq} numerically, the spin analogue
of the wave function formalism (see, \textit{e.g.}, Ref.~\cite{Duan:2005cp})
provides  an  intuitive way of understanding the results of 
our simulations. The wave function of a neutrino $\psi_\nu$ in
Eq.~\eqref{eq:wv-func} can be mapped into a Neutrino Flavor
Iso-Spin (NFIS) vector $\sB$ using the Pauli matrices ${\bm \sigma}$:
\begin{subequations}
\label{eq:nfis-def}
\begin{align}
\sB_{\nuI{\alpha}} &\equiv \psi_{\nuI{\alpha}}^\dagger \frac{\bm \sigma}{2} 
\psi_{\nuI{\alpha}}
= \frac{1}{2} \left(\begin{array}{c}
2 \mathrm{Re}(a^* b) \\
2 \mathrm{Im}(a^* b) \\
|a|^2 - |b|^2
\end{array}\right); \\
\sB_{\anuI{\alpha}}
&\equiv (\sigma_y\psi_{\anuI{\alpha}})^\dagger \frac{\bm \sigma}{2} 
(\sigma_y\psi_{\anuI{\alpha}})
= -\frac{1}{2} \left(\begin{array}{c}
2 \mathrm{Re}(a b^*) \\
2 \mathrm{Im}(a b^*) \\
|a|^2 - |b|^2
\end{array}\right). 
\label{eq:nfis-antinu}
\end{align}
\end{subequations}
Note that the extra $\sigma_y$ in Eq.~\eqref{eq:nfis-antinu}
transforms $\overline{\bf 2}$ of SU(2),
the fundamental representation of antiparticles,
into ${\bf 2}$, the fundamental representation of particles.
As a result, $\sB_{\anuI{\alpha}}$ transforms in the
same way as $\sB_{\nuI{\alpha}}$ under rotations. 
We also note the NFIS's $\sB_\nu$ defined in Eq.~\eqref{eq:nfis-def}
have constant magnitude 1/2. For a neutrino $\nuI{\alpha}$, 
$\varsigma_{\nuI{\alpha} z}=+1/2$ ($-1/2$) for
the pure $\nu_e$ ($\nu_\tau$) state, where $\varsigma_{\nuI{\alpha} z}$
is the third component of the NFIS. For an antineutrino $\anuI{\alpha}$, 
$\varsigma_{\anuI{\alpha} z}=+1/2$ ($-1/2$) 
for the pure $\bar\nu_\tau$ ($\bar\nu_e$) state.

The NFIS $\sB_\nu(q,\vartheta)$ for either a neutrino
or an antineutrino obeys the equation of motion
\begin{equation}
\begin{split}
\frac{\ud}{\ud t}\sB_\nu(q,\vartheta) &= 
\sB_\nu(q,\vartheta) \times \Big[\Hb^\eff(q) \\
&\quad+ \frac{1}{2\pi R_\nu^2}
\sum_{\nu^\prime}\int\mu(\vartheta,\vartheta^\prime)
\sB_{\nu^\prime}(q^\prime,\vartheta^\prime)\\
&\quad\times f_{\nu^\prime}(q^\prime)
\frac{L_{\nu^\prime}}{\langle E_{\nu^\prime}\rangle}
\,\ud(\cos\vartheta^\prime)\ud q^\prime\Big],
\end{split}
\label{eq:nfis-eom}
\end{equation}
where $q$ and $\vartheta$ are the magnitude and polar angle
of the momentum of the neutrino,
$\Hb^\eff$ is an effective field, and
$\mu(\vartheta,\vartheta^\prime)$ is the coupling coefficient
between $\sB_\nu(q,\vartheta)$ and 
the background neutrino $\sB_{\nu^\prime}(q^\prime,\vartheta^\prime)$ with
\begin{equation}
\mu(\vartheta,\vartheta^\prime) \equiv
-2\sqrt{2}\GF(1-\cos\vartheta\cos\vartheta^\prime).
\label{eq:mu-nu}
\end{equation}
The summation index $\nu^\prime$ in Eq.~\eqref{eq:nfis-eom} runs over
$\nuI{e}$, $\nuI{\tau}$, $\anuI{e}$, and $\anuI{\tau}$.
According to Eq.~\eqref{eq:nfis-eom}, the motion of a NFIS in
flavor space is analogous to that of a magnetic spin
which simultaneously precesses
around a ``magnetic field'' $\Hb^\eff$ and the other ``spins''.
The ``magnetic field'' $\Hb^\eff$ is composed of two components
in our case, 
\begin{equation}
\Hb^\eff(q) = \muV{}(q) \HV + \muA\HA.
\label{eq:Heff}
\end{equation}
In Eq.~\eqref{eq:Heff} $\HV$ stems from neutrino mass difference
and can be written as
\begin{equation}
\HV \equiv -\basef{x}\sin2\theta+\basef{z}\cos2\theta,
\label{eq:HV}
\end{equation}
where $\basef{x(y,z)}$ are the orthogonal unit vectors in flavor space
corresponding to $\sigma_{x(y,z)}$. Here we define 
\begin{equation}
\muV{}(q) \equiv \pm \frac{\delta m^2}{2 q}, 
\label{eq:muV}
\end{equation}
where the plus sign is for neutrinos
and the minus sign is for
antineutrinos. With these definitions,
neutrinos possess positive (negative) ``magnetic moments'' 
$\muV{}$ and antineutrinos possess negative (positive) ones
if $\delta m^2>0$ ($\delta m^2 <0$). Because neutrinos 
can have different energies, $\muV{}$ varies from $-\infty$ to $+\infty$.
The second term in Eq.~\eqref{eq:Heff} is induced by matter
(neutrino-electron forward scattering), and we can write
\begin{align}
\HA &\equiv -\basef{z} \nb Y_e
\intertext{and}
\muA &\equiv \sqrt{2}\GF.
\end{align}

Before we show the results of our simulations,
we shall estimate the ``MSW resonance radius'' $\rMSW$ for a neutrino 
with a typical energy. The MSW resonance condition would be
\begin{equation}
\Delta \cos2\theta = A(\rMSW)
\end{equation}
if we ignore the neutrino-neutrino flavor-diagonal potential $B$.
We will take $|\delta m^2|=3\times 10^{-3}\,\mathrm{eV}^2$,
the atmospheric value, and we will take the 
effective $2\times 2$ vacuum mixing angle to be $\theta=0.1$.
Note that this value is well below the experimental limit on $\theta_{13}$.
For these parameters,
the MSW resonance radius of a neutrino in the case of normal mass
hierarchy ($\delta m^2 > 0$)
or an antineutrino in the case of inverted mass hierarchy ($\delta m^2 < 0$) 
with energy $E_\nu=10$ MeV
is $\rMSW \simeq 127$ and 59 km for $S=140$
and 250, respectively. 
We shall also estimate the radius for significant  neutrino flavor
transformation if neutrinos
and antineutrinos are in the ``synchronization'' mode \cite{Pastor:2001iu}.
When neutrinos are in the synchronization mode, all the NFIS's behave
as one ``magnetic spin'' with 
\begin{equation}
\pm\frac{\delta m^2}{2\Esync}\equiv
\langle\muV{}\rangle 
=\frac{\displaystyle \sum_\nu \frac{L_\nu}{\langle E_\nu \rangle}\int \muV{}(q) \varsigma_{\nu z}f_\nu(q)\,\ud q}%
{\displaystyle \sum_\nu \frac{L_\nu}{\langle E_\nu \rangle}\int \varsigma_{\nu z}f_\nu(q)\,\ud q},
\label{eq:Esync-def}
\end{equation}
where we have assumed all the NFIS's are aligned or antialigned
with $\basef{z}$. Because $\Esync$ is positive, the sign of
the first term (left hand side) of Eq.~\eqref{eq:Esync-def} should be chosen
to be the same as that of the product $\delta m^2\langle\muV{}\rangle$.
If $\langle\muV{}\rangle >0$
and $\delta m^2 > 0$, all the neutrinos and antineutrinos go 
through the same conversion process
as a neutrino of energy $\Esync$ at $\rMSW(\Esync)$.
Similarly, if $\langle\muV{}\rangle > 0$ and $\delta m^2 < 0$,
all the neutrinos and antineutrinos go 
through the same conversion process
as an antineutrino of energy $\Esync$ at $\rMSW(\Esync)$.
Neutrino flavor transformation is suppressed for other
synchronization scenarios.
For the parameters we have chosen, we find that $\langle\muV{}\rangle >0$
if $\delta m^2 > 0$, and $\langle\muV{}\rangle <0$
if $\delta m^2 < 0$. The characteristic energy of the synchronization
mode is $E_{\mathrm{sync}} \simeq 2.47$ MeV for both cases.
Therefore, in the synchronized mode
neutrinos and antineutrinos should transform simultaneously
at $r_{\mathrm{MSW}}(E_{\mathrm{sync}}) \simeq 80$  and 37 km
for $S=140$ and 250, respectively, if $\delta m^2 > 0$.
These neutrinos/antineutrinos would experience
 very little flavor conversion if $\delta m^2 < 0$. 

A special case of synchronized behavior is the Background
Dominant Solution (BDS) \cite{Fuller:2005ae} where the
NFIS's rotate in the plane spanned by $\basef{x}$ and $\basef{y}$
in flavor space. One of the necessary conditions
for the BDS with large-scale simultaneous
neutrino and antineutrino flavor transformation 
is that the flavor off-diagonal neutrino background potential $B_{e\tau}$
dominates. To see this condition more clearly, we define the effective
net $\nuI{e}$ number density along the radial trajectory as
\begin{widetext}
\begin{subequations}
\label{eq:nue-eff}
\begin{align}
n_{\nuI{e}}^\eff &= 
\int (1-{\bf \hat{z}} \cdot {\bf \hat{q}^\prime})
\,\ud n_{\nuI{e}}({\bf q^\prime}) \ud q^\prime
-\int (1-{\bf \hat{z}} \cdot {\bf \hat{q}^\prime})
\,\ud n_{\anuI{e}}({\bf q^\prime}) \ud q^\prime \\
&=\frac{\geoD(r/R_\nu)}{2\pi R_\nu^2}
\left(\frac{L_{\nuI{e}}}{\avgE{e}}-\frac{L_{\anuI{e}}}{\avgaE{e}}\right)\\
&=(1.66\times 10^{32} \,\mathrm{cm}^{-3})
\left[1-\sqrt{1-\left(\frac{R_\nu}{r}\right)^2}\right]^2
\left(\frac{10\,\mathrm{km}}{R_\nu}\right)^{2}
\left[
\frac{L_{\nuI{e}}/(10^{51}\,\mathrm{erg/s})}{\avgE{e}/(10\,\mathrm{MeV})}
-\frac{L_{\anuI{e}}/(10^{51}\,\mathrm{erg/s})}{\avgaE{e}/(10\,\mathrm{MeV})}
\right] .
\end{align}
\end{subequations}
\end{widetext}
Note that $|B_{e\tau}|=\sqrt{2}\GF |n_{\nuI{e}}^\eff|$ and $B=0$ if the BDS
obtains. (Obviously, if no flavor transformation has occurred, $B_{e\tau}=0$
and $B=\sqrt{2}\GF n_{\nuI{e}}^\eff$.)
 We plot $n_{\nuI{e}}^\eff(r)$ together with
$n_e(r)$ used in our simulations in Fig.~\ref{fig:nl-r}. 
The neutrino background potential
will dominate the matter potential on the radial trajectory if
$n_{\nuI{e}}^\eff>n_e$, which corresponds to a radius as low as
$r\sim 13$ km for the parameters we have chosen.

For numerical simplicity,
we have fixed $Y_e=0.4$ in our simulations. Of course, the value of $Y_e$
actually varies with the radius and is affected by $\nu_e$ and $\bar\nu_e$
fluxes through the weak interactions 
\cite{Fuller:1992aa,Qian:1993dg}
\begin{subequations}
\label{eq:nu-interactions}
\begin{align}
\nu_e + n &\rightleftharpoons p + e^-,\\
\bar\nu_e + p &\rightleftharpoons n + e^+.
\end{align}
\end{subequations}
The rates of these processes can also be affected
by weak  magnetism corrections \cite{Horowitz:1999aa,Horowitz:2002aa}.
In the numerical simulations presented below we have not included
neutrino/antineutrino flavor transformation feedback through these
processes on $Y_e$. This is an important aspect of the physics of the
supernova environment which we leave to a subsequent paper.

Models for \textit{r}-process nucleosynthesis can be sensitive to the
value of $Y_e$ in the region where $T \gtrsim 0.1$ MeV
\cite{Qian:1993dg,Qian:1994wh}.
For our chosen density profile, $T=0.1$ MeV occurs at
$r \sim 139$ and 78 km for
entropy per baryon $S=140$ and 250, respectively.
These values of radius are well outside our simple estimates
for where conventional MSW, synchronization, or BDS-like
flavor conversion could occur.
The numerical results to be discussed in the next
section will give us a much better idea of where large-scale
neutrino flavor transformation actually occurs.

\section{Multiangle Numerical Simulations%
\label{sec:multiangle}}

In Sec.~\ref{sec:numerical-scheme} we discuss our numerical calculations,
 and point out two potential pitfalls 
in any multiangle simulation. In Sec.~\ref{sec:results} we
show the main results from our multiangle simulations.
For the simulation results  presented in this section
we have taken
$|\delta m^2| = 3\times 10^{-3}\,\mathrm{eV}^2$, $\theta = 0.1$, 
$L_{\nuI{e}} = L_{\anuI{e}} =L_{\nuI{\tau}} = L_{\anuI{\tau}} = 10^{51}$ erg/s
 and $S=140$ unless otherwise stated.

\subsection{Numerical Scheme%
\label{sec:numerical-scheme}}

We have developed two independent sets of
numerical codes using different computer languages.
We have used them to provide cross checks to obtain consistent results.
Both codes employ a large multidimensional array of neutrino
wave functions $\psi_{\nuI{\alpha}}(E_\nu,\cos\vartheta_0)$ 
and $\psi_{\anuI{\alpha}}(E_\nu,\cos\vartheta_0)$ and evolve them
simultaneously following the scheme outlined in Sec.~\ref{sec:background}.
Each code employs an adaptive step size control mechanism,
but the two codes  have different ways of
 estimating errors and adjusting step sizes.
The energy bins are chosen to have equal sizes 
for convenience in comparing neutrino energy spectra at
different radii.
The angle bins are determined in such a way that each bin has
the same size in $\cos\vartheta$ at radius $R_\mathrm{bin}$.
In most cases we have taken $R_\mathrm{bin}=R_\nu$, the
neutrino sphere radius. Note that the angle bins have different
sizes in $\cos\vartheta$ if $r\neq R_\mathrm{bin}$.

At the basic level, both codes obtain $\psi_\nu(l+\delta l)$
from $\psi_\nu(l)$ by using the following equation as a first step:
\begin{widetext}
\begin{subequations}
\label{eq:solution}
\begin{align}
\psi_\nu(l+\delta l) &\simeq \exp(-\ui H \delta l) \psi_\nu(l) \\
&=\frac{1}{\lambda}\begin{pmatrix}
\lambda\cos(\lambda \delta l)-\ui h_{11}\sin(\lambda \delta l) &
-\ui h_{12} \sin(\lambda \delta l)\\
-\ui h_{12}^* \sin(\lambda \delta l) &
\lambda\cos(\lambda \delta l)+\ui h_{11}\sin(\lambda \delta l)
\end{pmatrix} \psi_\nu(l),
\end{align}
\end{subequations}
\end{widetext}
where $h_{11}$ and $h_{12}$ are the diagonal and off-diagonal elements
of the Hamiltonian $H$, and 
\begin{equation}
\lambda\equiv\sqrt{h_{11}^2+|h_{12}|^2}.
\end{equation}
[Although not written out explicitly,
the Hamiltonian $H$ and its elements in the above equations 
have dependence on both the
 Affine parameter $l$ and trajectory angle $\vartheta$, as can be inferred
from Eqs.~\eqref{eq:schroedinger-eq} through \eqref{eq:Hnunu-multiangle}.]
We note that Eq.~\eqref{eq:solution} preserves the
unitarity of $\psi_\nu$ automatically. We also note that
Eq.~\eqref{eq:solution} becomes exact if $H$ is independent of
spatial coordinate.
Therefore, the step sizes employed in our numerical codes are
not restricted by the size of $H$ but are restricted by
the rate of change of $H$.

\begin{figure}
\begin{center}
\includegraphics*[width=\myfigwid, keepaspectratio]{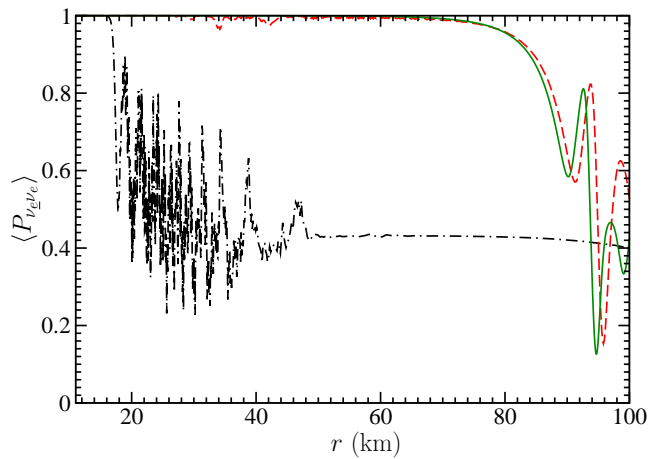}
\end{center}
\caption{\label{fig:P-r-cmpr}(Color online) 
Average $\nu_e$ survival probability 
$\langle\Pee(r)\rangle$ along the radial trajectory ($\cos\vartheta_0$=1)
with the normal mass hierarchy in different numerical schemes.
Here the average is done over the initial energy spectra of $\nuI{e}$.
The dot-dashed line uses 160 angle bins and error tolerance $10^{-5}$
in each step without the initial baryon density profile $\nb^\prime$.
The dashed and solid lines both include $\nb^\prime$ and 
employ error tolerance $10^{-10}$, but use 256 and 512 angle bins,
respectively. Calculations with 768, 1024 and 1407 angle bins
in different binning schemes
produce curves which fall on the solid line.}
\end{figure}

In the course of our work
we have discovered two
pitfalls which apply to any multiangle scheme.
Failure to avoid these pitfalls may lead to quantitatively or qualitatively
inaccurate results (see Fig.~\ref{fig:P-r-cmpr}).

The first potential problem has to do with the exponential term $\nb^\prime$
in the profile for the baryon density
[see Eq.~\eqref{eq:nb-profile-exp}]. The baryon density is
very high near the neutrino sphere when $\nb^\prime$
is included. This sometimes forces numerical schemes to employ
initially very small step sizes.
The numerical codes using the single-angle approximation
can generally drop $\nb^\prime$ without loss of accuracy
at large radius. These codes will of course run faster without $\nb^\prime$.
However, in  multiangle simulations, ignoring $\nb^\prime$
makes the background neutrino potential $B$ much
bigger than the matter potential $A$
even at the neutrino sphere.
As a result, the evolution histories of neutrino flavors on all
trajectories are strongly coupled starting from the beginning.
This strong correlation among all trajectories also
forces small step sizes. In addition, without $\nb^\prime$ there is a
tendency for neutrinos to undergo flavor transformation
very close to the neutrino sphere. 
This behavior is suppressed if there is a large and dominant
matter potential $A$.
Including $\nb^\prime$ makes the matter potential $A(R_\nu)$ much bigger
and this helps keep neutrinos in their initial flavor states,
at least for the significant range of  neutrino/antineutrino energies
and for our chosen value of $|\delta m^2| \simeq \delta m^2_{\mathrm{atm}}$.
We also note that neutrinos on different
trajectories propagate through different distances.
A big matter potential breaks the correlation
between neutrinos on different trajectories and
lets them evolve independently for awhile.

These considerations can be cast in simpler, more physical terms.
In the relatively narrow region near the neutrino sphere
where $\nb^\prime$ dominates it has the effect of changing,
or ``resetting'', the neutrino wavefunctions relative to
what they would have been had we employed the unphysical
low-density profile all the way to the neutrino sphere.
In the latter unphysical case, neutrinos are in flavor
eigen states at the neutrino sphere, and the NFIS's are
perfectly aligned with each other yet slightly deviated
from the total effective field $\Hb^\eff$. 
The effects of this unphysical setup does not
go away quickly with increasing radius because the coupling
among the NFIS's (arising from neutrino-neutrino forward scatterings)
is so strong. If the exponential baryon density profile
$\nb^\prime$ is added, the overwhelming matter field $\Hb_e$
at the neutrino sphere not only makes the NFIS's more aligned
with $\Hb^\eff$, but also breaks the coupling of the NFIS's
propagating along different trajectories. In the short
distance where the matter field $\Hb_e$ dominates, the NFIS's
on different trajectories have traveled different distances
and so have developed different phases. At the radius where
$\nb^\prime$ becomes negligible, the NFIS's are effectively
``reset'' to a more physical condition than one would
obtain without $\nb^\prime$. 

The other pitfall is that one may use an insufficient
number of angle bins.
Assuming that there has been very little neutrino flavor conversion
close to the neutrino sphere where  $r\sim R_\nu$, we can write
\begin{widetext}
\begin{subequations}
\begin{align}
B(r,\vartheta) &\simeq 
\frac{\sqrt{2}\GF}{2\pi R_\nu^2} \left(
\frac{L_{\nuI{e}}}{\avgE{e}} -
\frac{L_{\anuI{e}}}{\avgaE{e}}
\right) 
\left[
1-\sqrt{1-\left(\frac{R_\nu}{r}\right)^2} - 
\frac{1}{2} \cos\vartheta \left(\frac{R_\nu}{r}\right)^2
\right], 
\label{eq:B-small-r}\\
B_{e\tau}(r,\vartheta) &\simeq 0.
\end{align}
\end{subequations}
For a small step size $\delta l$, one has
\begin{subequations}
\begin{align}
\psi_{\nuI{\alpha}}(l+\delta l) &\simeq \exp(-\ui H \delta l)  
\psi_{\nuI{\alpha}}(l)\\
&\simeq \begin{pmatrix}
e^{-\ui (A+B) \delta l} 
& -\ui \frac{\displaystyle\Delta \sin2\theta}{\displaystyle A+B} 
\sin[(A+B) \delta l] \\
-\ui \frac{\displaystyle\Delta \sin2\theta}{\displaystyle A+B} 
\sin[(A+B) \delta l] & e^{\ui (A+B) \delta l} 
\end{pmatrix}
\psi_{\nuI{\alpha}}(l),
\label{eq:U-dl}
\end{align}
\end{subequations}
\end{widetext}
where we have used the fact that $A+B\gg\Delta$ at $r\sim R_\nu$.
It is the off-diagonal elements of the transformation
matrix in Eq.~\eqref{eq:U-dl} that govern the exchange of
the two flavor components of a neutrino wave function.
These off-diagonal terms, we note, are oscillatory functions of
the $B$ potential and the step size $\delta l$, both of
which have angular dependence [see Eqs.~\eqref{eq:prop-distance}
and \eqref{eq:B-small-r}].
Physically, this oscillatory feature with respect
to angles is suppressed by
strong correlation among neutrinos on different trajectories.
Numerical codes without enough angular resolution, however,
could allow a spurious ``cross-talk'' between angle zones which
artificially strengthens flavor oscillations.
This unphysical feedback could produce
substantial neutrino flavor conversion even at low radius
in some numerical schemes.

\begin{figure*}
\begin{center}
$\begin{array}{@{}c@{\hspace{\myfigsep}}c@{}}
\includegraphics*[width=\myfigwid, keepaspectratio]{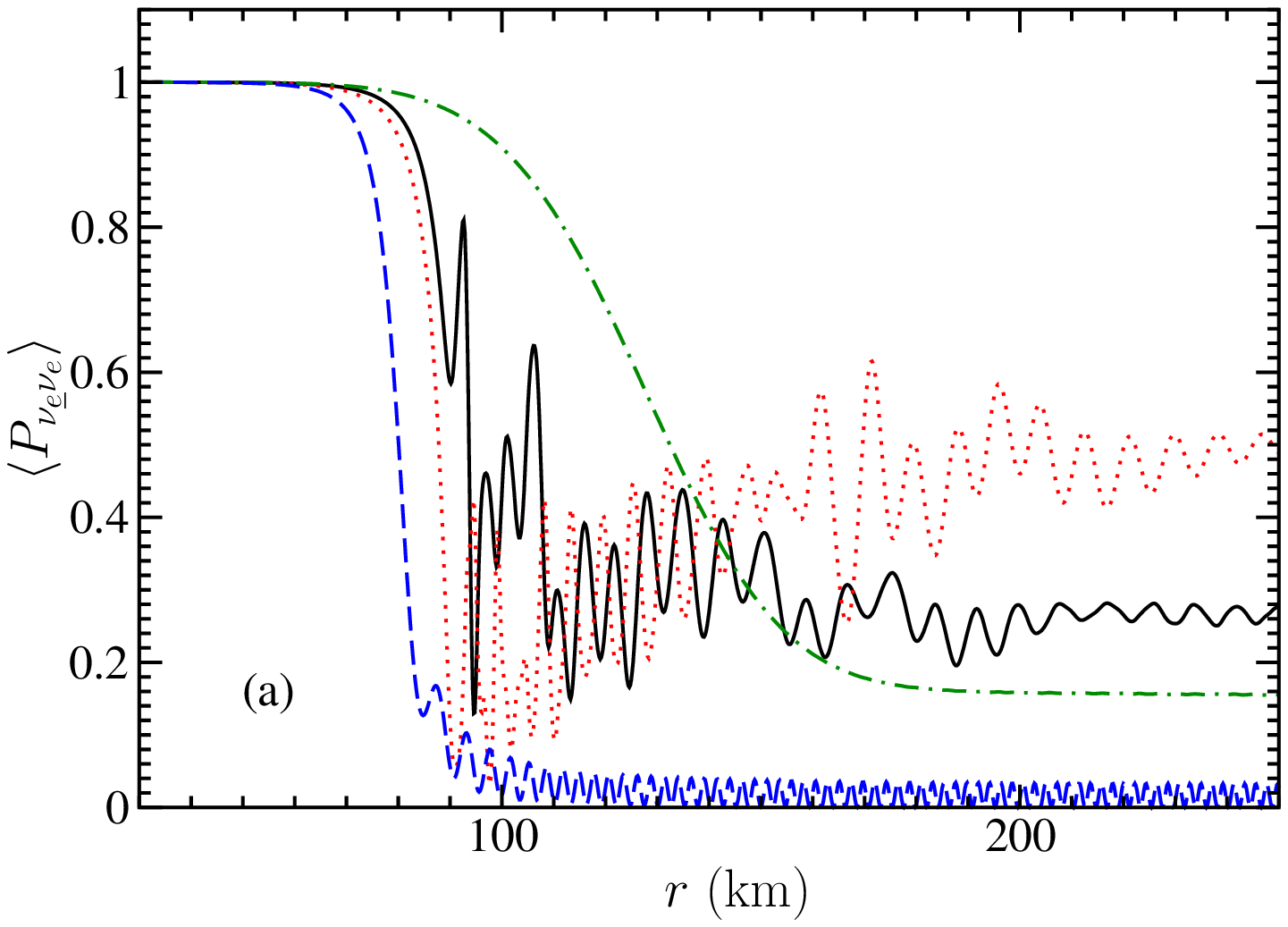} &
\includegraphics*[width=\myfigwid, keepaspectratio]{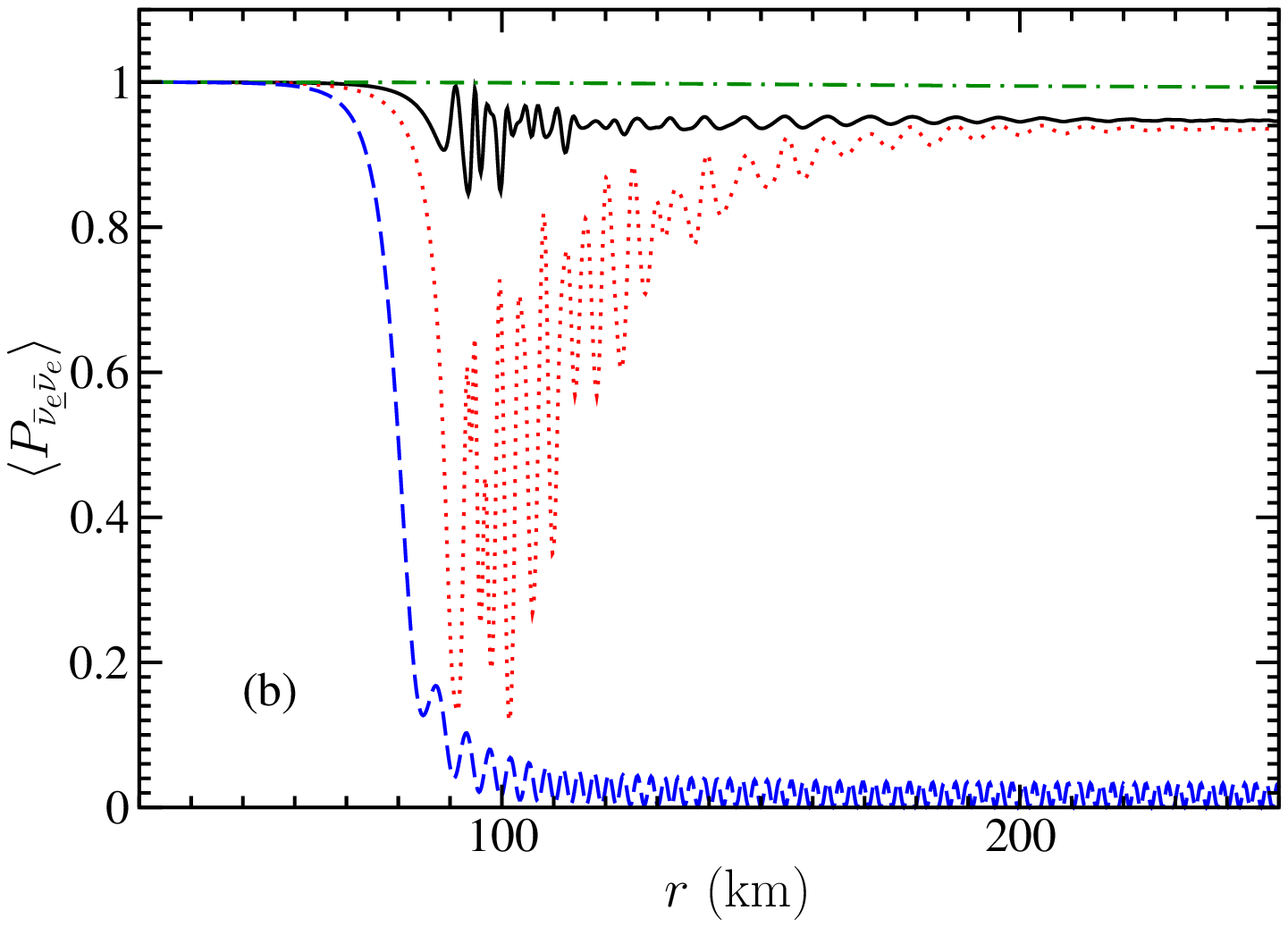} \\
\includegraphics*[width=\myfigwid, keepaspectratio]{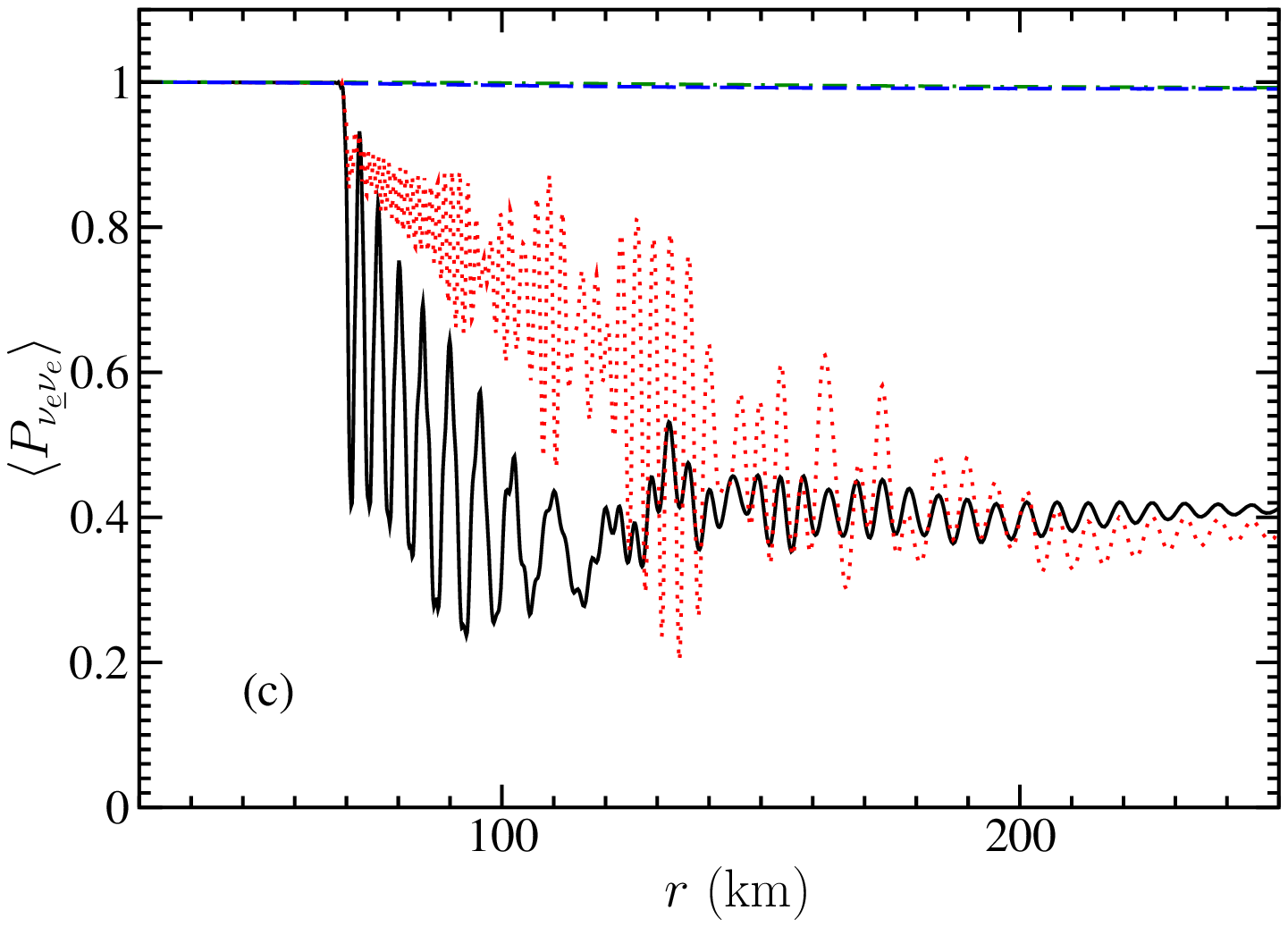} &
\includegraphics*[width=\myfigwid, keepaspectratio]{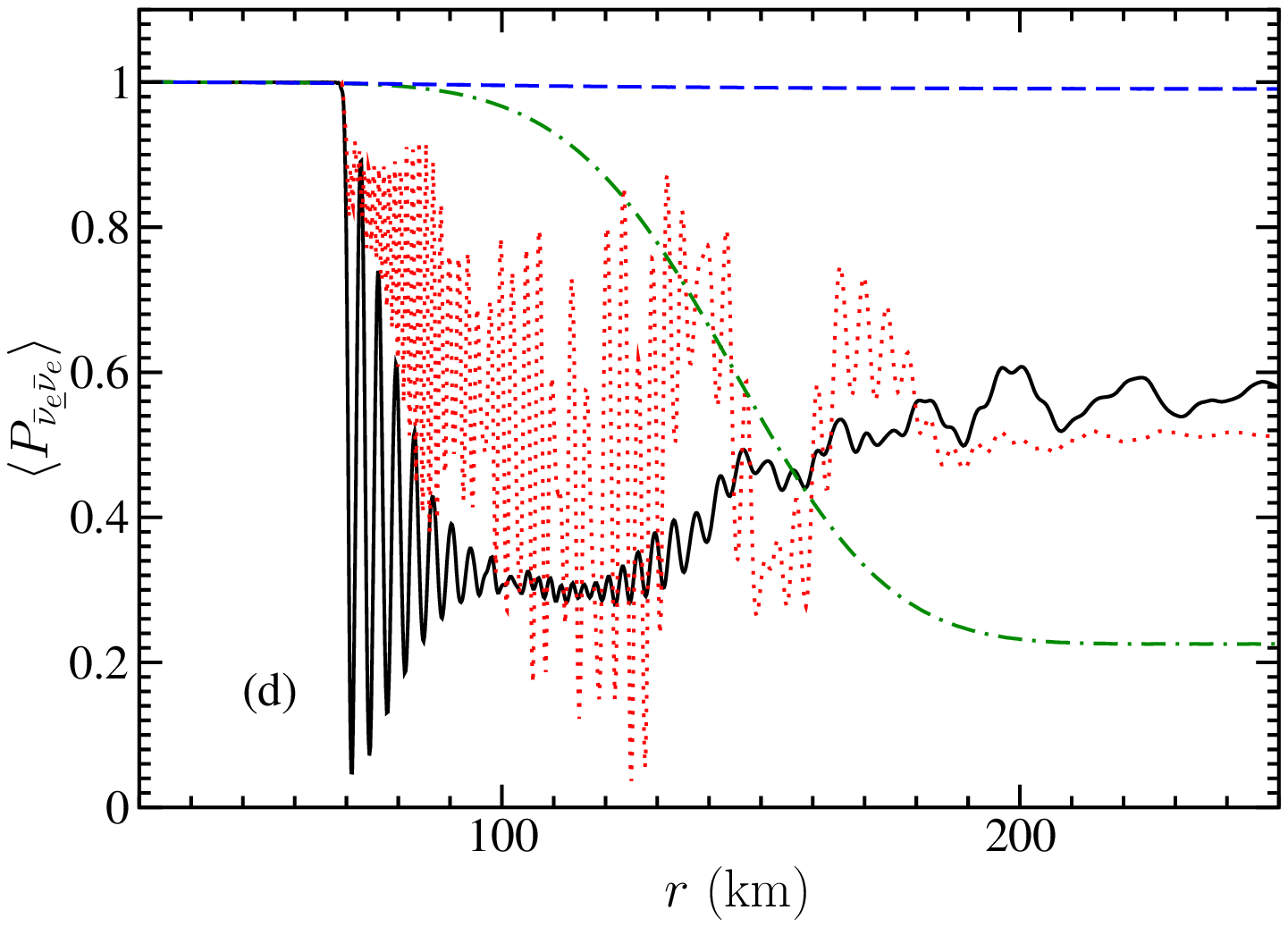}
\end{array}$
\end{center}
\caption{\label{fig:P-r}(Color online) 
Plots of average survival probability $\langle\Pee\rangle$ (left panels) and 
$\langle\Paeae\rangle$ (right panels) with the
normal (upper panels) and inverted (lower panels) neutrino mass
hierarchies, respectively. 
The solid and dotted lines give average survival probabilities 
along trajectories with $\cos\vartheta_0=1$
and $\cos\vartheta_0=0$, respectively, as computed in the 
multiangle simulations. The dot-dashed lines 
and the dashed lines characterize the limits where
neutrinos and antineutrino undergo flavor transformation
individually ($L_\nu=0$ and $A$ potential only)
and simultaneously (full synchronization), respectively.
The dashed line is not distinguishable from the dot-dashed line
in panel (c). }
\end{figure*}

In Fig.~\ref{fig:P-r-cmpr} we plot 
average survival probability $\langle\Pee(r)\rangle$ along
the radial trajectory with the normal mass hierarchy using different
numerical schemes (error tolerance, number of angle bins, \textit{etc.}).
Here $\Pee(r)$ is the probability for a $\nuI{e}$ to be a $\nu_e$ at radius $r$,
and the average is done over the initial energy distribution for $\nuI{e}$.
(As mentioned above, we use $\nuI{\alpha}$ and $\anuI{\alpha}$
to denote the neutrinos and antineutrinos that are emitted 
in flavor state $\alpha$ at the neutrino sphere.)
One sees that spurious neutrino
flavor transformation  (dot-dashed line) 
could occur at low radius with a combination of 
insufficient number of angle bins,
a loose error control, and neglect of $\nb^\prime$.
If we employ $L_\nu=10^{51}$ erg/s and choose a stringent
error tolerance ($\sim 10^{-10}$) at each step, we find that
it takes $\gtrsim 500$ angle bins in order to achieve convergence
and run-to-run consistency. 
Because the $B$ potential increases with neutrino luminosity,
we expect that even more angle bins would be required to obtain convergence
at larger neutrino luminosity.

Our numerical simulations generally employ $\gtrsim 500$ angle bins and 
$\gtrsim 500$ energy
bins for each neutrino species. Typically, our codes 
execute  $\gtrsim 10^5$ steps during each
production run. It is clear that multiangle simulations are only feasible 
using large-scale parallel computation.

\subsection{Simulation Results%
\label{sec:results}}

\begin{figure*}
\begin{center}
$\begin{array}{@{}c@{\hspace{\myfigsep}}c@{}}
\includegraphics*[width=\myfigwid, keepaspectratio]{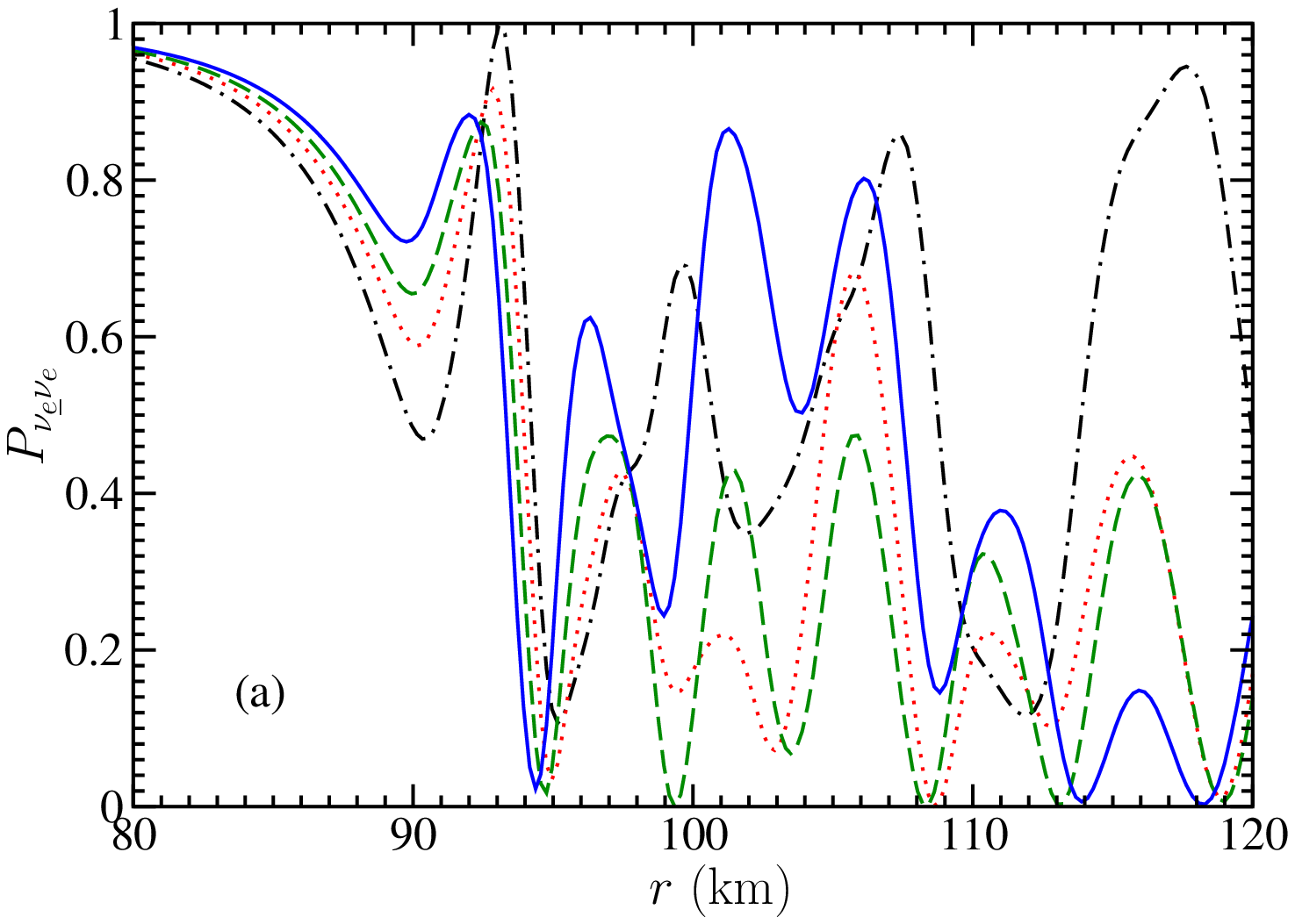} &
\includegraphics*[width=\myfigwid, keepaspectratio]{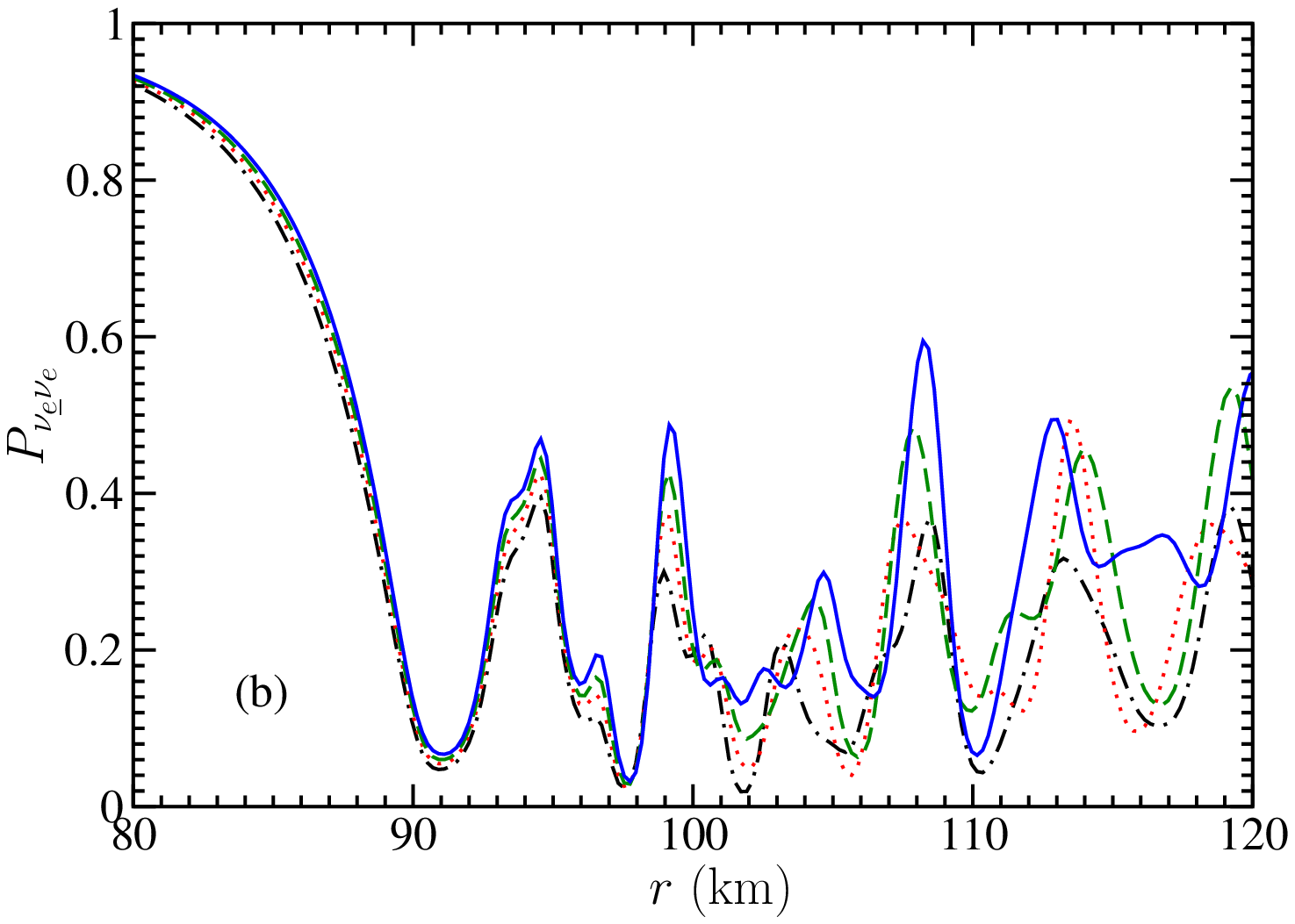}
\end{array}$
\end{center}
\caption{\label{fig:P-r-nue}(Color online) 
Plots of $\Pee(r)$ with the normal mass hierarchy. Panel (a) is
for the radial trajectory ($\cos\vartheta_0=1$), and (b) is for
the tangential trajectory ($\cos\vartheta_0=0$). The dot-dashed, 
dotted, dashed and solid lines are for $\nuI{e}$
of energies 6.95, 8.95, 10.95 and 14.95 MeV,
respectively.}
\end{figure*}

\begin{figure*}
\begin{center}
$\begin{array}{@{}c@{\hspace{\myfigsep}}c@{}}
\includegraphics*[width=\myfigwid, keepaspectratio]{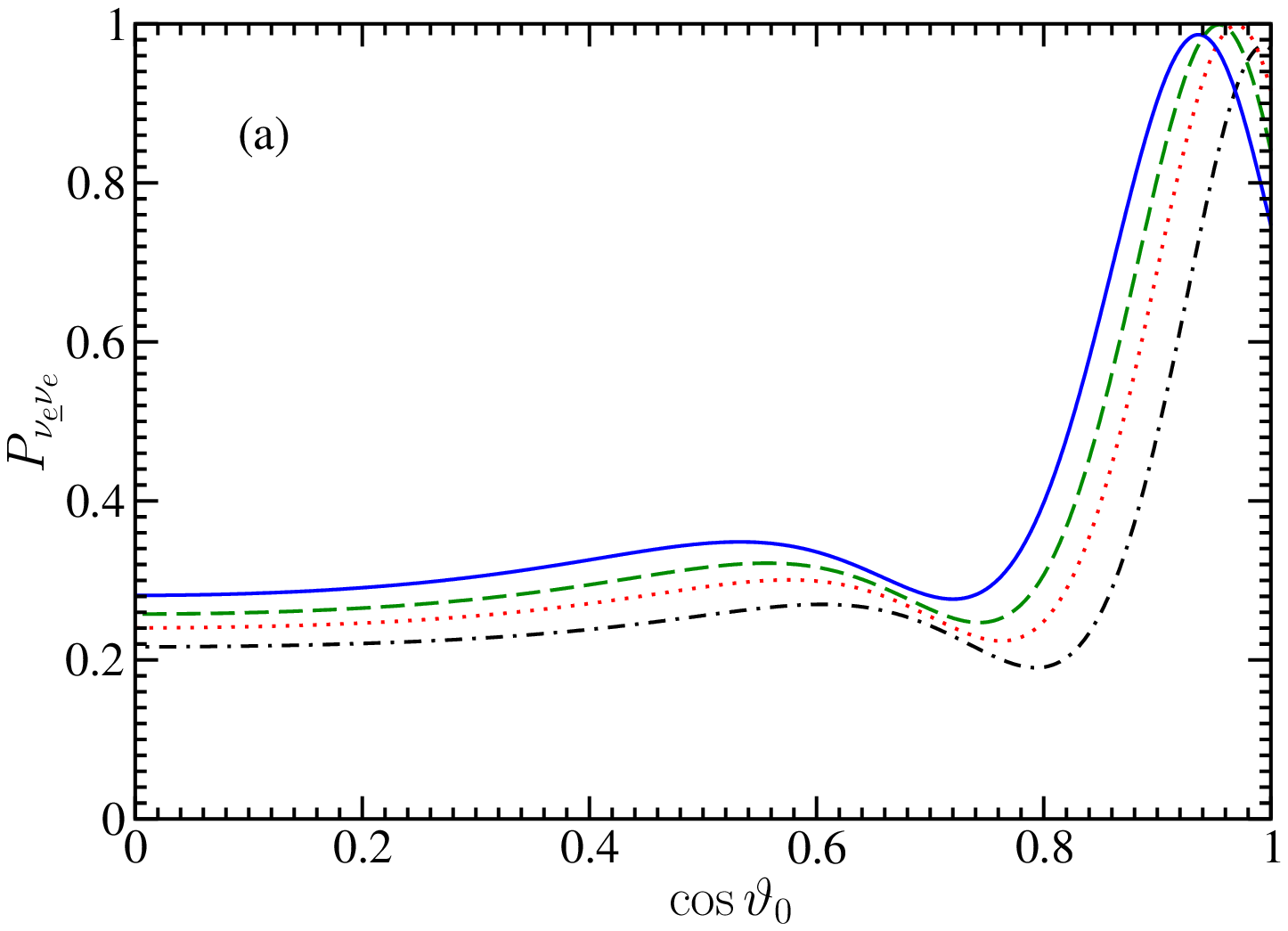} &
\includegraphics*[width=\myfigwid, keepaspectratio]{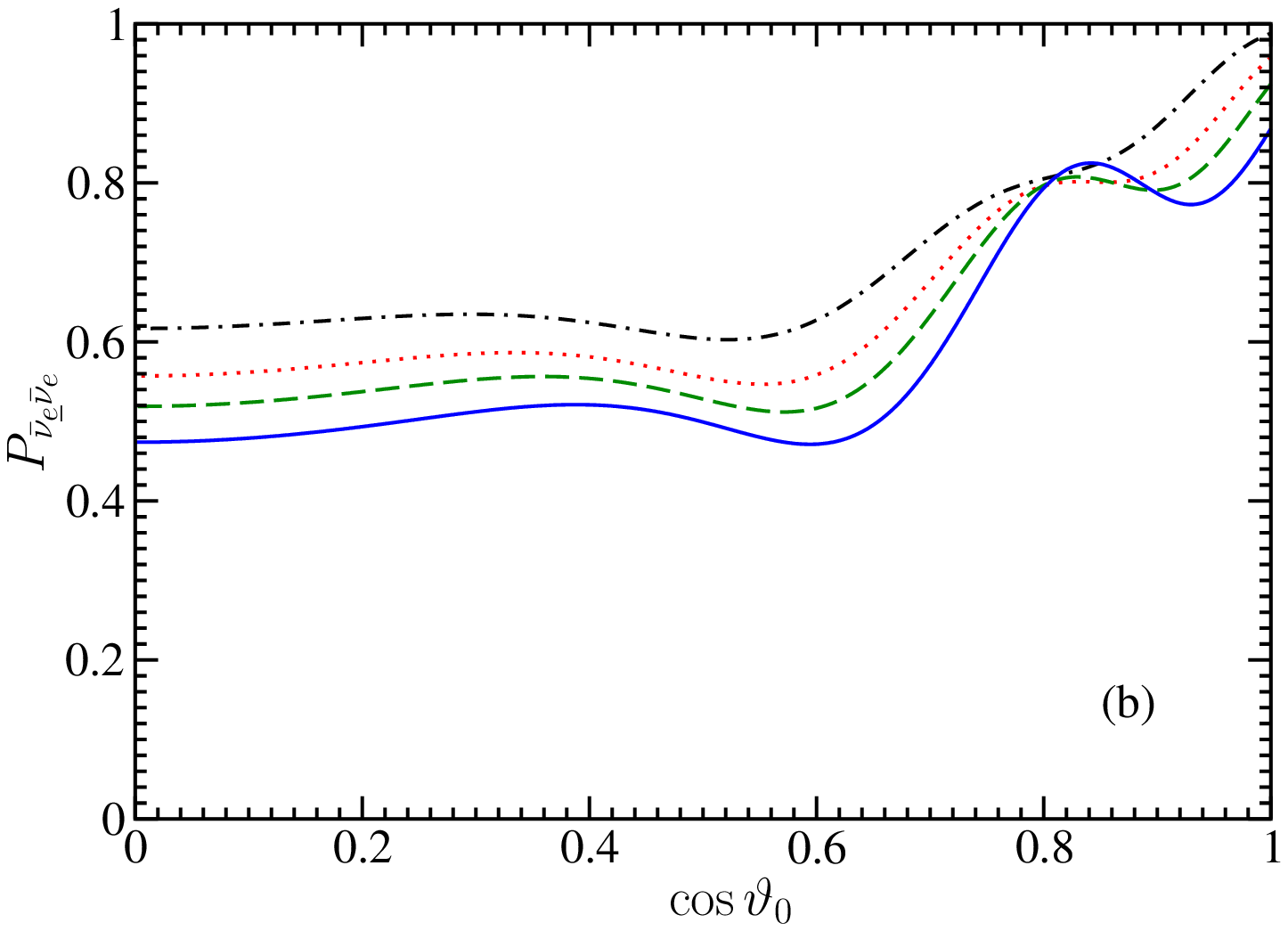} \\
\includegraphics*[width=\myfigwid, keepaspectratio]{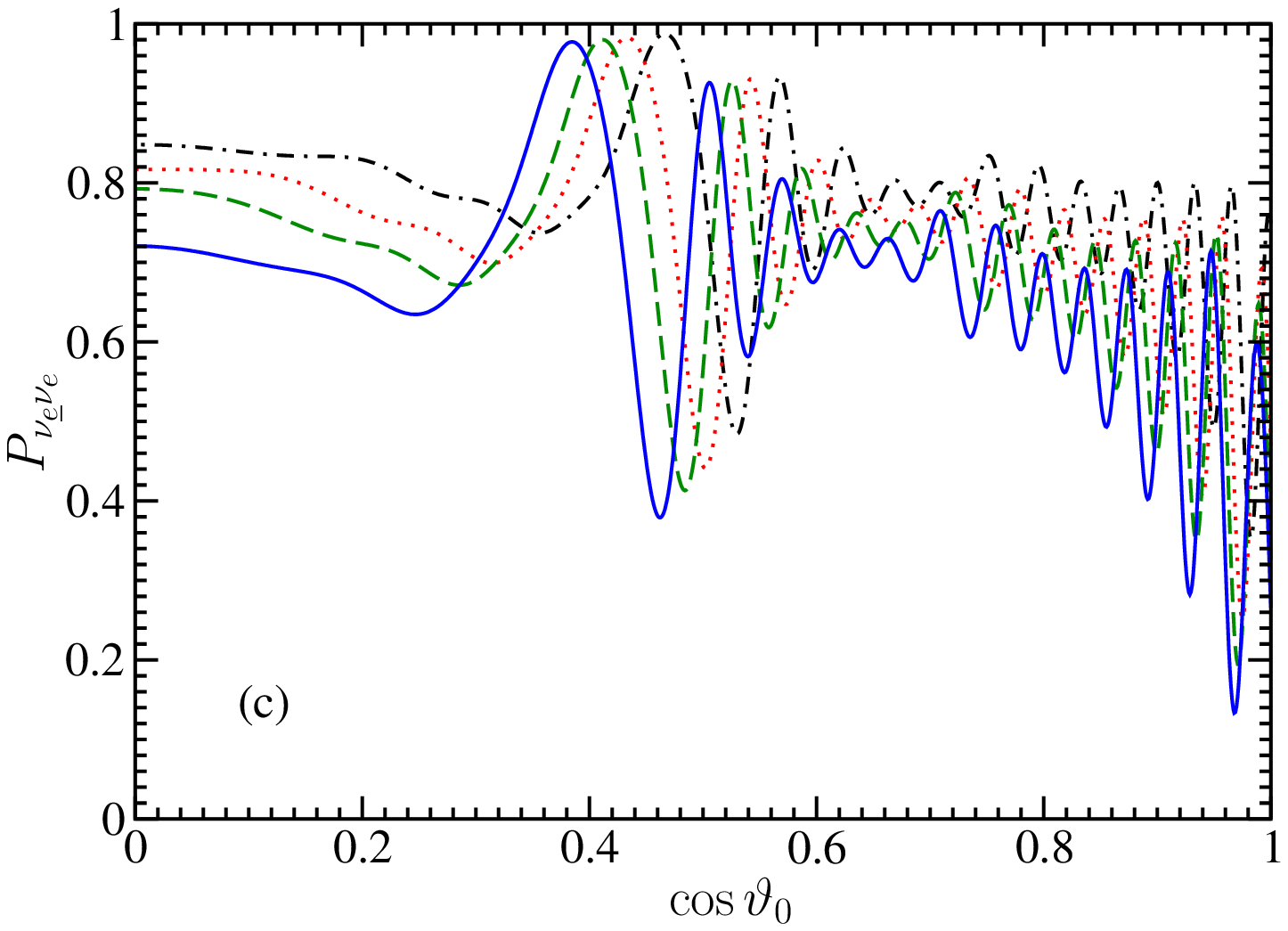} &
\includegraphics*[width=\myfigwid, keepaspectratio]{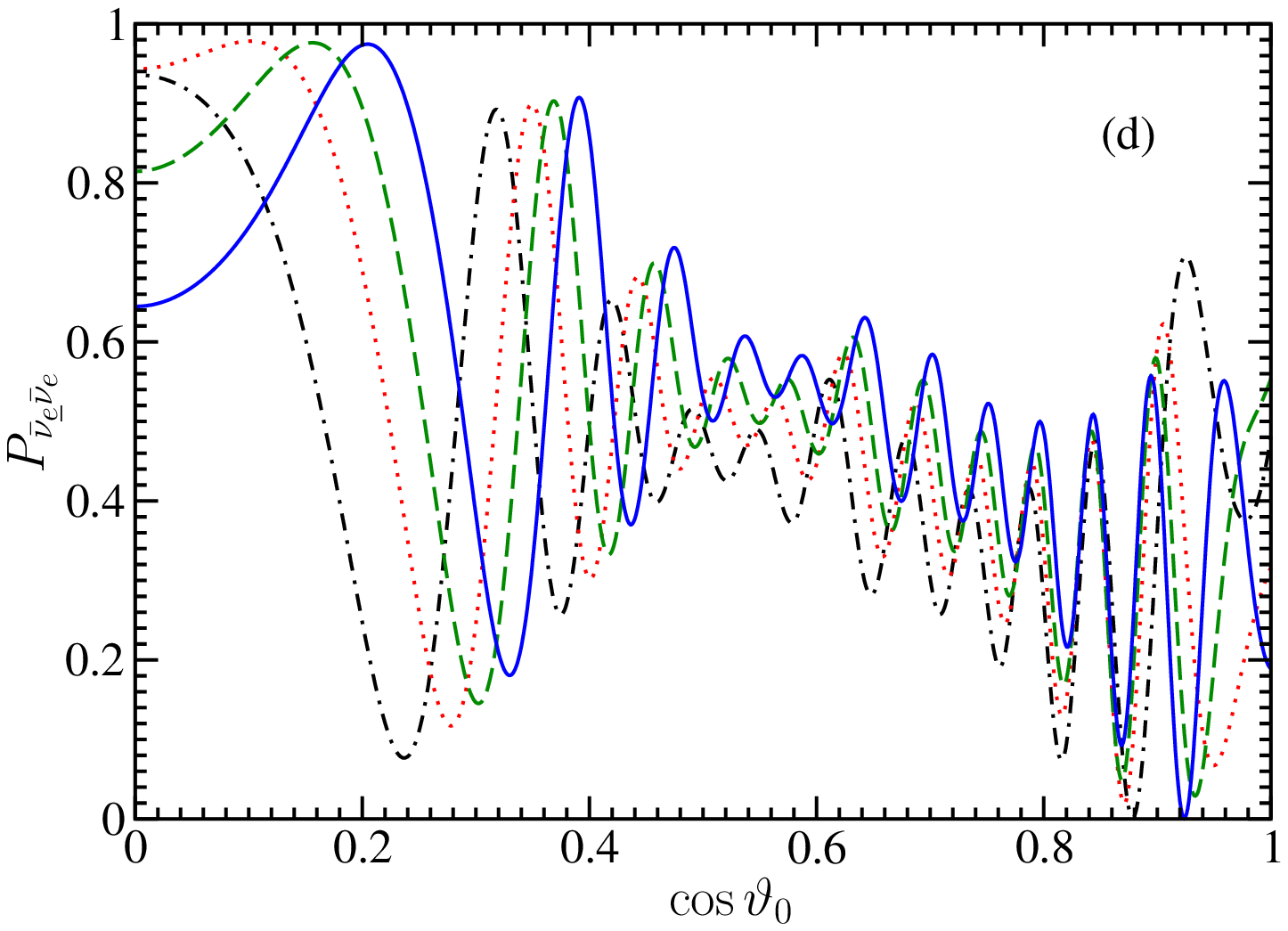}
\end{array}$
\end{center}
\caption{\label{fig:P-theta0}(Color online) 
Plots of $\Pee(\cos\vartheta_0)$ (left panels) and $\Paeae(\cos\vartheta_0)$
(right panels) at $r\simeq 92.85$ km for the normal mass hierarchy
(upper panels) and at $r\simeq 88.57$ km for the inverted mass
hierarchy (lower panels).
The dot-dashed, dotted, dashed and solid lines are for $\nuI{e}$
or $\anuI{e}$ of energies 6.95, 8.95, 10.95 and 14.95 MeV,
respectively.}
\end{figure*}

In Fig.~\ref{fig:P-r}(a), we plot $\langle\Pee(r)\rangle$ 
with the normal neutrino mass hierarchy ($\delta m^2 >0$)
on both the radial ($\cos\vartheta_0=1$) and 
tangential ($\cos\vartheta_0=0$) trajectories. 
For comparison, we also plot $\langle\Pee(r)\rangle$ for the $L_\nu=0$ 
($A$ potential only) case, which is obtained from the single-angle 
simulation by setting $L_\nu=0$.
The $L_\nu=0$ case corresponds to the limit where 
neutrinos go through MSW resonances independently of each other.
In the full synchronization limit, 
all neutrinos and antineutrinos undergo flavor transformation in
the same way as does a $\nuI{e}$ with energy $\Esync$ 
in the standard MSW mechanism. 
Using only the matter potential, we have calculated
$\Pee$ for a $\nuI{e}$ with energy $\Esync$ propagating along
the radial trajectory. The result is
shown in Fig.~\ref{fig:P-r}(a).
The results of our simulations are clearly different
from those in the $L_\nu=0$ 
and full synchronization limits. In particular, 
our simulation has $\langle\Pee(r)\rangle$
crossing the 1/2 line later than in the synchronization case, but
earlier than in the $L_\nu=0$ case. We also note that 
$\langle\Pee(r)\rangle$ oscillates and even bounces back after an initial
decrease.

\begin{figure*}
\begin{center}
$\begin{array}{@{}c@{\hspace{\myfigsep}}c@{}}
\includegraphics*[width=\myfigwid, keepaspectratio]{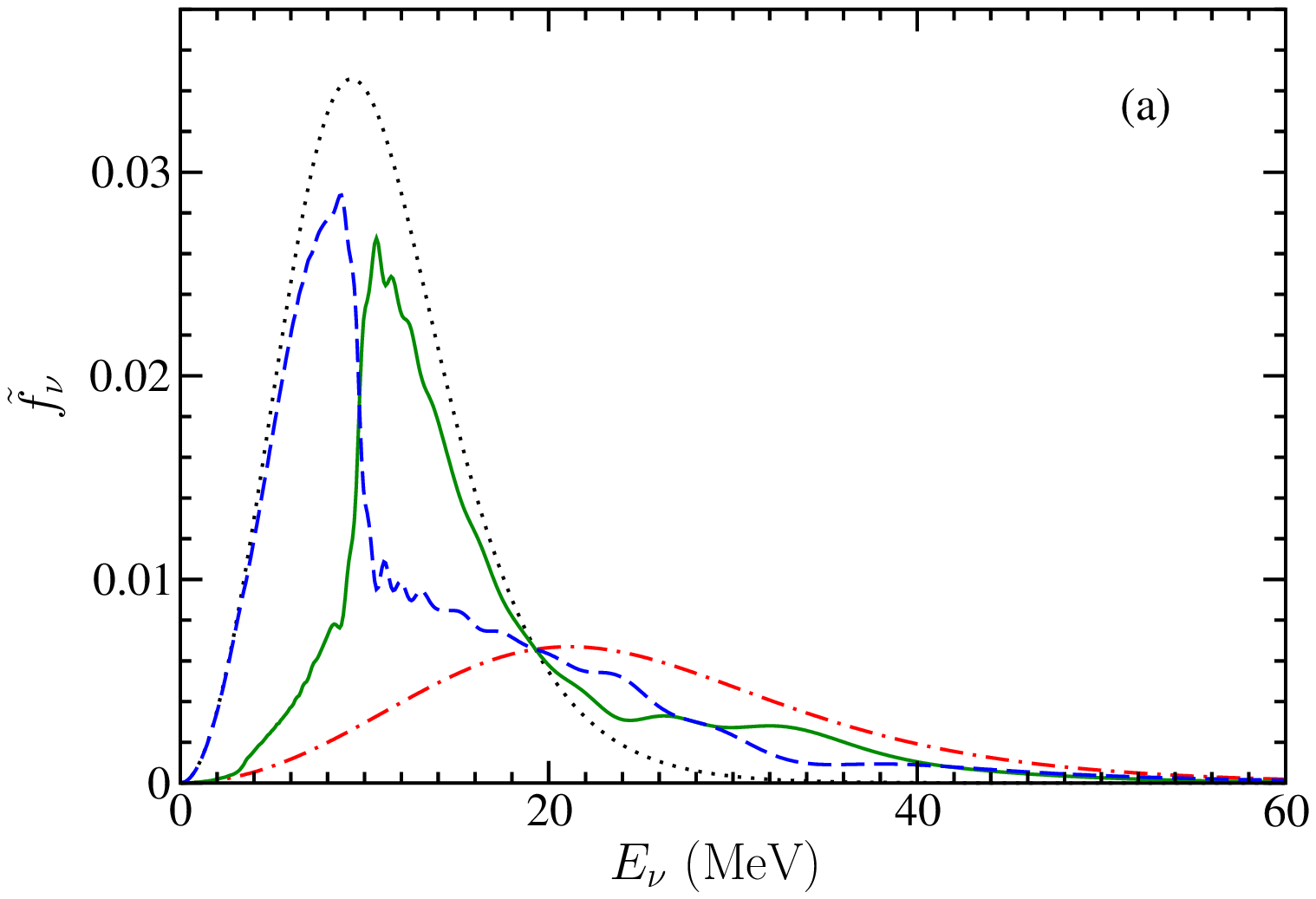} &
\includegraphics*[width=\myfigwid, keepaspectratio]{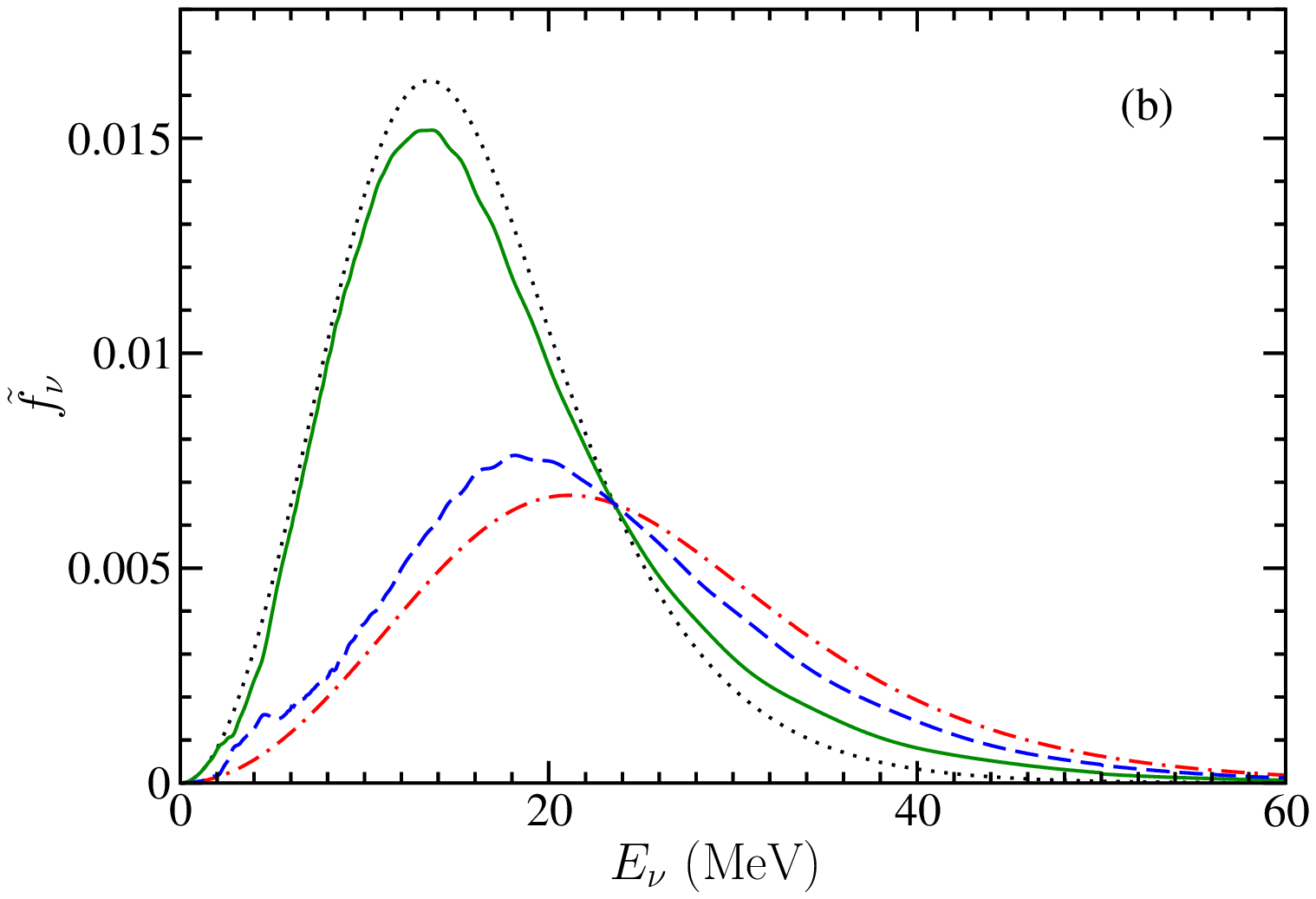} \\
\includegraphics*[width=\myfigwid, keepaspectratio]{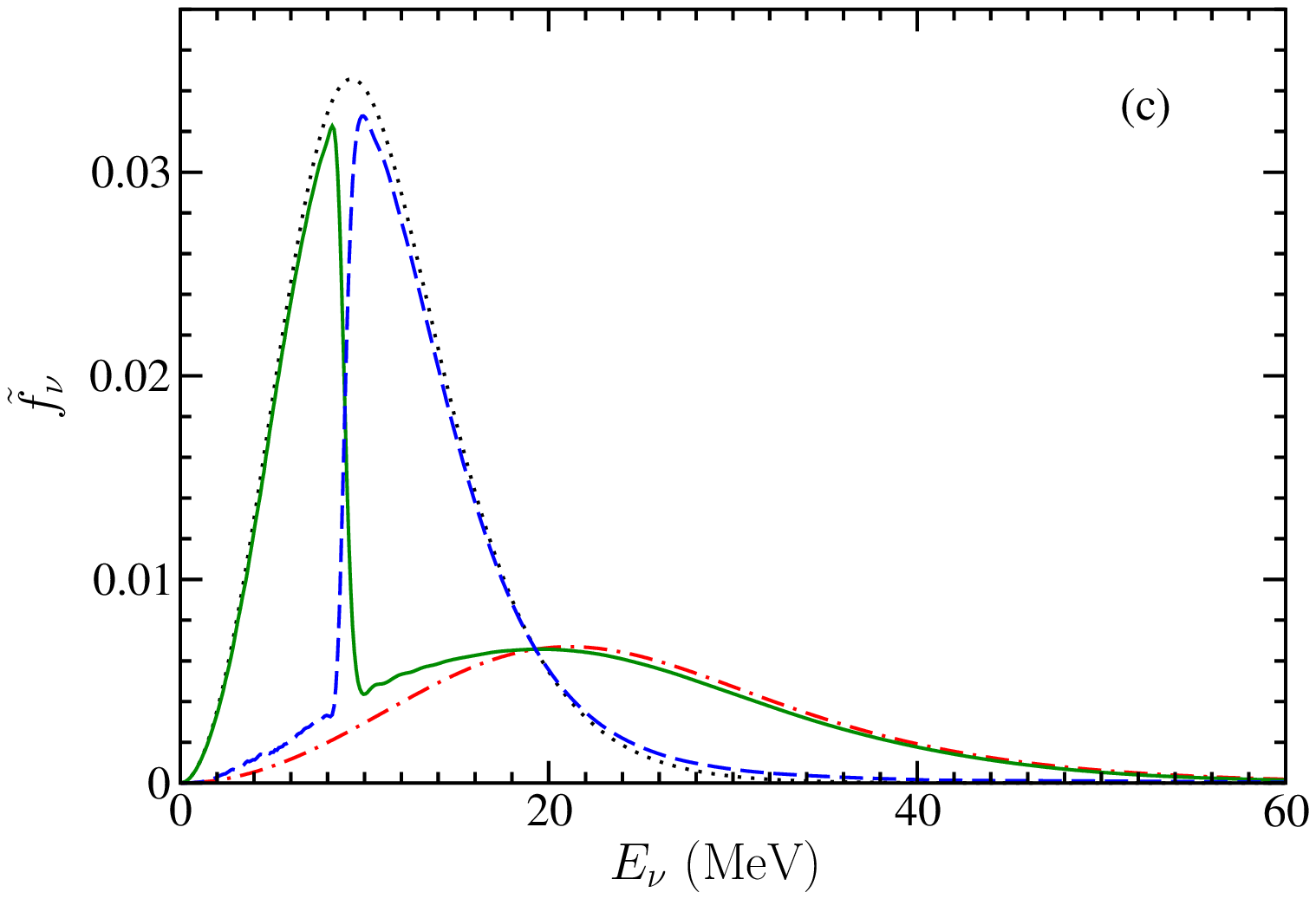} &
\includegraphics*[width=\myfigwid, keepaspectratio]{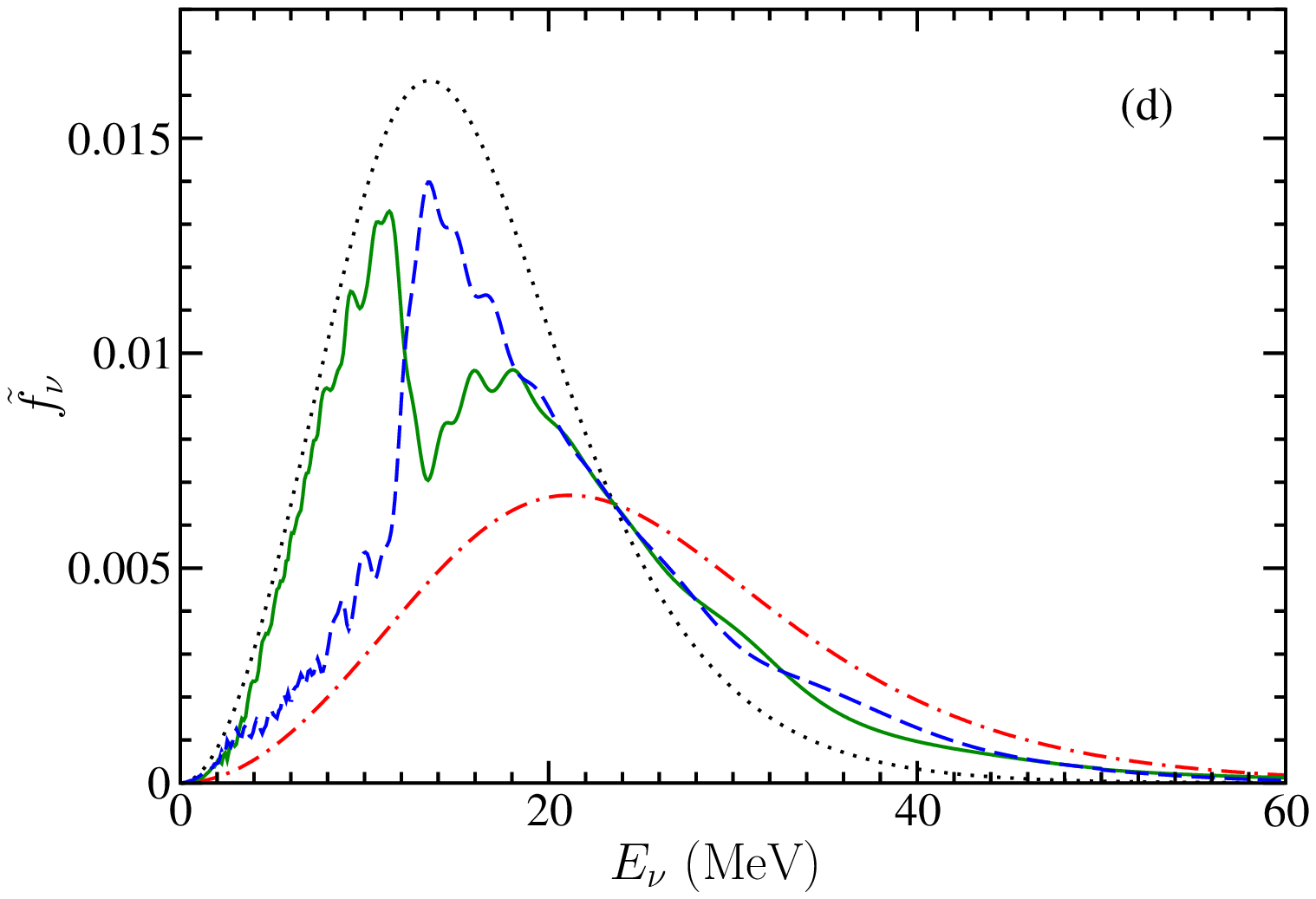}
\end{array}$
\end{center}
\caption{\label{fig:spectra}(Color online) 
Change of energy spectra of neutrinos (left panels) and antineutrinos
(right panels) with the normal (upper panels) and inverted (lower panels)
neutrino mass hierarchies. The dotted and dot-dashed lines are the
spectra of neutrinos (antineutrinos) in the electron and tau
flavors, respectively, at $r=R_\nu$, and the solid and dashed lines
are the corresponding spectra at $r=250$ km.}
\end{figure*}

In Fig.~\ref{fig:P-r}(b) we plot $\langle\Paeae(r)\rangle$ 
in the normal neutrino mass hierarchy scenario.
The conventional MSW conversion of $\bar\nu_e$ is suppressed 
if $\delta m^2 >0$, which is illustrated dramatically by the $L_\nu=0$
case. In the case of full synchronization, $\bar\nu_e$
will be converted into $\bar\nu_\tau$ in exactly the same way 
as $\nu_e$ is converted into
$\nu_\tau$. The results of our simulations are again
like neither of these limits. Unlike the $L_\nu=0$ case,
the actual values of $\langle\Paeae(r)\rangle$  may substantially
decrease at some values of radius, and unlike the full synchronization case,
$\langle\Paeae(r)\rangle$ oscillates and bounces 
back to nearly unity at large radius.

The results of the inverted neutrino mass hierarchy ($\delta m^2<0$)
are more surprising. These are plotted in panels (c) and (d) of
Fig.~\ref{fig:P-r}. The full synchronization
limit predicts no flavor conversion for both $\nu_e$
and $\bar\nu_e$, and the $L_\nu=0$ limit predicts that
only antineutrinos will be converted. Our simulation
finds substantial conversion of both $\nu_e$ and $\bar\nu_e$.
Furthermore, this phenomenon occurs at a radius even smaller than
that expected in the full synchronization limit with $\delta m^2>0$.
Again, we note that in the inverted mass hierarchy scenario
both $\langle\Pee(r)\rangle$ and $\langle\Paeae(r)\rangle$
oscillate after flavor transformation starts.

In Fig.~\ref{fig:P-r-nue}
we plot $\Pee(r)$ for $\nuI{e}$ with a few characteristic
energies on both the radial and tangential trajectories.
We have employed the normal mass hierarchy in this calculation.
One sees that the $\Pee(r)$ curves have similar trends 
with radius over most of the $\nuI{e}$  energy range considered.
This is especially true for the values of radius where neutrino flavor
transformation has just become significant and for the tangential
trajectory.

The results presented in Fig.~\ref{fig:P-r-nue} lead us to conclude
that the flavor
transformation histories of neutrinos on different trajectories
can be very different. To illustrate this point more clearly,
we plot in Fig.~\ref{fig:P-theta0}(a)
$\Pee(\cos\vartheta_0)$ at $r\simeq 92.85\,\mathrm{km}$  
for $\nuI{e}$ with specified energies,
and employing the normal mass hierarchy. Indeed the values of
$\Pee(\cos\vartheta_0)$ vary with angle, especially around
$\cos\vartheta_0=1$. Moreover, the trend of $\Pee(\cos\vartheta_0)$ 
with angle is similar  for $\nuI{e}$ with different energies
over most of the energy range considered. This again demonstrates the
collective feature of the neutrino flavor transformation 
in the hot bubble.

In Fig.~\ref{fig:P-theta0}(b) we
plot the corresponding antineutrino survival
probability $\Paeae(\cos\vartheta_0)$.
This also shows angular dependence and collective
flavor transformation. In Fig.~\ref{fig:P-theta0}(c--d),
we plot $\Pee(\cos\vartheta_0)$ and $\Paeae(\cos\vartheta_0)$ 
with the same parameters as in
panels (a--b) but at $r\simeq 88.57$ km and with the inverted
mass hierarchy. It is interesting to see that,
in addition to the features pointed out for panels (a--b),
in the inverted mass hierarchy case 
both $\Pee(\cos\vartheta_0)$ and $\Paeae(\cos\vartheta_0)$
oscillate over most of the range of $\cos\vartheta_0$.

In these simulations,  significant neutrino flavor transformation
ends at $r\sim 230$ km (Fig.~\ref{fig:P-r}). 
To see how the energy spectra of neutrinos and antineutrinos
have been altered by flavor transformation, 
in Fig.~\ref{fig:spectra}(a) we plot
both $\tilde{f}_{\nuI{e}}(E)$ and $\tilde{f}_{\nuI{\tau}}(E)$
at the neutrino sphere, and
$\tilde{f}_{\nu_e}(E)$ and $\tilde{f}_{\nu_\tau}(E)$ 
at $r=250$ km. Here we have employed the normal mass hierarchy
and we take $\tilde{f}_\nu(E)$ to be proportional to both $f_\nu(E)$
and the flux of $\nu$ [\textit{e.g.}, 
$\tilde{f}_{\nu_e}(E)\propto f_{\nu_e}(E)
\sum_\alpha\int P_{\nuI{\alpha}\nu_e}\ud n_{\nuI{\alpha}}$
and
$\tilde{f}_{\nuI{e}}(E)\propto f_{\nuI{e}}(E)L_{\nuI{e}}/\avgE{e}$], such that
\begin{subequations}
\begin{align}
1 &= 
\sum_\alpha\int [\tilde{f}_{\nu_\alpha}(E)+\tilde{f}_{\bar\nu_\alpha}(E)]\,\ud E\\
&= 
\sum_\alpha\int [\tilde{f}_{\nuI{\alpha}}(E)+\tilde{f}_{\anuI{\alpha}}(E)]\,\ud E.
\end{align}
\end{subequations}
Here the scheme for angle-averaging the energy spectra is simply the angle
dependence in the neutrino flux ``seen'' by a nucleon at radius $r$. 
As a result, the
angle-averaged spectra shown are those appropriate for use in the weak
interaction rates.
It is interesting
to see that most of the low-energy ($E_\nu\lesssim 9.5$ MeV) $\nu_e$ 
are converted into $\nu_\tau$, while a significant fraction of high
energy $\nu_e$ survive. 
We also plot the corresponding energy spectra of $\bar\nu_e$
and $\bar\nu_\tau$ in Fig.~\ref{fig:spectra}(b). The energy
spectra of antineutrinos are changed very little in
the normal mass hierarchy scenario. The energy
spectra of neutrinos and antineutrinos in the inverted mass
hierarchy scenario are plotted in 
Fig.~\ref{fig:spectra}(c) and (d), respectively. 
In these figures,
both $\nu_e$ and $\bar\nu_e$ swap spectra with
$\nu_\tau$ and $\bar\nu_\tau$, respectively, over a significant energy range.

The numerical results that we have presented
cannot be explained easily
by the conventional MSW mechanism or by synchronization.
We will try to develop some insight into, and understanding of these
results in the following section.

\section{Single-Angle Simulations and Phenomenological Analysis%
\label{sec:analysis}}

To understand the numerical results obtained from
the multiangle simulations, we have
re-examined the numerical simulations using the single-angle
approximation with similar setups and initial conditions.
We found that almost all the interesting
features seen in the multiangle simulations are also
present in the single-angle simulations, though they
can differ in a quantitative sense. The
simulations performed using the single-angle approximation 
do not have the numerical difficulties that are
the hallmark of the multiangle ones, and they
require fewer computational resources. 
Most importantly, the single-angle simulations produce
results qualitatively similar to those in the multiangle simulations,
and yet do not
involve  complicated entanglement of  neutrino
flavor transformation on different trajectories.
They are therefore easier to understand.
In Sec.~\ref{sec:bipolar} we will try to
explain some of the results presented in Sec.~\ref{sec:results}
with the help of these simplified
calculations. In Sec.~\ref{sec:start-collective} we will
study how the onset of large-scale collective neutrino
flavor transformation is related to the neutrino
luminosity $L_\nu$. We will comment on the validity
of the single-angle approximation at the end
of this section.

Unless otherwise stated,
all the simulations discussed in this section
have the same parameters as those in Sec.~\ref{sec:multiangle},
\textit{i.e.}, $|\delta m^2| = 3\times 10^{-3}\,\mathrm{eV}^2$,
$\theta = 0.1$, $L_\nu = 10^{51}$ erg/s
and $S=140$, but are based on the single-angle approximation.

\begin{figure*}
\begin{center}
$\begin{array}{@{}c@{\hspace{\myfigsep}}c@{}}
\includegraphics*[width=\myfigwid, keepaspectratio]{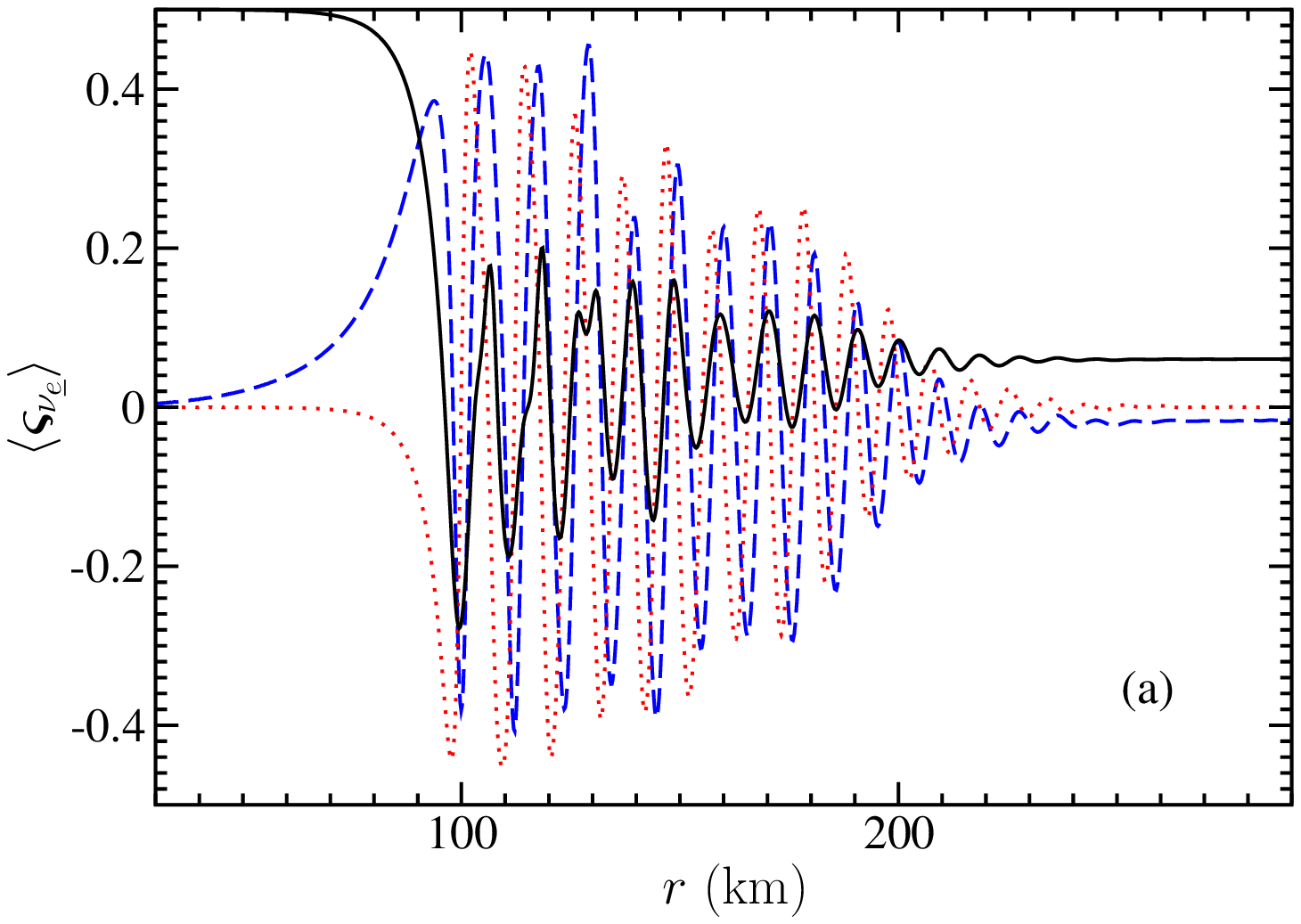} &
\includegraphics*[width=\myfigwid, keepaspectratio]{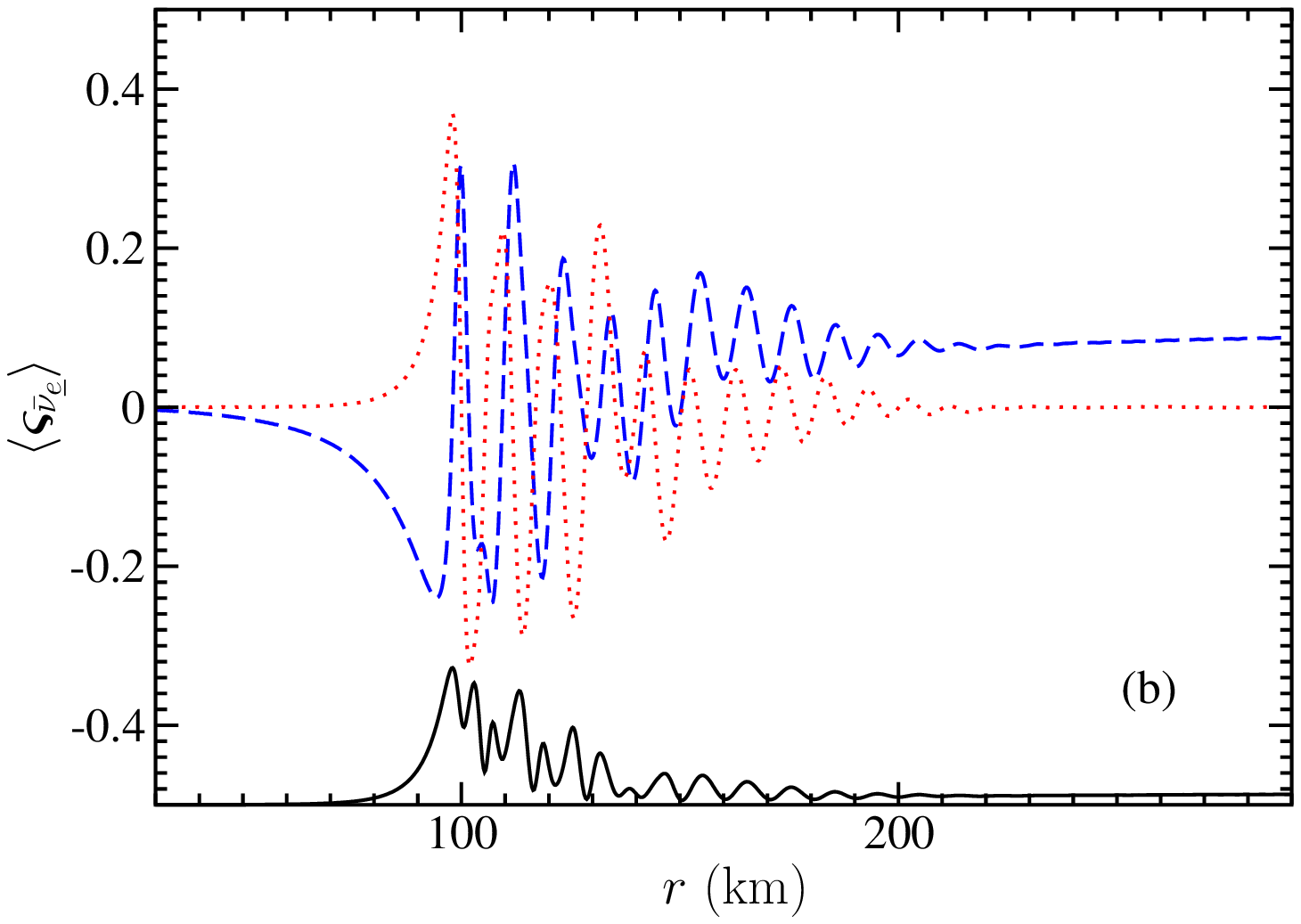} \\
\includegraphics*[width=\myfigwid, keepaspectratio]{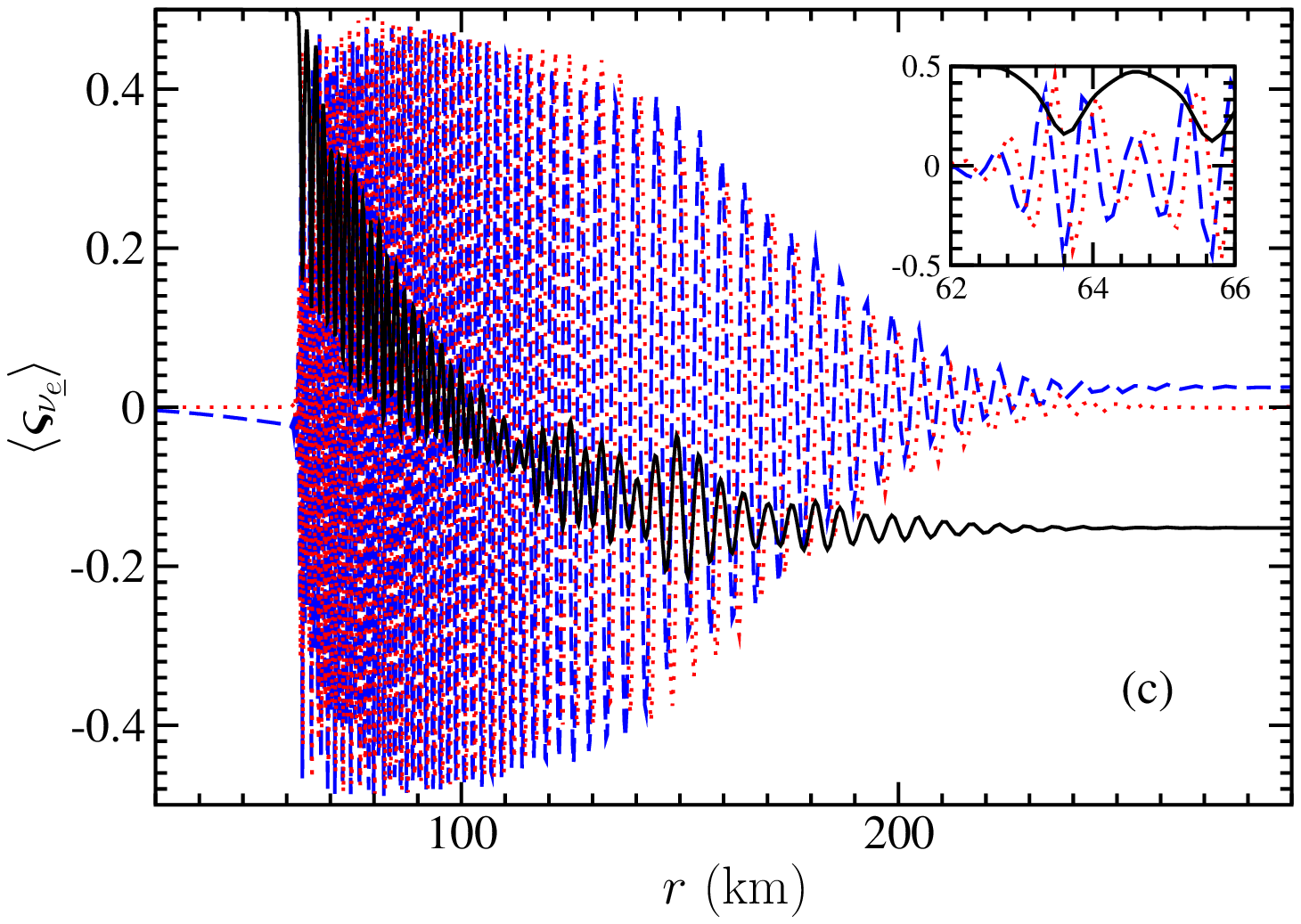} &
\includegraphics*[width=\myfigwid, keepaspectratio]{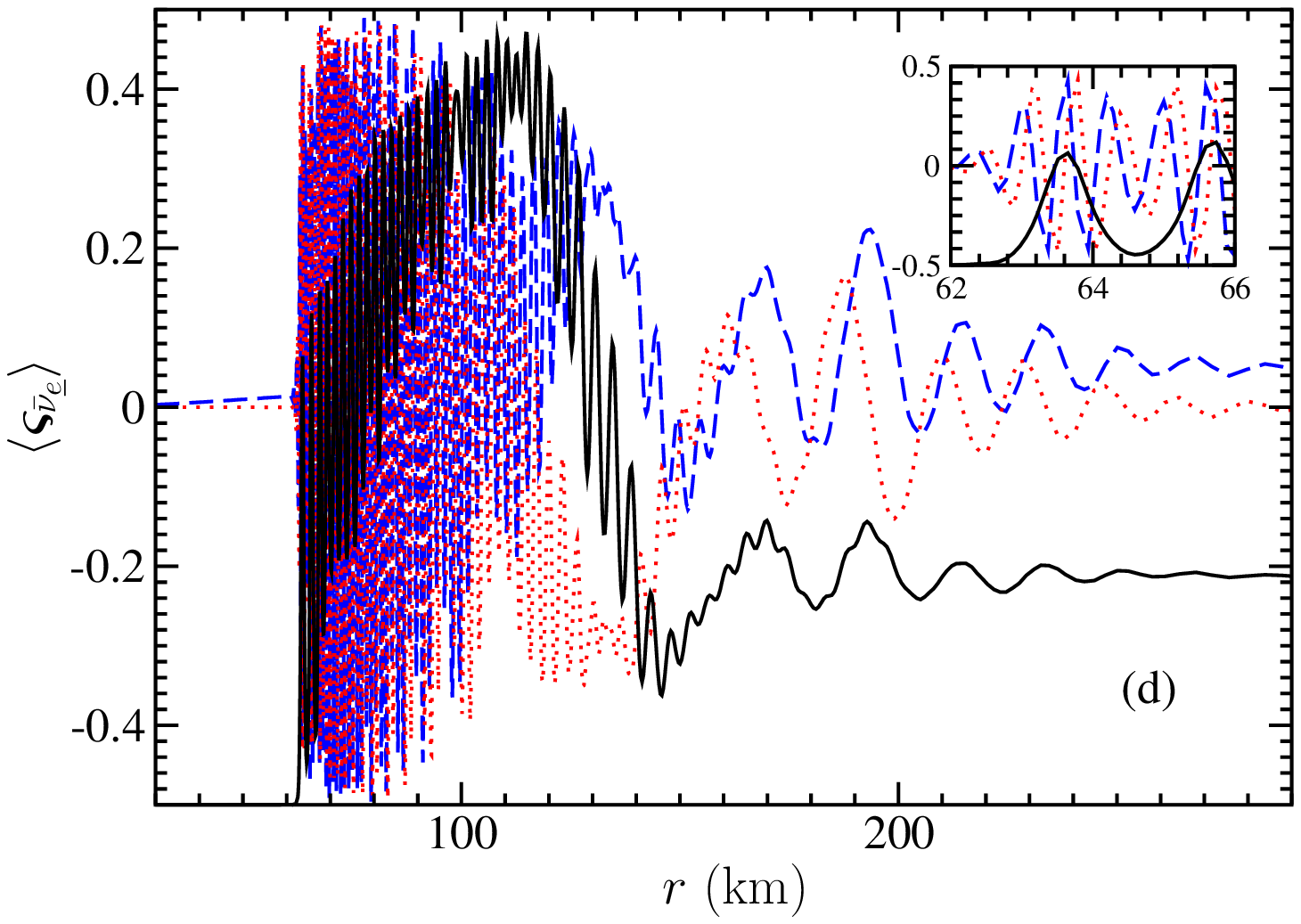}
\end{array}$
\end{center}
\caption{\label{fig:s-r}(Color online) 
Plots of $\langle\sB_{\nuI{e}}(r)\rangle$ (left panels) and 
$\langle\sB_{\anuI{e}}(r)\rangle$ (right panels) 
with the normal (upper panels) and inverted
(lower panels) mass hierarchies, respectively.
The dashed, dotted and solid
lines represent $\langle\varsigma_x(r)\rangle$, 
$\langle\varsigma_y(r)\rangle$ 
and $\langle\varsigma_z(r)\rangle$,
respectively. Note that 
$\langle\Pee\rangle=1/2+\langle\varsigma_{\nuI{e}z}\rangle$ 
and $\langle\Paeae\rangle=1/2-\langle\varsigma_{\anuI{e}z}\rangle$.
The insets in panels (c) and (d) are blowups of the corresponding
plots in the range $62\,\mathrm{km}\leq r\leq 66\,\mathrm{km}$.
These are single-angle calculation results.}
\end{figure*}

\subsection{Neutrino Flavor Transformation in the Bi-Polar Mode%
\label{sec:bipolar}}

The novel features of neutrino flavor transformation in the hot bubble
region are easier to understand in the formalism of
NFIS (Neutrino Flavor Iso-Spin) \cite{Duan:2005cp} than in the
traditional formalism of the wave functions. 
In Fig.~\ref{fig:s-r}, we plot $\langle\varsigma_x(r)\rangle$, 
$\langle\varsigma_y(r)\rangle$ and $\langle\varsigma_z(r)\rangle$,
the three components of the average NFIS's in flavor space, for
$\nuI{e}$ and $\anuI{e}$ in both the scenarios with
a normal mass hierarchy and with an inverted mass hierarchy.
(The three components of the NFIS's are averaged over the initial
neutrino or antineutrino energy spectra.)
We note that the probability for a neutrino or antineutrino
initially in the $\alpha$ flavor state
to be in the  electron flavor state is related to $\varsigma_z$ by
\begin{subequations}
\label{eq:P-sz}
\begin{align}
P_{\nuI{\alpha}\nu_e} &= \frac{1}{2} + \varsigma_{\nuI{\alpha}z}, \\
P_{\anuI{\alpha}\bar\nu_e} &= \frac{1}{2} - \varsigma_{\anuI{\alpha}z}.
\end{align}
\end{subequations}
Comparing Fig.~\ref{fig:s-r} with Fig.~\ref{fig:P-r},
one sees that the results of single-angle
simulations are qualitatively the same as those obtained 
in the full multiangle simulations. We also note that
in the region where neutrinos transform,
the NFIS's of both neutrinos and antineutrinos
have large values of $\varsigma_x$ and $\varsigma_y$, and roughly precess
 around the $\basef{z}$ direction.
Because the densities of neutrinos and antineutrinos are also
large in this region, the $B_{e\tau}$ potential 
in Eq.~\eqref{eq:schroedinger-eq}
dominates, and both neutrinos and antineutrinos
are in a state similar to the Background Dominant
Solution (BDS) \cite{Fuller:2005ae}.

The numerical results clearly have  shown that neutrinos
and antineutrinos undergo some collective flavor
transformation in the hot bubble with the neutrino 
mixing parameters we have used. The collective modes of
flavor transformation that neutrinos may have in
the hot bubble, according to Ref.~\cite{Duan:2005cp},
are either the synchronization or the bi-polar type. 
FIGs.~\ref{fig:P-r} and \ref{fig:s-r} show that the collective
mode corresponding to the conditions and parameters used here
does not conform to the full synchronization limit.
Therefore, we will focus our discussion on the bi-polar
flavor transformation. Neutrinos
and antineutrinos can have substantial flavor
transformation simultaneously only through the bi-polar mode in the inverted
mass hierarchy scenario. This is seen in
our simulations. In general, the region where neutrinos transform
through the bi-polar mode is characterized by parameters
satisfying \cite{Duan:2005cp}
\begin{equation}
\epsilon \lesssim 
\kappa \lesssim
\frac{\langle E_\nu\rangle}{2\delta E_\nu},
\label{eq:bipolar-cond}
\end{equation}
where $\epsilon$ is a measure of the difference 
in the energy distribution functions of $\nuI{e}+\anuI{\tau}$
and $\anuI{e}+\nuI{\tau}$, the parameter
\begin{equation}
\kappa \equiv 
\frac{|\delta m^2|/2 \langle E_\nu\rangle}{2\sqrt{2}\GF n_\nu^\eff(L_\nu,r)}
\label{eq:kappa-def}
\end{equation}
gives the strength of background neutrino effect
through the effective single species neutrino number density $n_\nu^\eff$
[\textit{e.g.}, 
$n_\nu^\eff(L_\nu,r)\simeq D(r/R_\nu)(L_\nu/\langle E_\nu\rangle)/2\pi R_\nu^2$
for the radial trajectory],
and $\delta E_\nu$ is the characteristic width of the 
neutrino energy distribution.

Neutrinos are in the synchronization mode if $\epsilon \gtrsim \kappa$.
This corresponds to the high neutrino luminosity limit.
Neutrinos will transform individually through the
MSW mechanism if $\kappa \gtrsim \langle E_\nu\rangle/2\delta E_\nu$,
which is effectively the low neutrino luminosity limit.
Using the single-angle approximation, one has \cite{Duan:2005cp}
\begin{subequations}
\begin{align}
\kappa &=  
\frac{|\delta m^2| \pi R_\nu^2}{\sqrt{2}\GF L_\nu}
\left[1-\sqrt{1-(R_\nu/r)^2}\right]^{-2}\\
&\simeq 3.6\times 10^{-6} 
\left(\frac{|\delta m^2|}{3\times 10^{-3}\,\mathrm{eV}^2}\right)
\left(\frac{R_\nu}{10\,\mathrm{km}}\right)^2
\nonumber\\
&\quad\times\left(\frac{10^{51}\,\mathrm{erg/s}}{L_\nu}\right)
\left[1-\sqrt{1-(R_\nu/r)^2}\right]^{-2}.
\label{eq:kappa-approx}
\end{align}
\end{subequations}
For simplicity, we assume that $\langle E_{\nuI{e}} \rangle$ and 
$\langle E_{\anuI{e}} \rangle$ characterize the energies
of $\nuI{e}+\anuI{\tau}$ and $\anuI{e}+\nuI{\tau}$,
respectively, and \cite{Duan:2005cp}
\begin{equation}
\epsilon \simeq
\frac{(\langle E_{\nuI{e}}\rangle -\langle E_{\anuI{e}}\rangle)^2}%
{2(\langle E_{\nuI{e}}\rangle^2+\langle E_{\anuI{e}}\rangle^2)}
\simeq 0.033.
\label{eq:epsilon}
\end{equation}
Using Eqs.~\eqref{eq:bipolar-cond},
\eqref{eq:kappa-approx} and \eqref{eq:epsilon}, we estimate
that neutrino flavor transformation exits the synchronization mode
and enters the bi-polar mode (Bi-polar Starting)
at $\rBS\sim 73$ km
for the parameters we have used. Taking 
$\delta E_\nu/\langle E_\nu\rangle \simeq T_\nu/\langle E_\nu\rangle\simeq 1/4$,
 we estimate 
that collective neutrino flavor transformation ends 
(Bi-polar Ending) at $\rBE\sim 202$ km. 
Beyond this point conventional MSW flavor transformation takes over.

Although these values are crude estimates based
on oversimplified assumptions, we find that they roughly match
the region where neutrinos and antineutrinos transform simultaneously
in the inverted mass hierarchy scenario. Therefore we conclude
that collective neutrino flavor transformation observed
in our full numerical simulations is indeed of the bi-polar type, as predicted
in Ref.~\cite{Duan:2005cp}.

\begin{figure*}
\begin{center}
$\begin{array}{@{}c@{\hspace{\myfigsep}}c@{}}
\includegraphics*[width=\myfigwid, keepaspectratio]{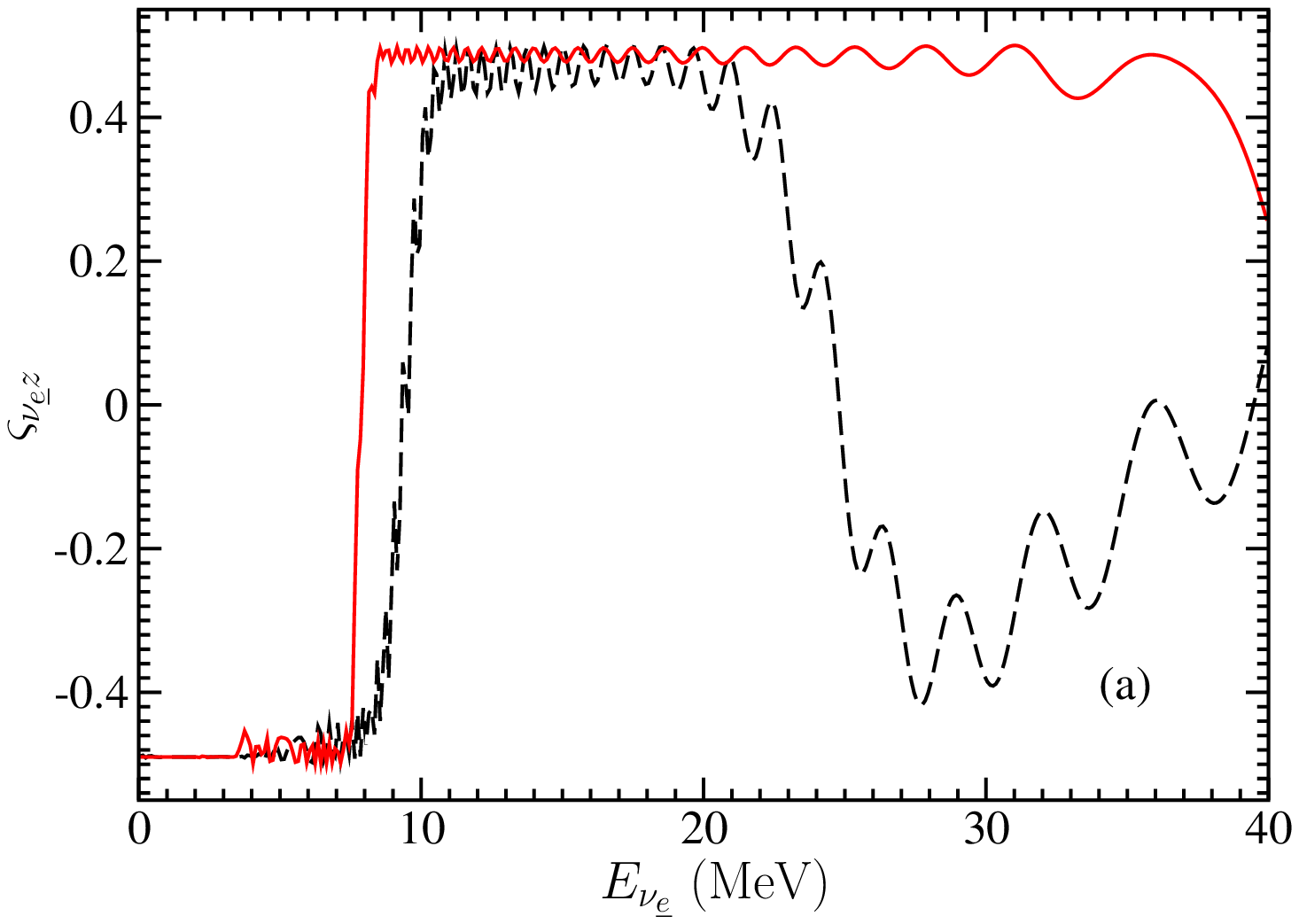} &
\includegraphics*[width=\myfigwid, keepaspectratio]{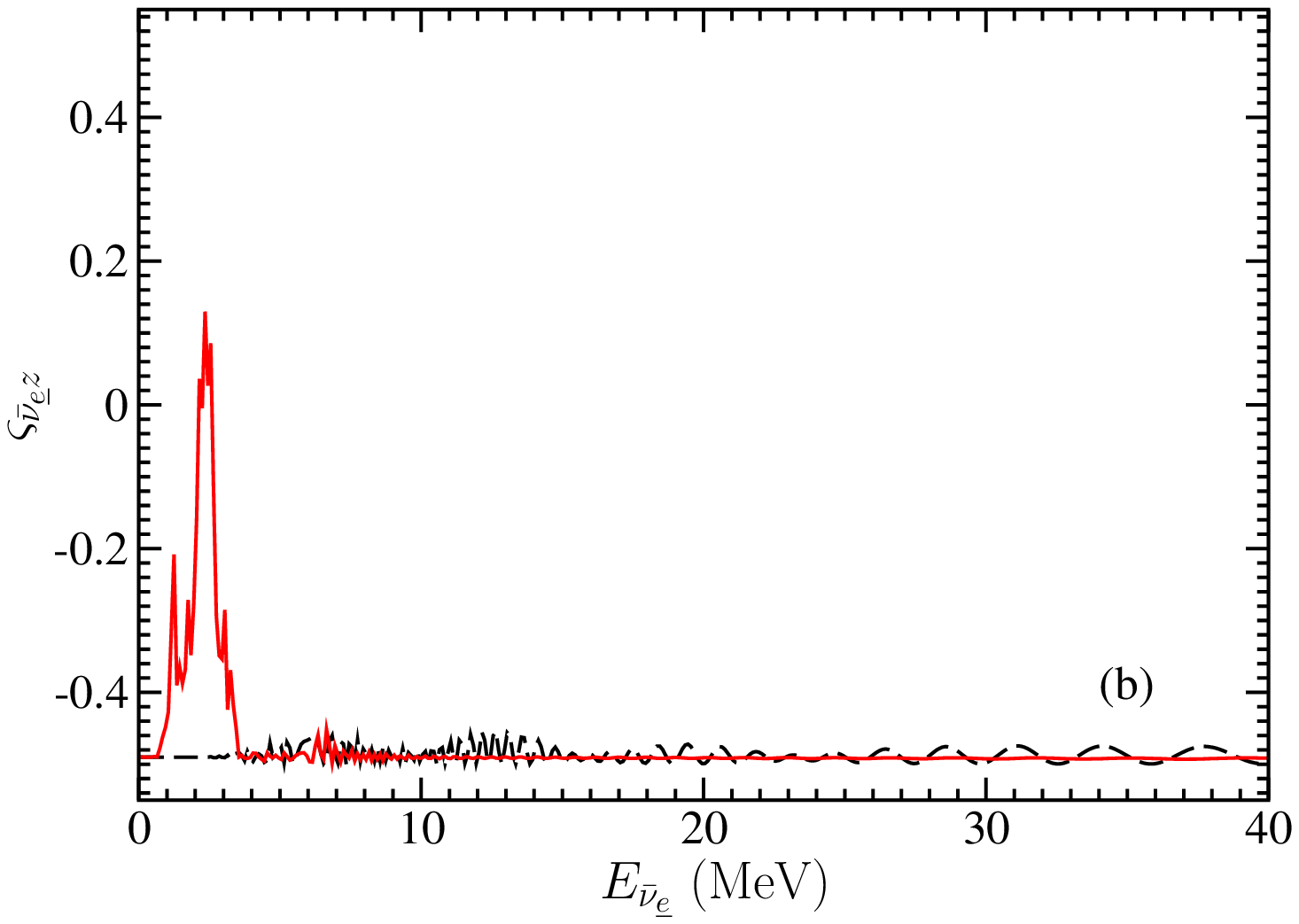} \\
\includegraphics*[width=\myfigwid, keepaspectratio]{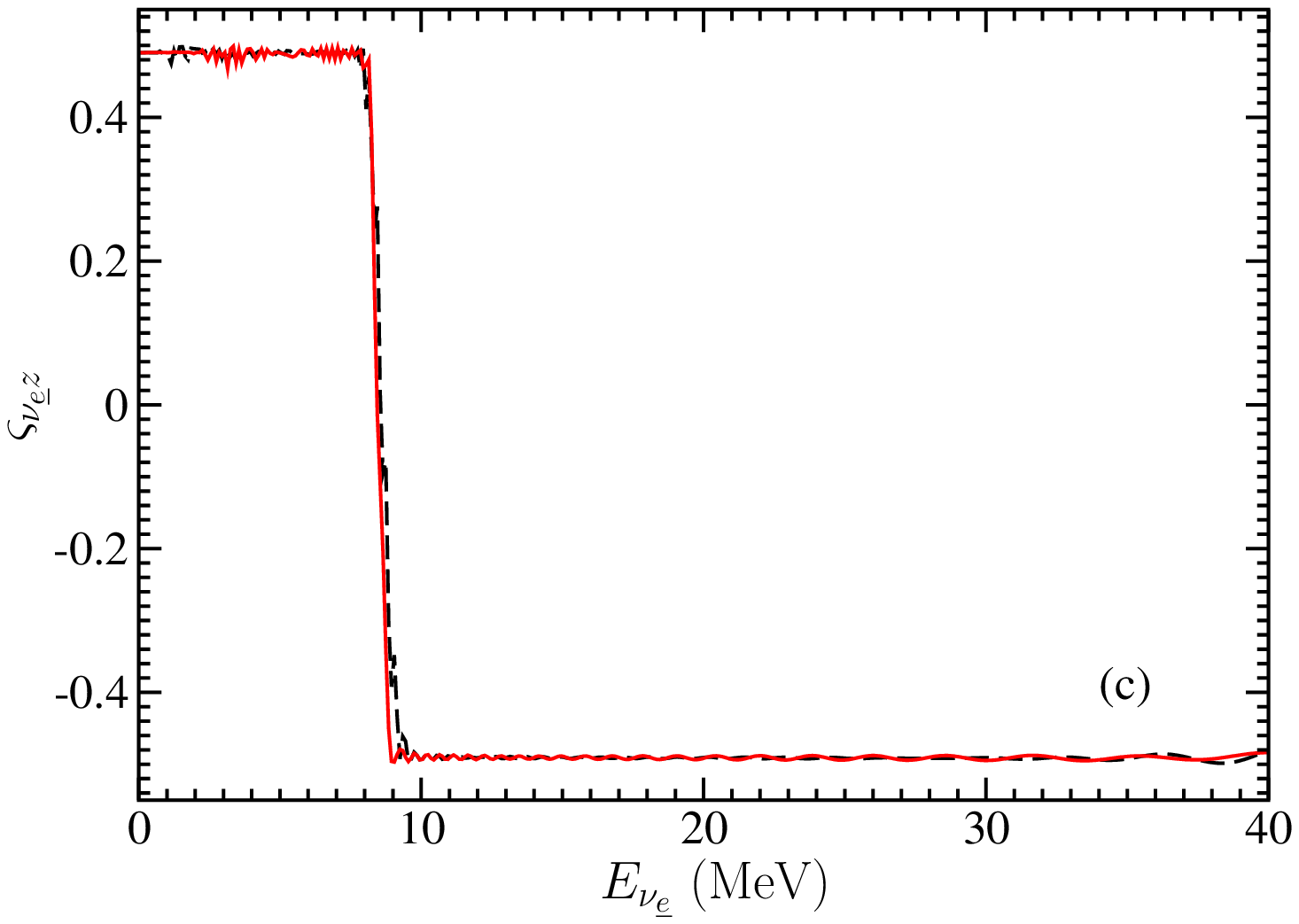} &
\includegraphics*[width=\myfigwid, keepaspectratio]{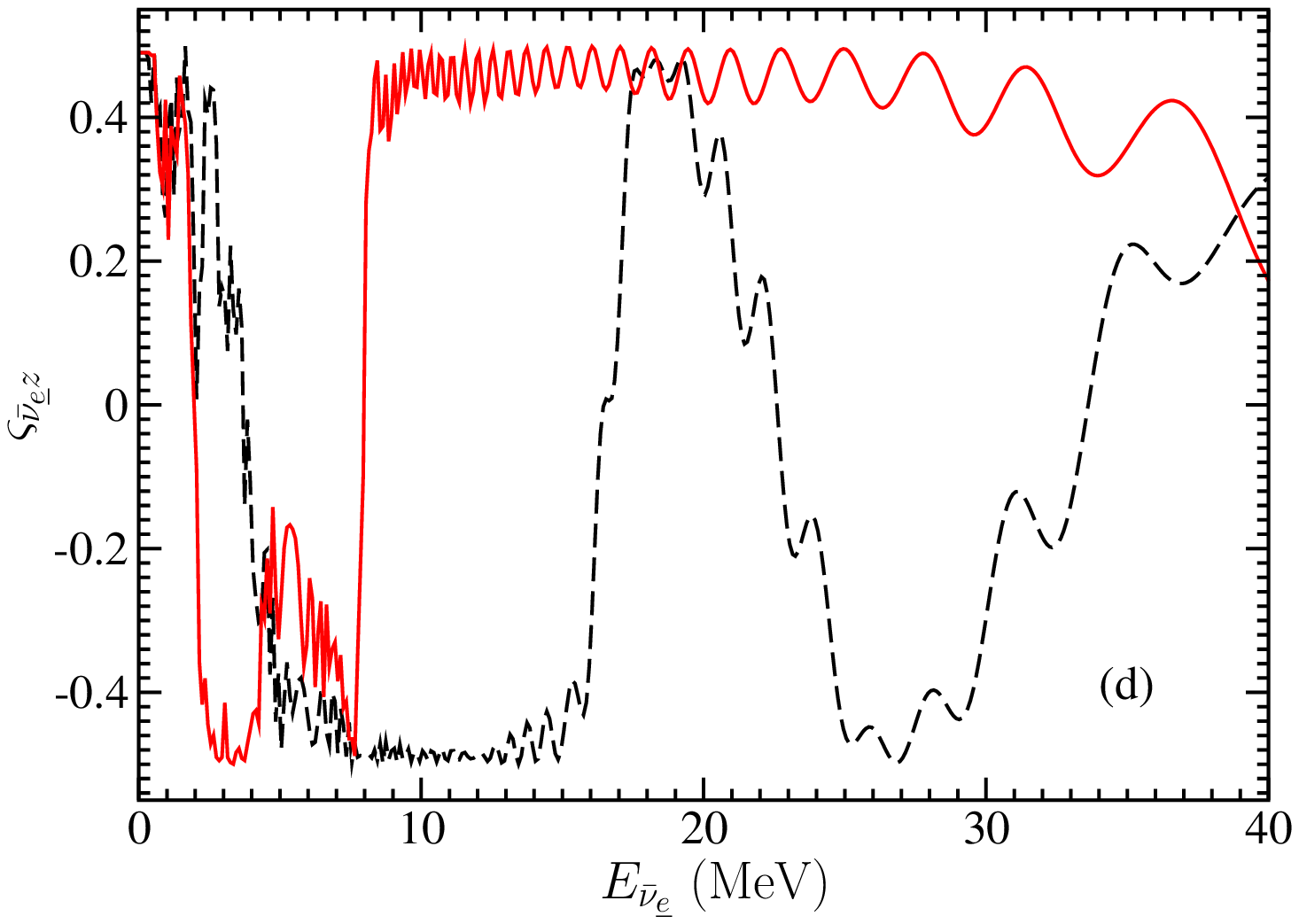}
\end{array}$
\end{center}
\caption{\label{fig:sz-E}(Color online) 
Plots of $\varsigma_{\nuI{e}z}(E_{\nuI{e}})$ (left panels) and 
$\varsigma_{\anuI{e}z}(E_{\anuI{e}})$
(right panels) for both the
normal (upper panels) and inverted (lower panels) mass hierarchies
at $r=400$ km.
The dashed and solid lines are for $L_\nu=10^{51}$ and $5\times 10^{51}$
erg/s, respectively. These are single-angle calculation results.}
\end{figure*}

An interesting feature of
 neutrino flavor transformation in the bi-polar mode is
that the transformation is not completely suppressed by the large
matter potential $A$ in Eq.~\eqref{eq:schroedinger-eq}
if $\delta m^2 <0$.
From Fig.~\ref{fig:P-r} and Fig.~\ref{fig:s-r}, one sees
that the flavor transformation with $\delta m^2<0$ may occur at values of radius
even smaller than those predicted in the synchronization limit
with $\delta m^2>0$.
Although collective neutrino flavor transformation
in the bi-polar mode has been studied in
the zero and large matter potential limits \cite{Duan:2005cp},
an analytical or semianalytical analysis has
yet to be performed to show how 
neutrinos transform in the bi-polar mode as the
matter potential $A$ decreases and approaches 
the vacuum potential $\Delta$.

In Fig.~\ref{fig:sz-E}(a) we plot $\varsigma_{\nuI{e}z}(E_{\nuI{e}})$
in the normal mass hierarchy scenario at 400 km for the cases 
with $L_\nu=10^{51}$
and $5\times 10^{51}$ erg/s. One immediately sees that there is
a rather sharp transition edge 
at $\EC \simeq 9.5$ and 7.9 MeV for $L_\nu=10^{51}$
and $5\times 10^{51}$ erg/s, respectively. Noting that
$\Pee = 1/2 + \varsigma_{\nuI{e}z}$, one sees that
$\nu_e$ with energy below $\EC$ are almost fully converted into
$\nu_\tau$, while $\nu_e$ with energy above $\EC$ but below
another threshold $\EH$ mostly survive.
The threshold $\EH$ is roughly at $\sim 22$ and 40 MeV for
$L_\nu=10^{51}$ and $5\times 10^{51}$ erg/s, respectively. 
Because $\Ptt(E)=\Pee(E)$, there
exists a similar transition edge for $\nuI{\tau}$.
This difference in the flavor transformation of the neutrinos of low 
and high energies
is responsible for the partial swap of the spectra of $\nu_e$
and $\nu_\tau$ seen in Fig.~\ref{fig:spectra}(a). 
We also plot the
corresponding values of $\varsigma_{\anuI{e}z}(E_{\anuI{e}})$ 
in Fig.~\ref{fig:sz-E}(b). Knowing that
$\Paeae = 1/2 - \varsigma_{\anuI{e}z}$, one
sees that most of the $\bar\nu_e$ survive. This is also true for
$\bar\nu_\tau$. 

\begin{figure*}
\begin{center}
\includegraphics*[width=0.8 \textwidth, keepaspectratio]{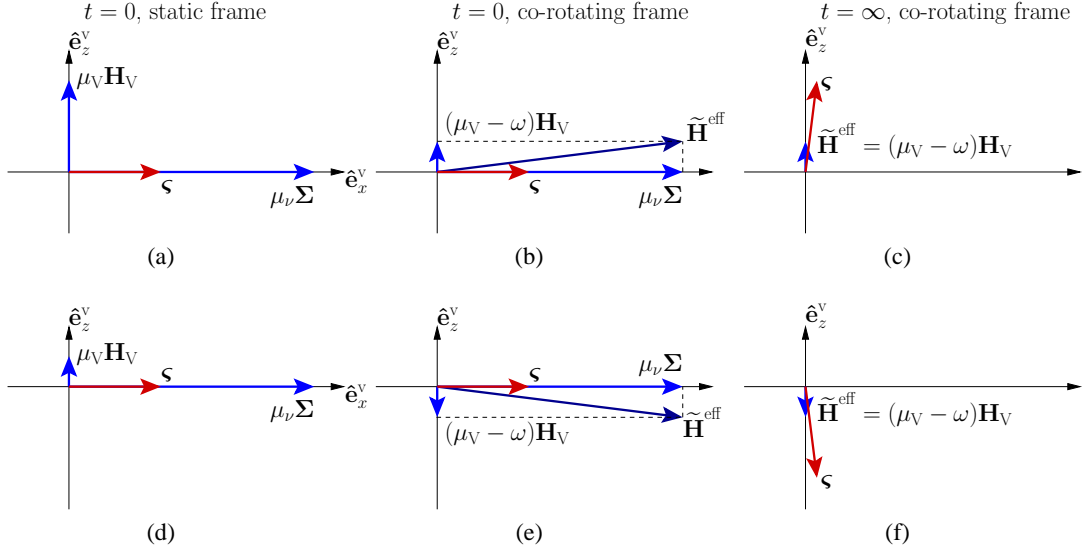}
\end{center}
\caption{\label{fig:rot}(Color online) 
The toy scenario explaining the evolution of NFIS $\sB$. 
The NFIS $\sB$ can be viewed as a ``magnetic spin'' which
 is coupled with a constant field
$\HV$ and a field $\bm{\Sigma}$ rotating with angular frequency $\omega$. 
We actually consider here two ``magnetic spin'' couplings,
one with``magnetic moment'' $\muV{}$ and the other with $\mu_\nu$.
The vector $\mu_\nu\bm{\Sigma}$ rotates in the $\basev{x}$--$\basev{y}$
plane in the clockwise sense when viewed from above, looking in the
$-\basev{z}$ direction. The ``magnetic spin'' 
$\sB$ is aligned initially 
with the dominant field $\bm{\Sigma}$ at time $t=0$ [panels (a) and (d)].
The problem is easily solved in the corotating frame 
where $\bm{\Sigma}$ is not rotating. In this corotating frame,
the ``magnetic spin'' $\sB$ rotates around the effective field
$\bm{\widetilde{\Hb}}^\eff=(\muV{}-\omega)\HV+\mu_\nu\bm{\Sigma}$
[panels (b) and (e)]. If $\bm{\Sigma}$ slowly reduces its length to zero,
the angle between the spin $\sB$ and the effective field 
$\bm{\widetilde{\Hb}}^\eff$ is constant (adiabatic process), and
$\sB$ ends up almost aligned with $\HV$ at $t=\infty$ 
if $\muV{}-\omega>0$ [panel (c)]
or almost antialigned with $\HV$ if $\muV{}-\omega<0$ [panel (f)].}
\end{figure*}

We plot $\varsigma_{\nuI{e}z}(E_{\nuI{e}})$ 
and $\varsigma_{\anuI{e}z}(E_{\anuI{e}})$ 
with the inverted mass hierarchy
in Fig.~\ref{fig:sz-E}(c) and (d), respectively. There also
we see a transition edge with $\EC \simeq 8.5$ MeV for
$\nu_e$, which is similar to that with the normal mass hierarchy,
but reversed in direction. We note that $\EC$ is essentially the same
for both $L_\nu=10^{51}$ and $5\times 10^{51}$ erg/s
in the case of an inverted mass hierarchy.
This transition edge results in the partial swap of the spectra of $\nu_e$
and $\nu_\tau$ shown in Fig.~\ref{fig:spectra}(c).
The behavior of $\anuI{e}$ is more complicated. Roughly speaking,
$\bar\nu_e$ with energy below some threshold $\EL$ or
between $\EM$ and $\EH$ are mostly converted into $\bar\nu_\tau$,
where $\EL\simeq 3$ and 1.8 MeV, $\EM \simeq 16.5$ and 8 MeV,
and $\EH \simeq 20$ and 40 MeV for $L_\nu=10^{51}$ and 
$5\times 10^{51}$ erg/s, respectively. 

As mentioned above,
we do not have an analytical or semianalytical analysis
of the bi-polar mode flavor transformation in the general cases.
Nevertheless, we propose a tentative explanation of
the main features in Fig.~\ref{fig:sz-E} as follows.

The results shown in Fig.~\ref{fig:s-r} suggest that
the NFIS's of both neutrinos and antineutrinos
roughly rotate around $\basef{z}$ with a frequency
$\omega$. In fact, in the limit that $A \ll \Delta$, the NFIS's
in the bi-polar mode rotate around the vacuum field
$\HV = \basev{z} \equiv -\basef{x}\sin2\theta + \basef{z}\cos2\theta$
[see Eq.~\eqref{eq:HV}],
which is close to $\basef{z}$ if $\theta\ll 1$.

To give a rough feel for the behavior of such a system,
let us study a toy scenario where $\sB(E_\nu)$ is coupled
to both $\HV$ and another field $\bm{\Sigma}(t)$, which rotates
in the plane perpendicular to $\HV$.
Thus the equation of motion for $\sB(E_\nu)$ can be written as
\begin{equation}
\begin{split}
\frac{\ud}{\ud t} \sB(E_\nu) &= \sB(E_\nu) \times
[\muV{}(E_\nu)\HV  \\
&+\mu_\nu\Sigma(t)(\basev{x}\cos\omega t
-\basev{y}\sin\omega t)],
\end{split}
\label{eq:rot-field}
\end{equation}
where $\mu_\nu$ is a coefficient, and
$\basev{x}$, $\basev{y}$ and $\basev{z}$ are a set of
 orthogonal unit vectors  with $\basev{x}\times\basev{y}=\basev{z}$.
Suppose that at $t=0$, we have $|\mu_\nu\Sigma(t=0)|\gg|\muV{}(E_\nu)\HV|$
and $\sB(E_\nu)$ is aligned or antialigned with $\bm{\Sigma}(t=0)$.
We want to find  the configuration of $\sB(E_\nu)$ as
$\Sigma(t)$ slowly decreases toward zero. Eq.~\eqref{eq:rot-field}
turns out to be very simple in a corotating frame in
which $\bm{\Sigma}(t)$ is fixed in one direction, say ${\bf \hat{w}}$.
(Ref.~\cite{Duan:2005cp} points out the utility of the corotating frame.)
The equation of motion of  $\sB(E_\nu)$ in this corotating frame is
\begin{subequations}
\label{eq:eom-rot-frame}
\begin{align}
\frac{\ud}{\ud t} \tilde{\sB}(E) 
&= \tilde{\sB}(E_\nu) \times \widetilde{\Hb}^\eff\\
&= \tilde{\sB}(E_\nu) \times [(\muV{}(E_\nu)-\omega)\HV + 
\mu_\nu\Sigma(t){\bf \hat{w}}],
\end{align}
\end{subequations}
where a vector with a tilde symbol is the same as that without
but viewed in the corotating frame.
As $\Sigma(t)$ decreases, $\widetilde{\Hb}^\eff$
rotates from the direction of $\mu_\nu{\bf \hat{w}}$ to 
that of $(\muV{}(E_\nu)-\omega)\HV$.
If this process is slow enough, $\tilde{\sB}(E_\nu)$ stays
aligned or antialigned with $\widetilde{\Hb}^\eff$,
depending on the initial conditions, and will be either
aligned or antialigned with $\HV$ when $\Sigma$ approaches zero. We define
\begin{equation}
\varpi(E_\nu) \equiv \mu_\nu [\muV{}(E_\nu)-\omega] 
[\sB(E_\nu)\cdot\bm{\Sigma}]_{t=0}.
\label{eq:varpi}
\end{equation}
One can check that $\tilde{\sB}(E_\nu)$, and therefore $\sB(E_\nu)$, 
will be aligned with $\HV$ as
$t\rightarrow\infty$ if $\varpi>0$, and will be
antialigned with $\HV$ if $\varpi<0$. 
There can be a sharp transition in the orientation of $\sB$
at energy $E_\nu=\EC$, where $\muV{}(\EC)=\omega$.
The general features of this toy problem are shown in Fig.~\ref{fig:rot}.

This  analysis applies to collective neutrino flavor 
transformation in the hot bubble if (1) neutrinos and antineutrinos
are in the collective mode even in the region where $A\ll\Delta$,
and (2) the frequency of rotating NFIS's varies significantly more slowly
than the neutrino density $n_\nu$. In this case, $\bm{\Sigma}(t)$
corresponds to the rotating total NFIS, which decays as
the neutrino density goes down with increasing radius.
Because $\nuI{e}$
dominates in number over other neutrinos and antineutrinos,
the factor $[\sB(E_\nu)\cdot\bm{\Sigma}]_{t=0}$ in Eq.~\eqref{eq:varpi}
is essentially the scalar product of the NFIS of the neutrino
in question and that of the total $\sB_{\nuI{e}}$, which is
positive for $\nuI{e}$ and negative for $\anuI{e}$.
For the normal mass hierarchy ($\delta m^2>0$), one
has $\omega >0$ [note this behavior in Fig.~\ref{fig:s-r}(a) and (b)].
Noticing that $\mu_\nu<0$ [Eq.~\eqref{eq:mu-nu}], one finds that
$\varpi$ is always negative for $\sB_{\anuI{e}}(E_{\anuI{e}})$, which will be
antialigned with $\HV\simeq\basef{z}$ in the end, as
we have seen in Fig.~\ref{fig:sz-E} (b).
One has $\varpi < 0$ for $\sB_{\nuI{e}}(E_{\nuI{e}})$ if $E_{\nuI{e}}<\EC$ and 
$\varpi > 0$  if $E_{\nuI{e}}>\EC$, where
\begin{equation}
\EC = \left|\frac{\delta m^2}{2\omega}\right|.
\label{eq:EC-omega}
\end{equation}
We see that $\varsigma_{\nuI{e}z}$ is either approximately $-1/2$ 
or $+1/2$, depending on whether $E_{\nuI{e}}$ is less than
or greater than $\EC$. This behavior can be seen in Fig.~\ref{fig:sz-E} (a). 
For the inverted mass hierarchy
($\delta m^2<0$), one has $\omega<0$
[see the small insets in Fig.~\ref{fig:s-r}(c) and (d)].
One finds that $\varpi$ is always positive for $\sB_{\anuI{e}}(E_{\anuI{e}})$,
and will be roughly aligned with $\basef{z}$ in the end,
as we see in Fig.~\ref{fig:sz-E} (d). 
For $\sB_{\nuI{e}}(E_{\nuI{e}})$, one has $\varpi > 0$ 
if $E_{\nuI{e}}<\EC$, and $\varpi < 0$  if $E_{\nuI{e}}>\EC$. Therefore, the
corresponding $\varsigma_{\nuI{e}z}(E_{\nuI{e}})$ transitions from $+1/2$
to $-1/2$ as $E_{\nuI{e}}$ increases and crosses $\EC$,
as we see in Fig.~\ref{fig:sz-E} (c). 

The above reasoning is, however, based on an idealized case.
In  reality, some NFIS's of high energy
may never be locked
into a collective bi-polar mode with other NFIS's under some
conditions. Some NFIS's of moderate energy
may start to peel away from the bi-polar mode
in the  region where the matter potential $A$ is comparable
to $\Delta$. In addition, some NFIS's of high energy may go
 through the conventional MSW conversion after the collective
mode breaks down. Our guess is that
$\nu_e$ and $\bar\nu_e$ of energy $E_\nu>\EH$ in 
Fig.~\ref{fig:sz-E} (a) and (d) have $A\gtrsim\Delta$ when
the collective mode breaks down, and they are 
at least partially converted through the MSW mechanism. The $\bar\nu_e$
with $E_\nu<\EL$ in Fig.~\ref{fig:sz-E} (d) never enter the
bi-polar mode, and are converted to $\bar\nu_\tau$ through
the MSW or synchronization mechanisms. The $\bar\nu_e$
with energies between $\EL$ and $\EM$ may have 
complicated flavor evolution histories which quite early
cease to follow the bi-polar mode.

Our argument becomes more accurate at high neutrino
luminosity. With larger $L_\nu$, more low-energy neutrinos
and antineutrinos join the bi-polar flavor transformation,
and more of them are locked into this collective mode until
$A\lesssim\Delta$. As a result, the threshold energies $\EL$
and $\EM$ decrease, and $\EH$ increases as $L_\nu$ goes up.
This is indeed the case as one can see from the comparison
of the simulations with $L_\nu=10^{51}$ and $5\times 10^{51}$ erg/s
(Fig.~\ref{fig:sz-E}).

We have assumed $\omega$ to be a constant in our idealized
analysis. This is not the case in reality. From Fig.~\ref{fig:s-r}
one sees that $|\omega|$ slowly decreases with radius. 
If $L_\nu$ is large enough, neutrinos and antineutrinos 
will be in the bi-polar mode
even at values of radius where the matter potential $A(r)$ is negligible.
We expect $\omega$ to be a function of $\delta m^2$, $\theta$, $f_\nu(E_\nu)$
and the local neutrino density $n_\nu^\eff(L_\nu, r)$,
but to be independent of $S$, $Y_e$, \textit{etc.}.
We note that neutrinos and antineutrinos start to deviate from
the collective mode behavior at some radius $\rC$
as $\tilde{\sB}$ adiabatically rotates away from the direction
of $\mu_\nu\widetilde{\bm{\Sigma}}$.
Further, we note that the value of $\EC$ should be determined from $\omega(\rC)$
using Eq.~\eqref{eq:EC-omega}.
One can attempt to estimate $\rC$ ($\lesssim\rBE$) from
Eqs.~\eqref{eq:bipolar-cond} and \eqref{eq:kappa-def}
directly, resulting in the condition
\begin{equation}
\kappa(\rBE)=\frac{|\delta m^2|}{4\sqrt{2}\GF}
\frac{1}{n_\nu^\eff(L_\nu, \rBE) \langle E_\nu\rangle} \simeq
\frac{\langle E_\nu\rangle}{2\delta E_\nu}.
\label{eq:n-nu-rBE}
\end{equation}
The value of $\rBE$ derived from Eq.~\eqref{eq:n-nu-rBE}
is an overestimate of $\rC$. We have seen in Fig.~\ref{fig:s-r}
that collective flavor transformation ceases at $r\gtrsim\rBE$.
However, at $r\simeq\rC$, all the NFIS's  begin to slightly
deviate from alignment, but are more or less still following the
 collective mode.
Nevertheless, we expect $\kappa(\rC)$, like $\kappa(\rBE)$, to be
 determined by $f_\nu(E_\nu)$ only. 
As a result, $n_\nu^\eff(L_\nu, \rC)$,
and thus $\omega(\rC)$ and $\EC$, are actually independent
of $L_\nu$, if $L_\nu$ is large enough.

We have calculated the energy spectra
of neutrinos at $r=400$ km using the single-angle approximation
with $S=140$ and 250, and $L_\nu=10^{51}$ and $5\times 10^{51}$
erg/s. The values of $\EC$ in most of the cases
agree well with each other for the same neutrino mass hierarchy.
The value of $\EC$ in the case with $S=140$ and $L_\nu=10^{51}$ erg/s 
is different from those in the other three cases for the normal mass hierarchy
[see, \textit{e.g.}, Fig.~\ref{fig:sz-E}(a)] because $L_\nu$ is
not large enough, or equivalently, the baryon density profile is
not sufficiently condensed toward the surface of the neutron star.
We also note that
$\varsigma_{\nuI{e}z}(E_{\nuI{e}})$ is not a strict step function,
but has a transition region of finite width.
The transition region in the normal mass hierarchy scenario
overlaps with that in the inverted mass hierarchy scenario,
which seems to suggest that the values of $|\omega(\rC)|$ are at least
similar in these two cases.

\subsection{Onset of Collective Neutrino Flavor Transformation%
\label{sec:start-collective}}

\begin{figure*}
\begin{center}
$\begin{array}{@{}c@{\hspace{\myfigsep}}c@{}}
\includegraphics*[width=\myfigwid, keepaspectratio]{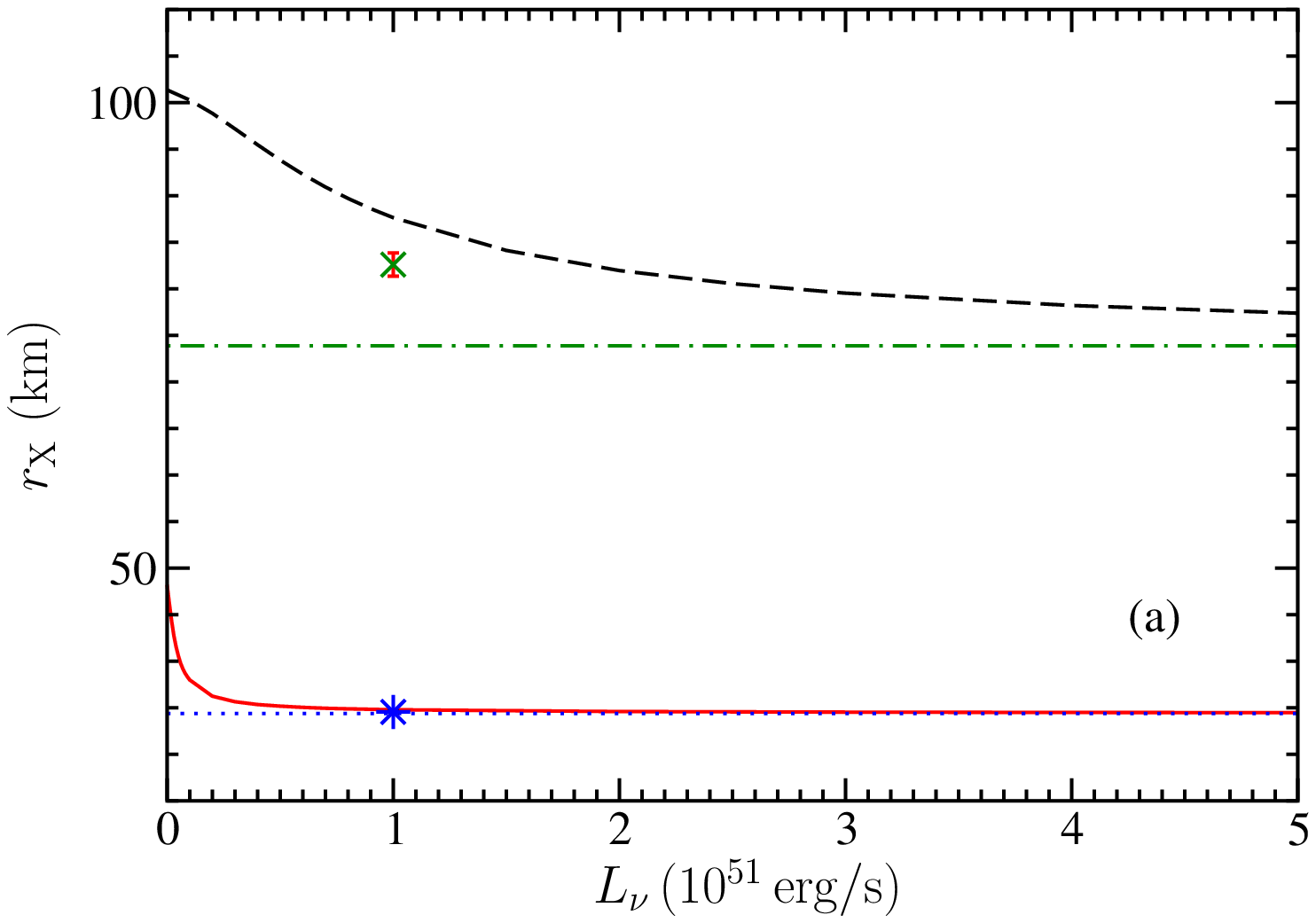} &
\includegraphics*[width=\myfigwid, keepaspectratio]{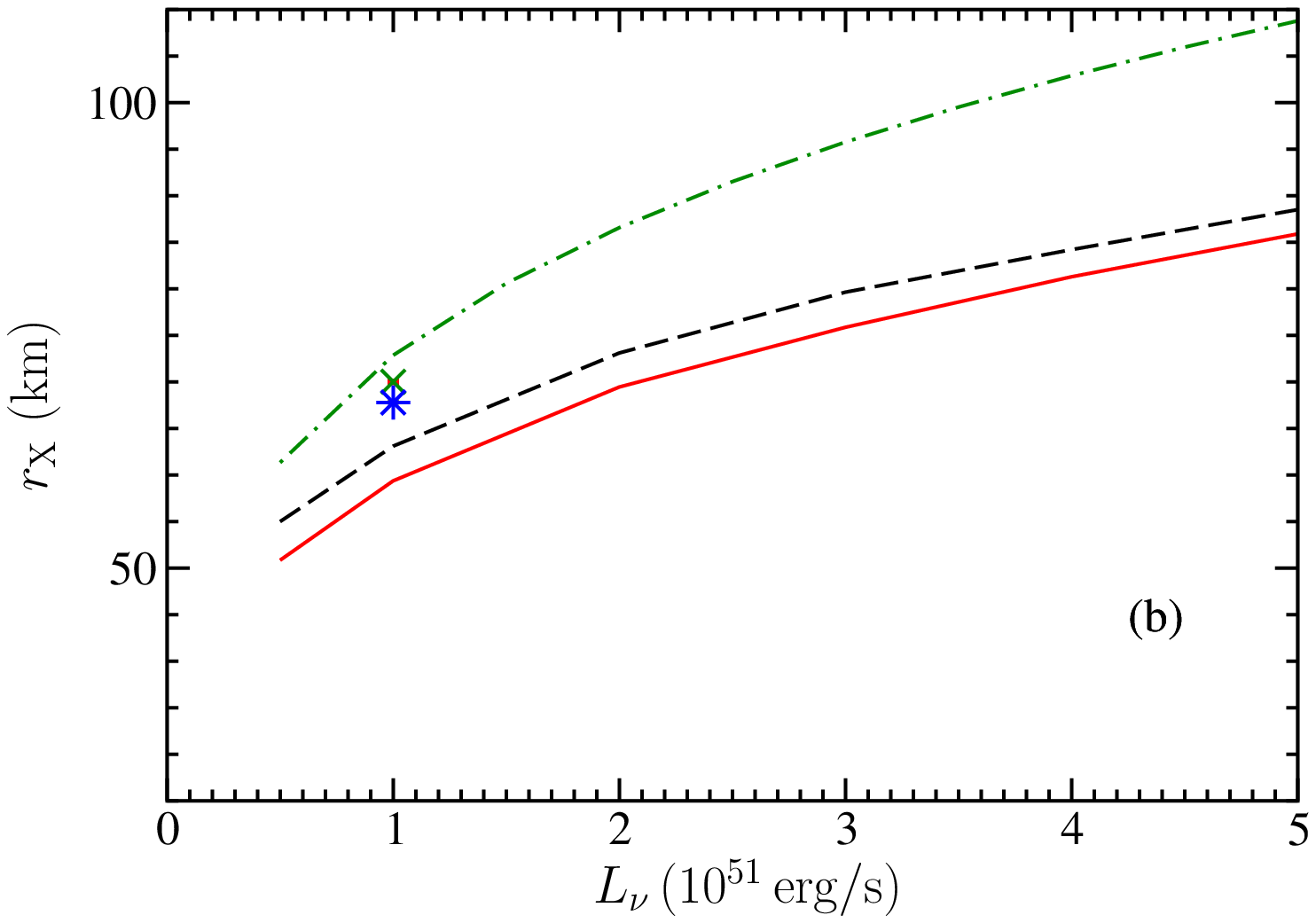}
\end{array}$
\end{center}
\caption{\label{fig:rX-L}(Color online) 
Plots of $\rX(L_\nu)$ for both the normal
[panel (a)] and inverted [panel (b)] neutrino mass hierarchies. The dashed
and solid lines are based on  single-angle simulations with 
$S=140$ and 250, respectively. The cross and star symbols 
are based on the average of $\rX$ on the radial and tangential
trajectories in our multiangle simulations with $S=140$ and 250,
respectively. The error bars associated with the cross and star symbols
indicate the range of  values of $\rX$ on different trajectories.
These are too small to be visible in most of the cases. 
For the normal mass hierarchy case [panel (a)], 
$\rX$ asymptotically approaches $\rXo(\Esync)$ with large $L_\nu$, 
which is 73.9 (dot-dashed lines) and 34.4 km (dotted lines)
for $S=140$ and 250, respectively.
The dot-dashed line in panel (b) represents a crude estimate for where
the collective neutrino flavor transformation exits the synchronization
mode and enters the bi-polar mode [Eq.~\eqref{eq:n-nu-rBS}].}
\end{figure*}

The radius where significant neutrino flavor transformation starts can be very
important for nucleosynthesis and for estimates of the expected
late-time neutrino signal \cite{Schirato:2002tg,Tomas:2004gr,Yoshida:2005uy}.
We define $\rX$ as the radius
where $\langle \Pee\rangle$ falls just below
0.9. In Fig.~\ref{fig:rX-L}(a),
we plot $\rX(L_\nu)$ for the cases with $S=140$
and 250 in the normal mass hierarchy scenario
based on our single-angle and multiangle simulations.
For both  entropy values, $\rX(L_\nu)$ in single-angle simulations
monotonically decreases as $L_\nu$ increases.
As a comparison, we also plot the corresponding values of
$\rXo(\Esync)$ in Fig.~\ref{fig:rX-L}(a). Here $\Esync\simeq 2.47$
MeV is the characteristic neutrino energy for the full synchronization mode,
and $\rXo(E_\nu)$ is the radius where a
$\nuI{e}$ with energy $E_\nu$  has $\Pee=0.9$
in the standard MSW mechanism. 
One sees that the values of $\rX(L_\nu)$
asymptotically approach $\rXo(\Esync)$.
This is not a surprise. According to
Eq.~\eqref{eq:bipolar-cond}, neutrinos are in the synchronization
mode if $n_\nu^\eff(L_\nu, r)$ is large. In turn, 
$n_\nu^\eff(L_\nu, r)$ increases with 
increasing $L_\nu$ at a fixed radius $r$. As $L_\nu$ increases,
more and more low-energy neutrinos and antineutrinos are locked
into the synchronization mode, and the characteristic neutrino energy
of the synchronization mode
decreases and asymptotically approaches $\Esync$.
One also sees that for the same $L_\nu$, the radius $\rX(L_\nu)$ 
is much closer to $\rXo(\Esync)$ 
in the $S=250$ case than in the case with
$S=140$. This is because with larger $S$
the baryon density profile is more condensed toward the 
neutrino sphere, and $\rX(L_\nu)$, like $\rXo(\Esync)$,
 is smaller in this case. Therefore,
$n_\nu^\eff(L_\nu, \rX)$ is larger with a larger $S$ but the same $L_\nu$,
and the synchronization is more complete.

It is interesting to see that the values of $\rX(L_\nu)$ obtained from
the multiangle simulations all fall between those from the single-angle
simulations and $\rXo(\Esync)$. 
Comparing Eq.~\eqref{eq:Hnunu-multiangle}
with \eqref{eq:Hnunu-single-angle}, one can see that the single-angle
approximation uses $n_\nu^\eff(L_\nu, r)$ on the radial trajectory.
Note that $n_\nu^\eff(L_\nu, r)$ 
has smaller values on the radial trajectory than 
it does on any other trajectory.
On average, values of $n_\nu^\eff(L_\nu, r)$ are larger in the multiangle
simulations than in the single-angle ones at the same radius $r$.
At the same time, the full synchronization mode obtains
when $n_\nu^\eff(L_\nu, r)\rightarrow\infty$.
Thus $\rX(L_\nu)$ computed from single-angle calculations
gives upper bounds on the actual $\rX(L_\nu)$, and
$\rXo(\Esync)$ gives a lower bound on this quantity.

In Fig.~\ref{fig:rX-L}(b), we plot the numerical values 
of $\rX(L_\nu)$ in our single-angle
and multiangle simulations for the cases with $S=140$
and 250, respectively, and employing the inverted mass hierarchy.
One sees that the values of $\rX(L_\nu)$
monotonically increase with $L_\nu$.
In addition, they are not very sensitive to the value of $S$.
To explain this phenomenon,
we note that $\rBS$, the radius where the neutrinos exit
the synchronization mode and enter the bi-polar mode, can be estimated 
from the condition
\begin{equation}
\frac{|\delta m^2|}{4\sqrt{2}\GF}
\frac{1}{n_\nu^\eff(L_\nu, \rBS) \langle E_\nu\rangle} \simeq
\frac{(\langle E_{\nuI{e}}\rangle -\langle E_{\anuI{e}}\rangle)^2}%
{2(\langle E_{\nuI{e}}\rangle^2+\langle E_{\anuI{e}}\rangle^2)}
\label{eq:n-nu-rBS}
\end{equation}
[see Eqs.~\eqref{eq:bipolar-cond}, \eqref{eq:kappa-def} and
\eqref{eq:epsilon}]. Clearly $n_\nu^\eff(L_\nu, \rBS)$
depends only on $\delta m^2$, $\theta$ and $f_\nu(E_\nu)$.
Once these parameters are specified, $n_\nu^\eff(L_\nu, \rBS)$ is fixed.
As a result, $\rBS(L_\nu)$ must increase with $L_\nu$ 
for a fixed neutrino density $n_\nu^\eff(L_\nu, \rBS)$.
In addition, $\rBS(L_\nu)$ depends only on $\delta m^2$, $\theta$,
$f_\nu(E_\nu)$ and $L_\nu$, and is independent
of $S$, $Y_e$, \textit{etc.}.
We plot the estimated values of $\rBS(L_\nu)$ 
determined from Eq.~\eqref{eq:n-nu-rBS} in Fig.~\ref{fig:rX-L}(b).
These indeed increase with $L_\nu$. However,
the estimated values of $\rBS(L_\nu)$ are always larger than $\rX(L_\nu)$.
This is because we have made many simplifications in
deriving Eq.~\eqref{eq:n-nu-rBS}. In particular,
we have assumed that the ``magnetic moments'' of the NFIS's
in the opposite directions are $\delta m^2/2\langle E_{\nuI{e}}\rangle$
and $\delta m^2/2\langle E_{\anuI{e}}\rangle$. This is
a very crude approximation. According to Ref.~\cite{Duan:2005cp},
the flavor conversion of neutrinos and antineutrinos in the bi-polar mode
is suppressed very little by the matter potential
in the scenario with the inverted mass hierarchy.
This is contrary to a contemporary false belief that a
large matter potential always strongly suppresses 
neutrino flavor transformation. Therefore, $\rX(L_\nu)$ should
roughly trace the actual values of $\rBS(L_\nu)$,
\begin{equation}
\rX(L_\nu) \simeq \rBS(L_\nu).
\end{equation}
We conclude that, like $\rBS(L_\nu)$, $\rX(L_\nu)$ with the inverted 
mass hierarchy has little dependence on
$S$ or $Y_e$, and increases monotonically with $L_\nu$.
Because the single-angle approximation uses the smallest value
of $n_\nu^\eff$ among all trajectories, it gives lower bounds
on the actual values of $\rX(L_\nu)$. This is clear in Fig.~\ref{fig:rX-L}(b).

We note that $\rBS(L_\nu)$ is the same for both the normal and
inverted mass hierarchies. Using this information, we can
estimate whether significant neutrino flavor transformation for the
normal mass hierarchy case begins in the bi-polar mode or not.
Comparing panel (a) with 
panel (b) in Fig.~\ref{fig:rX-L},  we note that 
for the normal mass hierarchy, neutrinos and antineutrinos
start flavor transformation in the synchronization mode for $S=250$.
They begin flavor transformation through the bi-polar mode
for $S=140$ when $L_\nu$ is less than a few times $10^{51}$ erg/s.

\begin{figure}
\begin{center}
\includegraphics*[width=\myfigwid, keepaspectratio]{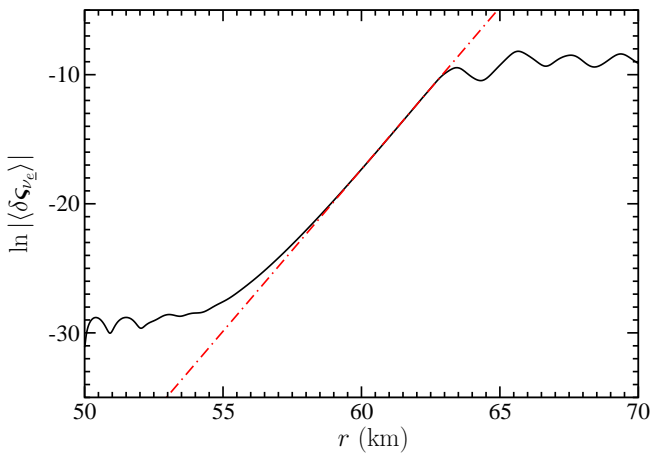}
\end{center}
\caption{\label{fig:chaos} (Color online) 
The difference between two almost identical systems grows
exponentially in the region where collective
neutrino flavor transformation changes from the synchronized
mode to the bi-polar mode in the inverted neutrino mass hierarchy case.
The solid line shows the exponential growth of
$|\langle\delta\sB_{\nuI{e}}\rangle|$ (difference between the
energy-averaged $\nuI{e}$ NFIS's of the two systems) as a function of radius 
in the transition region. The dot-dashed line is a linear fit 
to $\ln|\langle\delta\sB_{\nuI{e}}(r)\rangle|$. 
This line has slope $\sim 2.5/\mathrm{km}$.}
\end{figure}

For the inverted neutrino mass hierarchy, our simulations show
that chaotic behavior in neutrino flavor transformation
can occur in the narrow region of radius
where the collective behavior transitions from the synchronized
to the bi-polar mode. To study this, we manually added random
perturbations of order $~10^{-12}$
in $\sB_{\nuI{e}}(E)$ at radius $r=50$ km (the region
just below the synchronized-to-bi-polar transition for
the particular case with $S=140$ and $L_\nu=10^{51}$ erg/s).
We follow the evolution of $\delta \sB_\nu(E)$, the difference
between the NFIS's of neutrino $\nu$ with energy $E$
in the perturbed and unperturbed cases. We find that
$|\delta \sB_\nu(E)|$ for neutrinos of all species and energies 
grows with the same exponential factor in the transition region 
\begin{equation}
|\delta\sB_\nu(E)| \propto \exp\left(2.5\frac{r}{\mathrm{km}}\right).
\end{equation}
In Fig.~\ref{fig:chaos} we plot $\ln|\langle\delta\sB_{\nuI{e}}\rangle|$
as a function of radius $r$. Note that the difference 
$|\langle\delta\sB_{\nuI{e}}\rangle|$
between the energy-averaged $\nuI{e}$ NFIS's of the 
perturbed and unperturbed cases
is $\sim5\times 10^8$ times larger
at the radius where the system is fully in the
bi-polar mode than in the region
 before the synchronized-to-bi-polar transition.
This chaotic behavior obviously causes difficulty
in accurately simulating neutrino flavor transformation. However,
we have performed several computations with different numerical
schemes, all of which show qualitatively similar results. Therefore our 
analysis and conclusions are not affected by this behavior.
At this point we do not know whether this behavior reflects true
chaos or the appearance of a critical point in the neutrino/antineutrino
system.

It is appropriate to comment on the validity of the single-angle
approximation at this point. The traditional single-angle
approximation picks the radial trajectory as the representative
trajectory, which turns out to have the smallest $n_\nu^\eff$. As a result,
the neutrino background effect tends to be underestimated. A slightly
better approximation would be to average over all trajectories.
This ``averaged'' case would have changed the geometric factor
$\geoD(r/R_\nu)$ in Eq.~\eqref{eq:Psi} to
\begin{equation}
\begin{split}
\geoD(r/R_\nu) &= \frac{1}{2}
\left[1-\sqrt{1-\left(\frac{R_\nu}{r}\right)^2}\right]\\
&\quad\times
\left[1-\sqrt{1-\left(\frac{R_\nu}{r}\right)^2}+\frac{1}{2}
\left(\frac{R_\nu}{r}\right)^2\right].
\end{split}
\tag{\ref{eq:Psi}$^\prime$}
\end{equation}
Even with this improvement, we would not expect to be able to simulate
 the complicated entanglement of neutrino flavor
transformation among different trajectories.
The single-angle approximation is
only accurate when neutrinos on different trajectories have
the same flavor transformation histories. 
This does not seem to be the case for the
bi-polar  collective transformation.
As a result, it will be necessary to use multiangle simulations
 to accurately gauge, \textit{e.g.}, the effect of
neutrino flavor transformation on $Y_e$.
Nevertheless, as we have demonstrated,
the numerical simulations using the single-angle approximation
are very useful as a means of  exploring the basic physics of
neutrino flavor transformation in the hot bubble. These
simple models do provide
simple checks on more complex and computationally-intensive
simulations.

\section{Conclusions%
\label{sec:conclusions}}

We have carried out large-scale multiangle simulations of neutrino flavor
transformation in the hot bubble employing the atmospheric neutrino 
mass-squared difference, $|\delta m^2|\simeq 3\times 10^{-3}\,\mathrm{eV}^2$,
and effective $2\times 2$ vacuum mixing angle $\theta=0.1$.
The numerical results we have presented support previous conjecture
on the existence of collective neutrino flavor transformation
of the bi-polar type in the supernova environment
\cite{Duan:2005cp}. Our simulations also show that both neutrinos
and antineutrinos can simultaneously undergo significant flavor conversion,
largely driven by flavor off-diagonal potentials, at 
values of radius much smaller than those expected 
from ordinary MSW. This is along the lines of what was
predicted in Ref.~\cite{Fuller:2005ae}.
We have found that this flavor transformation 
occurs in both the normal and inverted
neutrino mass hierarchy scenarios. 

For the normal mass hierarchy
case, the full synchronization limit gives a lower bound
on the radius where large-scale neutrino flavor transformation begins.
(Ref.~\cite{Pastor:2002we} was the first to point out that, 
since $\Esync$ is
much smaller than average neutrino energies, synchronized
flavor transformation modes operate closer to the neutrino sphere
than those driven by the MSW matter potential.)
Although an analytical analysis of neutrino flavor transformation
in the bi-polar mode has yet to be done, our numerical simulations
suggest that, for the normal mass hierarchy, the onset of bi-polar
type flavor transformation always occurs at 
values of radius larger
than those required in the full synchronization case. 
Our simulations also support
the prediction of large-scale neutrino flavor transformation in the inverted
mass hierarchy scenario \cite{Duan:2005cp}.
(Large-scale neutrino/antineutrino flavor transformation
with small mixing angles
in the inverted mass hierarchy case previously was seen in 
the early universe context \cite{Kostelecky:1993dm} and also
in the supernova context  \cite{Pastor:2002we}.)
We have found that this may occur at values of radius
even smaller than those seen in the full synchronization mode in the normal
mass hierarchy scenario. We have found that
single-angle simulations can be used to give a lower bound
on the radius where large-scale neutrino flavor
transformation occurs in  the inverted mass hierarchy scenario.

Our ``multiangle'' calculations are the first to include self-consistent
flavor evolution history entanglement on intersecting neutrino world
lines. Although we find that ``single-angle'' simulations in some
cases can give the correct \textit{qualitative} features of
large-scale neutrino and/or antineutrino conversion in the
late-time, hot bubble region, our simulations clearly show that
a \textit{quantitatively} correct treatment must include coupled
flavor development on different neutrino trajectories.
Furthermore, since the location where large-scale neutrino
and/or antineutrino flavor transformation begins in the supernova
envelope can be a crucial issue for supernova shock reheating
\cite{Fuller:1992aa}, \textit{r}-process nucleosynthesis
\cite{Qian:1993dg,Qian:1994wh,Sigl:1994hc,Qian:1995ua,Pastor:2001iu,Balantekin:2004ug},
and the supernova neutrino signal
\cite{Schirato:2002tg,Tomas:2004gr,Yoshida:2005uy}, it is
essential that simulations \textit{be} quantitatively as
accurate as possible.

The simulations we have presented focus on the late-time
supernova  environment, \textit{i.e.}, the regime
after the shock has been somehow re-energized. This
epoch is a leading candidate for the site of the
production of some or all of the \textit{r}-process
elements and will be a major focus of future neutrino
detectors/observatories should we be lucky enough
to catch a galactic core collapse event. Though our
simulations show that large-scale neutrino and antineutrino
flavor conversion can take place during this epoch for the
expected conditions of neutrino flux and entropy,
we must go further than we have in this paper to
produce quantitative predictions. There are three
principal reasons for this: (1) we do not as yet know
the matter density distribution above the proto-neutron
star to sufficient accuracy at any epoch; (2) we do not
know precisely the neutrino and antineutrino energy distributions and 
fluxes which are emergent from the proto-neutron star;
and (3) the matter composition (\textit{i.e.}, $Y_e$)
can be affected by any changes in the neutrino and antineutrino
spectra engendered by flavor transformation and we have not
put this feedback in the calculations  presented here.

On point (2), recent work on supernova models at
$\tpb < 1$ s suggests that additional channels for neutrino
scattering may weaken or dilute the effects of the charged
current opacities \cite{Keil:2002in,Raffelt:2003en}. This would
tend to make the $\nu_e$, $\bar\nu_e$, and 
$\nu_\mu\bar\nu_\mu\nu_\tau\bar\nu_\tau$ energy spectra more similar.
Of course, if the neutrino energy spectra and fluxes are identical
for all flavors, interconversion of these will have no astrophysical
effect. We note, however, that reliable neutrino transport
calculations at the late time we considered here do not exist,
and the core's composition and neutron excess is expected to change
considerably between $\tpb\simeq 1$ s and $\tpb\simeq 10$ s. Clearly,
this issue is critical for gauging the astrophysical effect
of neutrino flavor mixing.

It is well known that density fluctuations on short length scales and
other inhomogeneities can modify coherent neutrino flavor
evolution through MSW resonances
\cite{Sawyer:1990tw,Loreti:1995ae}. How these fluctuation-induced
modifications could manifest themselves in quantum flavor history
entanglement on intersecting neutrino trajectories is not known.
This issue may be closely related to the problem of calculating
neutrino transport and predicting the emergent neutrino energy spectra
in general \cite{Liebendoerfer:2002xn,Thompson:2002mw,Liebendoerfer:2003es,%
Mezzacappa:2004iz,Walder:2004ym} 
and to the inclusion of neutrino flavor mixing in the
core in particular \cite{Sawyer:2005jk,Strack:2005ux}. Though
our simulations are spherically symmetric, they do show that
the density and $Y_e$ profiles near the proto-neutron star surface
are important for obtaining the correct flavor evolution of the neutrino and
antineutrino fields, even well above the proto-neutron star.

These uncertainties aside, our calculations indicate that large 
modifications of the emergent neutrino and antineutrino energy spectra are
likely to occur over most of the range of expected thermodynamic and
neutrino emission parameters of relevance in the late-time supernova
environment. Furthermore, we have found that these modifications
could set in sufficiently deep in the supernova envelope to affect
$Y_e$ \cite{Qian:1993dg} and 
\textit{r}-process nucleosynthesis \cite{Fuller:1995ih,Qian:1996db}
through neutrino interactions.
However, we have not included charged-current weak interaction 
[Eq.~\eqref{eq:nu-interactions}] feedback in
the calculations presented here.

We have fixed  $Y_e$ and $g_{\mathrm{s}}$ in this work,
essentially to simplify the computations.
In future simulations we will remove these constraints and allow 
$Y_e$ and $g_{\mathrm{s}}$
to be calculated consistently with feedback from neutrino capture
reactions. However, we expect that the collective neutrino flavor 
transformation illustrated here will not be changed 
qualitatively with changing $Y_e$ and 
$g_{\mathrm{s}}$. The bi-polar neutrino flavor
transformation seen in our simulations is largely independent of the values
of $Y_e$ and $g_s$. For example, we have shown
that $\nu_e$ of energy smaller (larger) than
a critical energy $\EC$ could convert to other flavors if $\delta m^2 > 0$
($\delta m^2 < 0$). This critical energy $\EC$ asymptotically
approaches a limit
if $L_\nu$ is large enough, or equivalently, the electron density profile
is sufficiently 
condensed toward the proto-neutron star. The asymptotic limit of $\EC$
depends only on the neutrino mixing parameters and 
the initial energy spectra for neutrinos and antineutrinos.

Because the proto-neutron star
is neutron-rich, the initial $\nu_e$ energy spectrum may be softer
than those for neutrinos in other flavors. 
Our simulations suggest that $\nu_e$ and neutrinos in other flavors may 
swap the low-energy ($E_\nu<\EC$)  or high-energy ($E_\nu>\EC$)  parts 
of their spectra 
depending on the sign of $\delta m_{13}^2$. 
Note that this stepwise swapping is independent of the details of
the neutrino energy spectra. With other effects
correctly accounted for and a good signal from a galactic supernova, 
this phenomenon may
offer a unique probe of the neutrino mass hierarchy problem.

We have employed $2\times 2$ neutrino flavor mixing in our simulations.
It is possible to extend our codes to implement the neutrino mixing of
all three active flavors. However, we expect that neutrino
flavor transformation in the hot bubble region will not change
 much on inclusion of a third neutrino flavor. For one thing,
$\nu_\mu$ and $\nu_\tau$ are almost equally mixed in the hot bubble
because they experience the same weak interactions
 and $\theta_{23}\simeq \pi/4$.
For another, the two neutrino mass-squared differences,
$\delta m_{\mathrm{atm}}^2$ and $\delta m_\odot^2$, are
separated by over an order of magnitude. Taking 
$\delta m^2=\delta m_\odot^2\simeq 8\times 10^{-5}\,\mathrm{eV}^2$ and
$\theta = \theta_{12} \simeq 0.6$, we estimate that
the onset radius of large-scale neutrino flavor transformation in
the full synchronization limit is $\rXo(\Esync)\simeq 227$ km
for an entropy per baryon $S=140$. This location is almost outside the range
of  the simulation results presented in Sec.~\ref{sec:results}.

In summary, though many aspects of our calculations are
reasonable approximations 
at best (\textit{e.g.}, $2\times 2$ mixing, assumptions
of spherical symmetry, an infinitely thin neutrino sphere,
neutrino/antineutrino energy spectra of the Fermi-Dirac type, \textit{etc.}),
our computations do mark an important advance in that they self-consistently 
treat coupled neutrino flavor evolution on different trajectories.
We cannot claim generality for our conclusions. However, our
assumptions are reasonable, and our results are robust, and
so there is nothing to suggest that our results represent
an isolated case either.
Not only do our results 
show that a proper treatment of coupled neutrino trajectories is important,
but they also indicate that the measured neutrino mass-squared difference values
and mixing angles likely imply large-scale flavor conversion of neutrinos and
antineutrinos in astrophysically important regions in the post-explosion
supernova environment.

\begin{acknowledgments}
This work was supported in part by a UC/LANL CARE grant,
NSF grant PHY-04-00359,
the Terascale Supernova Initiative (TSI) collaboration's 
DOE SciDAC grant at UCSD, and 
DOE grant DE-FG02-87ER40328 at UMN.
This work was also supported in part by the LDRD Program
and Open Supercomputing at LANL, and by
the National Energy Research Scientific Computing Center through
the TSI collaboration using Bassi, and the San Diego Supercomputer 
Center through the Academic Associates Program using DataStar. 
We would like to thank A.~B.~Balantekin, S.~Bruenn, 
C.~Y.~Cardall, J.~Hayes,
W.~Landry, O.~E.~B.~Messer, A.~Mezzacappa, M.~Patel, and H.~Y\"{u}ksel
for valuable conversations.
We would especially like to thank G.~Raffelt for a careful reading
of the manuscript and many useful comments.
\end{acknowledgments}

\bibliography{ref}

\end{document}